\documentclass{article}
\usepackage{graphicx, amsthm, amsmath, xcolor, amssymb, subfiles, subfigure, booktabs, doi, multirow, adjustbox, orcidlink, algpseudocode, algorithm}
\usepackage[round]{natbib}
\usepackage[left=2.5cm, right=2.5cm, top=2.5cm, bottom=2.5cm]{geometry}
\usepackage{newpxtext}
\usepackage[bb=dsserif]{mathalpha}
\usepackage{hyperref}
\hypersetup{
  colorlinks   = true, %Colours links instead of ugly boxes
  urlcolor     = blue, %Colour for external hyperlinks
  linkcolor    = blue, %Colour of internal links
  citecolor   = black %Colour of citations
}

\newcommand{\btheta}{\boldsymbol{\theta}}
\newcommand{\bdelta}{\boldsymbol{\delta}}

\newcommand{\E}{\mathbb{E}}
\newcommand{\Sk}{\mathcal{S}_k}
\newcommand{\pa}{\mathcal{I}}

\newcommand{\ba}{\,|\,}
\newcommand{\bba}{\,\big|\,}
\newcommand{\ind}{\mathbb{1}}
\newcommand{\dd}{\mathrm{d}}
\newcommand{\bx}{\boldsymbol{x}}
\newcommand{\bN}{\boldsymbol{N}}
\newcommand{\Tn}{\mathcal{T}_n}
\newcommand{\wpa}{\widehat{\mathcal{I}}}

\newcommand{\dhsq}{d_H^{\mspace{2mu}2}}
\newcommand{\bigO}{\mathcal{O}}
\newcommand\smallO{
  \mathchoice
    {{\scriptstyle\mathcal{O}}}% \displaystyle
    {{\scriptstyle\mathcal{O}}}% \textstyle
    {{\scriptscriptstyle\mathcal{O}}}% \scriptstyle
    {\scalebox{.45}{$\scriptscriptstyle\mathcal{O}$}}%\scriptscriptstyle
}
\newcommand{\proglang}[1]{\textsf{#1}}

\newtheorem{theorem}{Theorem}
\newtheorem{lemma}{Lemma}

{\theoremstyle{definition}
\newtheorem{definition}{Definition}}
\makeatletter
\renewenvironment{proof}[1][\proofname]{%
  \par\pushQED{\qed}%
  \normalfont % Body will be upright and non-bold
  \topsep6\p@\@plus6\p@\relax
  \trivlist
  \item[\hskip\labelsep \normalfont\bfseries #1\@addpunct{.}]\ignorespaces
}{%
  \popQED\endtrivlist\@endpefalse
}
\makeatother

{\theoremstyle{definition}
\newtheorem{example}{Example}}
{\theoremstyle{definition}

}

\AtBeginEnvironment{example}{%
  \pushQED{\qed}%
}
\AtEndEnvironment{example}{\popQED\endexample}

\AtBeginEnvironment{excont}{%
  \pushQED{\qed}%
}
\AtEndEnvironment{excont}{\popQED\endexample}

{\theoremstyle{definition}
\newtheorem{remark}{Remark}}

\AddToHook{cmd/appendix/before}{
  \setcounter{equation}{0}
  
  \setcounter{lemma}{0}
  
  \setcounter{corollary}{0}
  
  \renewcommand{\thesection}{Appendix \Alph{section}}
  \renewcommand{\thesubsection}{\Alph{section}.\arabic{subsection}}
}

\title{\vspace{-2.0cm}\textbf{Random irregular histograms}}
\author{
    Oskar H{\o}gberg Simensen$^{1,2}$\footnote{Corresponding author. Email: \texttt{oskar.simensen@kit.edu}}\ \ \orcidlink{0009-0009-5056-7324}\and
    Dennis Christensen$^3$~\orcidlink{0000-0002-7540-7695} \and
    Nils Lid Hjort$^2$ \\
    \small $^1$Scientific Computing Center, Karlruhe Institute of Technology\\ 
    \small $^2$Department of Mathematics, University of Oslo\\
    \small $^3$Norwegian Defence Research Establishment (FFI)
}
\date{}

\begin{document}

\maketitle

\begin{abstract}
    We propose a new method of histogram construction, providing a fully Bayesian approach to irregular histograms. Our procedure applies Bayesian model selection to a piecewise constant model of the underlying distribution, resulting in a method that selects both the number of bins as well as their location based on the data in a fully automatic fashion. We show that the histogram estimate is consistent with respect to the Hellinger metric under mild regularity conditions, and that it attains a convergence rate equal to the minimax rate (up to a logarithmic factor) for H\"{o}lder continuous densities. Simulation studies indicate that the new method performs comparably to other histogram procedures, both for minimizing the estimation error and for identifying modes. A software implementation is included as supplementary material.

    \vspace{0.2cm}
\end{abstract}

\section{Introduction}
The much celebrated histogram is the earliest example of a nonparametric density estimator, and remains widespread in use even to this day. Although more efficient density estimators have been devised since, histograms have retained their popularity due to their simple nature and interpretability. The key difficulty encountered when drawing a histogram is that the appearance of the density estimate is sensitive to the choice of partition used to draw the histogram. If the partition is not chosen well, the quality of the resulting histogram will be rather poor, failing to resemble the underlying distribution at all. As a result, the question of designing automatic histogram procedures, where the partition is chosen based on the given sample, has attracted considerable interest in the statistics community; see \citet{scott2010histogram} for a review. However, this problem has turned out to be a difficult one, with no solution universally accepted as the best one.

Many of the histogram procedures proposed in the statistical literature simplify the problem of selecting the partition somewhat by exclusively considering regular partitions \citep{birge2006bins}, where the bins are of equal width, so that one only needs to choose the number $k$ of bins. However, even this simplified problem has not led to a canonical procedure to choose the number of intervals in the partition, as evidenced by the sheer number of different bin selection rules for regular histograms \citep{li2020essential}. 

Automatic irregular histogram methods provide a more flexible alternative to regular histograms by determining both the number of bins and their location based on the sample. The benefits of using optimally chosen cut points between the intervals to draw a histogram as opposed to a regular grid with the same number of bins are rather immediate, as the bin widths are in the irregular case able to adapt to the local behavior of the underlying density, offering more smoothing near modes and in the tails of the density, resulting in smaller estimation risks \citep{scott1992multivariate}. Although the adaptive nature of irregular histograms is an attractive feature from a theoretical perspective, \citet{birge2006bins} remark that the search for an appropriate set of cut points can lead to increases in the statistical risk of the procedure, which may result in worse density estimates in practice according to classical loss functions.

Although automatic irregular histogram procedures do not always yield smaller estimation risks than regular ones, they often have an advantage when it comes to detecting important features in a density such as modes. Automatic mode identification is a feat that cannot be achieved by regular histogram procedures designed to produce small risks with respect to classical loss functions; as shown by \citet{scott1992multivariate}, the asymptotically optimal bin width in terms of $\mathbb{L}_2$ risk results in an undersmoothed histogram if the goal is to automatically infer the modes of a density.
In contrast, our proposed irregular histogram procedure shows that there need not be a trade-off between low estimation error and automatic mode identification.

Despite the apparent advantages of irregular histogram procedures, they have not been as widely adopted as their regular counterparts \citep{davies2009automatic}. This is in no doubt in part due to the computational challenges encountered by most irregular histogram methods, which typically involves solving a more difficult optimization problem than that encountered in the regular case. Moreover, many previous proposals for irregular histograms depend on the selection of key tuning parameters, with no universal recommendation given for their default values, which has hindered their use by practitioners \citep{rozenholc2010irregular}. However, in recent years, there has been a renewed interest in irregular histogram procedures, with many different proposals appearing in the statistical literature that offer automatic histogram construction at speeds that make them an appealing alternative to regular histogram methods; see e.g.~\citet{davies2004densities,rozenholc2010irregular,li2020essential,mendizabal2023fast}.

Parallel to the development of the irregular histogram literature, the topic of tree-based Bayesian nonparametric density estimators has received much attention in recent years. Examples include the optional P{\'o}lya tree of \citet{wong2010optional} and the binary-partition based estimator of \citet{lu2013multivariate}, which have both been proven to perform well in a large-sample setting \citep{castillo2022optional,liu2023convergence}. Further proposals in this spirit include ensemble-estimators such as the Bayesian CART method of \citet{jeoung2023bart}. Although both irregular histograms and tree-based estimators are based on a piecewise constant model for the data-generating process, they typically have a different focus. Whereas for irregular histogram procedures, the aim is typically to provide a parsimonious account of an observed sample using a moderate number of bins, while tree-based methods mostly attempt to minimize prediction error with little regard for model parsimony.

The irregular histogram model proposed in this paper provides a fully Bayesian approach to histogram estimation. In particular, our method is based on finding the partition which maximizes the posterior probability under a piecewise constant model of the data-generating density, resulting in an automatic data-based rule for choosing the histogram partition. Using a combination of search heuristics and dynamic programming, our approach results in a bin selection rule that is very quick to compute, even for large datasets. Unlike regular Bayesian histogram models, the random irregular histogram method is shown to excel at automatic mode detection while achieving similar performance in terms of classical loss functions, making it an attractive choice for exploratory data analysis. A software implementation of the irregular random histogram estimator proposed in this article can be found in the \texttt{AutoHist.jl} \proglang{Julia} package, available as part of the general package registry \citep{simensen2025autohist}.
The source code used to create all the figures and tables presented in this article is available at the following GitHub repository: \url{https://github.com/oskarhs/Random-Histograms---Paper}.

The remainder of the paper is structured as follows. Section~\ref{sec:histogram_construction} gives an introduction to histogram construction, focusing on irregular methods. Section~\ref{seq:irreghist_model} introduces our irregular random histogram model and describes an algorithm to compute the approximate a posteriori most probable partition given an observed sample. Consistency and convergence rate results are given in Section~\ref{sec:asymptotics}. In Section~\ref{sec:simulation_study}, we describe a simulation study in which we compare the performance of our method with other state-of-the-art histogram procedures from the statistical literature. In Section~\ref{sec:examples} we illustrate the proposed density estimator by applying it to some real-world datasets. We conclude the article by discussing some extensions of the methodology presented here to hazard rate estimation and semiparametric regression.

\section{Histogram construction}
\label{sec:histogram_construction}
The histogram approach to nonparametric density estimation is based on a piecewise constant model of the data-generating density $f$. For ease of exposition, we present all models for densities on the unit interval, but note that the methodology presented here can be extended to other compact intervals via a suitable affine transformation.
For a given interval partition $\mathcal{I} = (\mathcal{I}_1, \mathcal{I}_2, \ldots, \mathcal{I}_k)$ of $[0,1]$, the piecewise constant approximation to the density $f$ is given by
\begin{equation}\label{eq:histogram_model}
    f\big(x\ba \mathcal{I}, \boldsymbol{\theta}\big) = \sum_{j=1}^k \frac{\theta_j}{|\mathcal{I}_j|}\ind_{\mathcal{I}_j}(x),\quad x\in [0,1],
\end{equation}
where $\boldsymbol{\theta}$ belongs to the $k$-simplex $\mathcal{S}_k = \big\{\boldsymbol{\theta}\in [0,1]^k\colon \sum_{j=1}^k \theta_j = 1\big\}$ and $\ind_\mathcal{A}$ denotes the indicator function for a set $\mathcal{A}$, so $\ind_\mathcal{A}(x) = 1$ if $x\in\mathcal{A}$ and 0 otherwise. Although the dimension of $\btheta$ depends on the given partition $\pa$, we generally omit this in the notation when the number of bins in the partition in question is clear from the context. For a given partition, estimation of the parameter $\boldsymbol{\theta}$ is usually done by maximizing the likelihood function based on $f$ or minimizing an estimated $\mathbb{L}_2$ loss, yielding the familiar bin proportions $\widehat{\theta}_j = N_j/n,$ where $N_j = \sum_{i=1}^n \ind_{\mathcal{I}_j}(x_i)$ is the number of observations falling into bin $\mathcal{I}_j$.
Although constructing a histogram for a given partition is a simple task, selecting a good partition based on the observed sample constitutes a much more difficult problem. The chosen partition controls the degree of smoothing, with larger intervals inducing a smoother density estimate at the cost of increased bias. Attempting to find a partition that strikes a good balance between smoothing and small bias is the basis of most automatic histogram procedures.

Most automatic histogram procedures are based on a decision-theoretic framework where the sample $\bx = (x_1,\ldots, x_n)$ has been generated by a true density $f_0$ and seek to achieve good performance in terms of the frequentist risk,
\begin{equation}\label{eq:frequentist_risk}
    R_n\big(f_0, \widehat{f}\;\big) = \E_{\bx \sim f_0}\Big\{\ell\big(f_0, \widehat{f}\;\big)\Big\},
\end{equation}
where $\ell\big(f_0, \widehat{f}\;\big)$ is a loss function measuring the quality of using the estimate $\widehat{f}$ as an approximation of $f_0$. Typical loss functions include those based on powers of the $\mathbb{L}_{\mspace{1mu}r}$ or Hellinger metrics, 
\begin{equation*}
    \big\rVert f_0 - \widehat{f}\,\big\lVert_r = \Big(\int_0^1 \big\{f_0(x) - \widehat{f}(x)\big\}^r\, \dd x\Big)^{1/r},\quad
    d_H\big(f_0, \widehat{f}\;\big) = \Big(\int_0^1 \Big\{\sqrt{f_0(x)} - \sqrt{\widehat{f}(x)}\Big\}^2\, \dd x\Big)^{1/2}.
\end{equation*}
On the other hand, approaches based on maximum likelihood attempt to minimize the Kullback--Leibler divergence,
\begin{equation*}
    K\big(f_0, \widehat{f}\;\big) = \int_0^1 f_0(x)\log \frac{f_0(x)}{\widehat{f}(x)}\, \dd x.
\end{equation*}

In general, the risk of an estimator will depend on the true density, and there is no procedure that universally yields the smallest possible risk for all densities. As such, most histogram procedures aim for small risks over a large class of densities $f_0$. The vast majority of regular histogram methods take this approach to histogram estimation by minimizing an asymptotic expression for the risk, a penalized likelihood, or a direct estimate of the risk \citep{davies2009automatic}. Irregular histogram methods based on risk minimization include proposals based on $\mathbb{L}_2$ leave-one-out cross-validation \citep{rudemo1982empirical,celisse2008leavepout} and the penalized maximum likelihood approach of \citet{rozenholc2010irregular}, which seeks to minimize an upper bound on the Hellinger risk. Not all histogram procedures follow the decision-theoretic paradigm; the approaches of \citet{davies2004densities,li2020essential} construct an irregular histogram based on approximating the empirical distribution function under a parsimony constraint.

The determination of the cut points of an irregular histogram constitutes a difficult problem from a computational perspective, and most irregular histogram procedures rely on various discretization schemes that specify a finite set of possible cut points between the intervals. The reduction in computational complexity brought about by discretization is in itself not enough to guarantee that the resulting optimization problem can be solved in a reasonable amount of time, even for datasets of modest size. As a result, many irregular histogram procedures are based on a selection criterion with a particular additive structure that enables the use of a dynamic programming algorithm \citep{kanazawa1988optimal}, efficient search heuristics \citep{mendizabal2023fast} or both \citep{rozenholc2010irregular}.

\section{A Bayesian approach to irregular histograms}
\label{seq:irreghist_model}
We now describe our proposed approach for constructing an irregular histogram. Assuming that we have observed an independent and identically distributed (i.i.d.)~sample $x_1, \ldots, x_n$ on the unit interval, our Bayesian modeling approach proceeds by specifying a parameterized density $f$ as in \eqref{eq:histogram_model} and prior distributions on the model parameters, $\pa, \btheta$. This choice of $f$ leads to the following expression for the joint density of the sample $\bx$,
\begin{equation}\label{eq:likelihood_sample}
    f(\bx\ba \pa, \btheta) = \prod_{i=1}^n \prod_{j=1}^k \left(\frac{\theta_j}{|\mathcal{I}_j|}\right)^{\ind_{\mathcal{I}_j}(x_i)} = \prod_{j=1}^k \left(\frac{\theta_j}{|\pa_j|}\right)^{N_j},
\end{equation}
which depends only on the sample through the interval counts $\bN = \big(N_1, \ldots, N_k\big)$.

In our model, we restrict our attention to partitions with interval endpoints belonging to a given finite set $\mathcal{T}_n = \big\{\tau_{n,j}\colon 0\leq j\leq k_n\big\}$, where $\tau_{n,0} = 0$, $\tau_{n,k_n} = 1$ and $\tau_{n,j-1} < \tau_{n,j}$ for all $j$, and $k_n$ is a sequence of positive integers, growing with the sample size $n$. Let $\mathcal{P}_{\mathcal{T}_n, k}$ denote the set of interval partitions of $[0,1]$ consisting of $k$ bins with endpoints in $\mathcal{T}_n$, and put $\mathcal{P}_{\Tn} = \cup_{k=1}^{k_n}\mathcal{P}_{\mathcal{T}_n,k}$. Since $\mathcal{P}_{\Tn}$ depends on $n$, the prior for $\pa$ must also necessarily do so, and we write $p_n(\pa)$ to make this dependence explicit. The prior distribution on the partitions is most easily described by introducing the number $k$ of bins as a further random variable, with prior distribution $k\sim p_n(k)$, supported on $\{1,2,\ldots, k_n\}$. Conditional on $k$, the prior distribution on the partitions $p_n(\pa \ba k)$ is the uniform distribution on $\mathcal{P}_{\Tn,k}$. Since $|\mathcal{P}_{\Tn,k}| = \binom{k_n-1}{k-1}$, it follows that the prior probability mass function~of $\pa$ is $p_n(\pa \ba k) = \binom{k_n-1}{k-1}^{-1}$. We note that if $\pa\in \mathcal{P}_{\Tn,k}$, then $p_n(\pa, k')$ is nonzero only for $k' = k$ and as such, we have the equality $p_n(\pa) = \sum_{k' = 1}^{k_n} p(k') \mspace{2mu} p_n(\pa\ba k') =  p_n(k)\mspace{2mu} p_n(\pa\ba k)$. Finally, as prior distribution for $\btheta\ba \pa$ we take a $k$-dimensional $\mathrm{Dir}(\boldsymbol{a})$ distribution, which has density
\begin{equation*}
    p(\btheta\ba \pa) = \frac{\Gamma\big(a\big)}{\prod_{j=1}^k \Gamma(a_j)}\prod_{j=1}^k \theta_j^{a_j-1},\quad \btheta \in \Sk,
\end{equation*}
where $\boldsymbol{a} = (a_1, \ldots, a_k) \in (0,\infty)^k$ and $a = \sum_{j=1}^k a_j$. In general, the parameters $a_j$ may depend on the partition $\pa$, although we omit this in the notation. We defer the discussion on how to choose $p_n(k)$ and $\boldsymbol{a}$ to Section~\ref{subsec:computational_details}.

The prior-model specification can thus be summarized as 
\begin{align}\label{eq:prior_model_hierarchical}
\begin{split}
    k &\sim p_n(k),\\
    \pa \ba k &\sim \mathrm{Unif}\big(\pa \bba \mathcal{P}_{\Tn, k}\big),\\
    \btheta \ba \pa &\sim \mathrm{Dir}(\btheta\ba \boldsymbol{a}),\\
    f(x\ba \pa, \btheta) &= \sum_{j=1}^k \frac{\theta_j}{|\pa_j|}\ind_{\pa_j}(x),\\
    x_i \ba f &\sim f.
\end{split}
\end{align}

The model \eqref{eq:prior_model_hierarchical} is similar to some Bayesian approaches to regular histogram estimation, which have been investigated from a theoretical perspective \citep{hall1988stochastic,scricciolo2007rates} and from a computationally oriented one \citep{knuth2019optimal}. What separates our proposal from these prior approaches is that we consider more than one partition for each value of $k$, resulting in a more flexible procedure due to the variable bin widths. Our approach also shares similarities with the Bayesian sequential partitioning model of \citet{lu2013multivariate}, the main difference being that their model is based on binary partitions, while ours relies on a sample-size dependent grid.

\subsection{Posterior distribution}
We now derive an expression for the posterior distribution of $\pa$.
As a first step, we show that the Dirichlet distribution is a conjugate prior for $f$, conditional on the chosen partition. Indeed, as the likelihood function takes the form in \eqref{eq:likelihood_sample}, it follows from Bayes' rule that for $\pa\in \mathcal{P}_{\Tn, k}$,
\begin{equation*}
    p(\boldsymbol{\theta}\ba \bx, \pa) \propto p(\btheta\ba \pa)\,f(\bx\ba \pa, \btheta) \propto 
    \prod_{j=1}^k \theta_j^{a_j-1}\prod_{j=1}^k \theta_j^{N_j} \propto 
    \mathrm{Dir}\big(\btheta\ba \boldsymbol{a} + \bN\big).
\end{equation*}

To find the posterior probability of $\pa$ up to proportionality, we note that by Bayes' rule,
\begin{equation*}
    p_n(\pa \ba \bx) \propto p_n(\pa)\, p(\bx\ba \pa) =
    p_n(\pa)\, \frac{p(\btheta\ba \pa)\,f(\bx\ba \pa, \btheta)}{p(\btheta\ba \bx, \pa)},
\end{equation*}
where the last equality follows from the relation $p(\btheta\ba \pa)\,f(\bx\ba \pa, \btheta) = p(\bx\ba \pa)\,p(\btheta\ba \bx, \pa)$.
Plugging in the expressions for our chosen prior distributions along with the posterior for $\btheta$, we arrive at the following expression for the posterior probability of $\pa$,
\begin{equation}\label{eq:posterior_prob_partition}
    p_n(\pa \ba \bx) \propto \frac{p_n(k)}{\prod_{j=1}^k |\pa_j|^{N_j}}\frac{\prod_{j=1}^k\Gamma(a_{j}+N_j)}{\prod_{j=1}^k \Gamma(a_{j})}\frac{\Gamma(a)}{\Gamma(a+n)}\binom{k_n-1}{k-1}^{-1},\quad \pa \in \mathcal{P}_{\Tn, k}.
\end{equation}

Our proposed Bayesian histogram estimator is based on the partition which minimizes the Bayes risk,
\begin{equation*}
    \wpa = \underset{\mathcal{J}\in \mathcal{P}_{\Tn}}{\arg\min}\:\E_{\pa, \btheta}\Big(\E_{\bx\sim f}\big\{\ell(\pa,{\mathcal{J}})\bba \pa, \btheta\big\} \Big),
\end{equation*}
where $\ell$ is a nonnegative loss function.
To derive the optimal model selection procedure in terms of the Bayes risk, we work with the 0-1 loss function,
\begin{equation*}
    \ell(\pa, \mathcal{J}) =  \begin{cases}
        1& \mathrm{if}\ \mathcal{J} = \pa,\\
        0& \mathrm{else}.
    \end{cases}
\end{equation*}

To find the minimizer of the Bayes risk, it suffices to minimize the posterior expected loss, $\E_{\pa}\big\{\ell(\pa, \mathcal{J})\bba \bx\big\}$ [cf.~\citet[p.~228]{lehmann2006theory}], which leads to the following rule for selecting the optimal partition
\begin{equation}\label{eq:MAP_partition}
    \wpa = \underset{\mathcal{I}\in \mathcal{P}_{\Tn}}{\arg\max}\: p_n(\mathcal{I}\ba \bx).
\end{equation}
We refer to $\wpa$ as the maximum a posteriori (MAP) partition. 

Since the histogram is first and foremost a graphical tool, providing point estimates of the density is of great interest. Having computed the maximizer of \eqref{eq:MAP_partition} we can construct a density estimate $\widehat{f}$ based on the conditional posterior $p\big(\btheta\ba \bx, \wpa\mspace{1mu}\big)$. For a given partition $\pa\in \mathcal{P}_{\Tn,k}$, the density $f$ can be parameterized in terms of the $k$-dimensional vector $\btheta$, and we proceed by estimating $\btheta$ by $\widehat{\btheta}$, the Bayes estimator under squared $\mathbb{L}_2$ loss conditional on $\pa$. Writing $f_{\pa, \bdelta}$ for the density given by $f(x\ba \pa, \bdelta)$ as in \eqref{eq:histogram_model} for $\bdelta\in \Sk$, we can expand the loss as
\begin{equation*}
    \big\lVert f_{\pa, \btheta} - f_{\pa, \bdelta}\big\rVert_2^2 = \sum_{j=1}^k \int_{\pa_j}\big\{f_{\pa, \btheta}(x) - f_{\pa, \bdelta}(x)\big\}^2\mspace{1mu} \dd x =
    \sum_{j=1}^k \int_{\pa_j} |\pa_j|^{-2}\big\{\theta_j -\delta_j\big\}^2\mspace{1mu}\dd x = \sum_{j=1}^k |\pa_j|^{-1}\big\{ \theta_j - \delta_j\big\}^2.
\end{equation*}
Since the Bayes estimator $\widehat{\btheta}$ of $\btheta$ minimizes the posterior expected loss, we need to solve the following optimization problem:
\begin{equation*}
    \underset{\bdelta\in \Sk}{\min}\: \E_{\btheta}\Big\{\big\lVert f_{\pa, \btheta} - f_{\pa, \bdelta}\big\rVert_2^2\mspace{1mu}\bba \bx\Big\} = \underset{\bdelta\in \Sk}{\min} \sum_{j=1}^k |\pa_j|^{-1}\E_{\btheta}\big\{ (\theta_j - \delta_j)^2\bba \bx\big\}.
\end{equation*}
This problem is most easily solved by first removing the restriction $\bdelta\in \Sk$, and showing that the unconstrained and constrained optima coincide. Since the summands of the above criterion are nonnegative, the problem then reduces to minimizing $\E_{\btheta}\big\{(\theta_j - \delta_j)^2\bba \bx\big\}$ with respect to $\delta_j$ for each $j$, the solution of which is $\widehat{\theta}_j = \E_{\btheta}\{\theta_j\ba \bx\}$ \citep[p.~228]{lehmann2006theory}. Using the linearity and monotonicity properties of expectations, one quickly verifies that $\widehat{\btheta}\in \Sk$, so it must also be the solution to the constrained problem. As the posterior distribution of $\btheta$ given $\pa\in \mathcal{P}_{\Tn,k}$ is a Dirichlet distribution, we can read off the posterior mean directly, yielding Bayes estimates
\begin{equation}\label{eq:posterior_mean_theta}
    \widehat{\theta}_j = \frac{a_j + N_j}{a+n} = \frac{a}{a+n}\frac{a_j}{a} + \frac{n}{a+n}\frac{N_j}{n},\quad j = 1,2,\ldots, k.
\end{equation}
Since the prior mean of $\theta_j$ is $a_j/a$, we see that the posterior mean is a convex combination of the prior mean and the maximum likelihood estimate $N_j/n$. If $a$ is small relative to $n$, then the data-based part dominates the estimate, and the estimated bin probabilities end up being close to the corresponding maximum likelihood estimates. Our Bayesian approach thus differs from the penalized likelihood approach mainly in the selection of the optimal partition and not in the estimation of bin probabilities conditional on the selected model. 

After computing the optimal partition and the corresponding Bayes estimates of the bin probabilities, a point estimate of the density $f$ can be obtained via
\begin{equation}\label{eq:bayes_histogram_estimator}
    \widehat{f}_{\wpa}(x) = \sum_{j=1}^k \frac{\widehat{\theta}_j}{\big\lvert \wpa_j\big\rvert}\ind_{\wpa_j}(x).
\end{equation}
We refer to the estimator \eqref{eq:bayes_histogram_estimator} as the \textit{Bayes histogram estimator}. The $\widehat{\theta}_j$ serve as estimates of the true bin probabilities $\theta_j = \int_{\pa_j} f(x)\, \dd x$. Although the form of \eqref{eq:bayes_histogram_estimator} is rather simple, the bin selection procedure itself is quite complex, as it requires finding the best partition according to \eqref{eq:posterior_prob_partition} among a field of $2^{k_n-1}$ candidate partitions. As argued by \citet{wand1997data}, the use of a complex bin selection procedure seems almost inevitable for a histogram method designed to perform well for a large class of densities. In fact, simulation studies show that most well working automatic regular and irregular histogram methods either solve an optimization problem over a set of possible partitions or estimate derivatives of the underlying density \citep{birge2006bins,davies2009automatic}.

\begin{remark}
    Although the primary focus of our article is the study of the Bayes histogram estimator, we note that full posterior inference for $f$ is available through the posterior distribution for $\pa$ given by \eqref{eq:posterior_prob_partition}. In particular, Markov chain Monte Carlo (MCMC) methods or importance sampling estimators as in \citet{lu2013multivariate} can be used to generate approximate posterior samples, which in turn can be used to estimate posterior quantities of interest, such as the posterior probability distribution of the number of modes in the density.
\end{remark}
\begin{remark}
    The estimator in \eqref{eq:bayes_histogram_estimator} is not the overall Bayes estimator of the density $f$ under the standard $\mathbb{L}_2$ loss as it does not minimize the Bayes risk. The minimizer of the posterior expected loss is rather the posterior mean of $f$,
    \begin{equation}\label{eq:grand_posterior_mean}
        \E_{f}\big\{f(x)\bba \bx\big\} = \sum_{\pa \in \mathcal{P}_{\Tn}} p_n(\pa\ba \bx)\,f\big(x\ba \pa, \widehat{\btheta}\big),
    \end{equation}
    which is a model averaging estimator, weighting each conditional posterior mean $\E_{f}\big\{f(x)\bba \bx, \pa\big\}$ according to the posterior probability of the corresponding partition $\pa$. Since the partition $\mathcal{J}$ which includes every grid point in $\mathcal{T}_n$ is a refinement of every $\pa\in \mathcal{P}_{\Tn}$, the estimator \eqref{eq:grand_posterior_mean} is also a piecewise constant density with cut points in $\Tn$, one could also consider using it as an estimate of the density. We do not pursue this approach here for two reasons.
    The first reason for favoring \eqref{eq:bayes_histogram_estimator} over the posterior mean is that it is quick to compute; the posterior mean is a sum of $2^{k_n-1}$ terms, making the evaluation of \eqref{eq:grand_posterior_mean} infeasible even for moderate values of $k_n$. Although simulation from the posterior $\pa\ba \bx$ is possible through MCMC, such a procedure may suffer from slowly mixing chains and requires performing convergence diagnostics, which makes the model fitting procedure rather involved compared to alternative automatic histogram methods. On the other hand, an approximation to the Bayes histogram estimator can be computed quickly using the methods of the next subsection without requiring any tuning from the user.
    Another reason for preferring the Bayes histogram estimator is that of model parsimony, as it depends only on the chosen partition and the corresponding interval probability estimates \eqref{eq:posterior_mean_theta}. Our simulations show that \eqref{eq:bayes_histogram_estimator} often yields quite good visual summaries of the underlying density with a relatively small number of bins. In contrast, the computation of $\E_{f}\big\{f(x)\bba \bx\big\}$ involves computing a weighted average of the estimated bin probabilities from many different models, which leads to a complicated expression for the resulting bin heights.
\end{remark}

\subsection{Computing the Bayes histogram estimator}
\label{subsec:computational_details}
The key observation that allows us to find the maximizer of $p_n(\pa\ba \bx)$ efficiently is that we can write the logarithm of the right hand side in \eqref{eq:posterior_prob_partition} as
\begin{equation}\label{eq:logposterior_partition}
    \Phi(\pa) + \Psi_n(k),\quad \pa \in \mathcal{P}_{\Tn, k},
\end{equation}
where $\Phi$ and $\Psi_n$ are given by
\begin{align*}
    \Phi(\pa) &= \sum_{j=1}^k \big\{\log \Gamma(a_j + N_j) - \log \Gamma(a_j) - N_j\log |\pa_j|\big\},\\
    \Psi_n(k) &= \log p_n(k) + \log \Gamma(a) - \log \Gamma(a+n) - \log \binom{k_n-1}{k-1}.
\end{align*}
$\Phi$ is additive with respect to the intervals of the partition in the sense that we can write $\Phi = \sum_{j=1}^k \Phi_0(\pa_j)$, where $\Phi_0(\pa_j)$ depends only on $\pa_j$. This allows for the use of the dynamic programming algorithm due to \citet{kanazawa1988optimal} which determines the maximizer of \eqref{eq:logposterior_partition} using $\mathcal{O}\big(k_n^3\big)$ floating-point operations.

Although the dynamic programming algorithm succeeds in identifying the MAP partition within a reasonable time budget for smaller datasets, the $\mathcal{O}\big(k_n^3\big)$ runtime is prohibitive for large values of $k_n$.
It is generally desirable to have $k_n$ increase at a rate close to $n$, as otherwise $\Tn$ can end up being too coarse, and the resulting histogram may miss key details in the data that would have been shown if a high-resolution grid was used.
To reduce the computational burden, we have implemented a greedy search heuristic similar to that of \citet{rozenholc2010irregular} to construct a reduced grid $\mathcal{Q}_n\subseteq \Tn$ whose cardinality $q_n$ grows more slowly with $n$. After computing $\mathcal{Q}_n$, we run the dynamic programming algorithm to maximize $p_n(\pa \ba \bx)$ over $\mathcal{P}_{\mathcal{Q}_n}$, and take $\wpa$ as the resulting maximizer. Combining these two heuristics allows us to let $k_n$ increase at a rate close to $n$, while retaining a fast implementation. Although this approach is not guaranteed to find the exact maximizer of $p_n(\pa\ba \bx)$ over $\Tn$, we have found that in practice the chosen partition typically yields a histogram close to the one based on the true MAP in terms of Hellinger loss in cases where exact evaluation is feasible.

\subsection{Prior elicitation and practical implementation}
\label{subsec:prior_elicitation}
The model specified by \eqref{eq:prior_model_hierarchical} leaves the statistician with some flexibility through the choice of the prior for $k$ and the hyperparameter $\boldsymbol{a}$. To have a fully automatic procedure, however, default choices for $p_n(k)$ and $\boldsymbol{a}$ are needed, which we provide in this section.

As prior for $k$, we have found that in practice using the uniform prior on $\{1,2,\ldots, k_n\}$ with $k_n = \lceil 4n/\log^2(n) \rceil$ appears to work well compared to other alternatives in simulations, as documented in Section~\ref{subsubsec:numbin_prior}. As such, we have used this uniform prior for $k$ as part of our default procedure.

The hyperparameters $\boldsymbol{a}$ of the Dirichlet prior for $\btheta\ba \pa$ can be chosen to center the prior distribution of $f$ on a particular reference density. If $g_0$ is a initial guess for the density $f$, we let
\begin{equation}\label{eq:prior_center}
    a_j = a\int_{\pa_j} g_0(x)\mspace{2mu} \dd x,\quad \pa \in \mathcal{P}_{\Tn, k}, \quad j = 1,2,\ldots, k,
\end{equation}
where $a > 0$ is a constant. Conditional on the partition $\pa$, we see that the prior mean of $f$ is
\begin{equation*}
    \E_{\btheta}\big\{f(x)\ba \pa\big\} = \sum_{j=1}^k \frac{\E_{\btheta}\big\{\theta_j\big\}}{\big\lvert \pa_j\big\rvert} \ind_{\pa_j}(x) = \sum_{j=1}^k \frac{a_j}{a\big\lvert \pa_j\big\rvert} \ind_{\pa_j}(x) = \sum_{j=1}^k |\pa_j|^{-1}\int_{\pa_j} g_0(y)\mspace{2mu}\dd y\, \ind_{\pa_j}(x),
\end{equation*}
which is the best piecewise constant approximation of the density $g_0$ based on the partition $\pa$, in the sense that it minimizes the $\mathbb{L}_2$ distance to $g_0$ among all piecewise constant densities on $\pa$.
To guarantee that this choice of $a_j$ always yields a proper posterior for $\btheta\ba \bx, \pa$, we must ensure that $a_j > 0$, which is true if the density $g_0$ is chosen to be positive almost everywhere in $[0,1]$. If prior information is not available, we can take $g_0$ equal to the uniform distribution in the unit interval, leading to the non-informative choice $a_j = a|\pa_j|$ for all $j$. We have made the choice $g_0 = \ind_{[0,1]}$ the default in the available software implementation of the random irregular histogram, and we have adopted this choice in all subsequent simulations and applications of the histogram method in this paper.

The choice of the parameter $a$ is more delicate, as it is difficult to directly assess its impact on the chosen partition $\wpa$. Through \eqref{eq:posterior_mean_theta}, smaller values of $a$ yield Bayes estimates $\hat{\theta}_j$ which are closer to the maximum likelihood estimate of $\theta_j$, depending more on the data and less on the prior mean. However, if $a$ is small relative to $n$ its effect on the chosen partition is typically of much greater importance than its influence on the estimated bin probabilities, since the data-based part of the estimate tends to dominate for moderate values of $n$, which is confirmed by the simulations in Section~\ref{subsubsec:concentration_prior}. Based on the results from these preliminary simulations, we suggest $a = 5$ as a reasonable default choice for $a$.

As presented, the random irregular histogram method applies to densities supported on a known closed interval. To apply the methodology to data with unknown support, the support of the data has to be estimated. Our preferred method in this regard is to rescale the data to the unit interval via the affine transformation $z_i = (x_i-x_{(1)})/(x_{(n)}-x_{(1)})$, apply the Bayesian histogram method to the $z_i$ and rescale the density estimates to the original scale, which is equivalent to the approach taken by most histogram procedures. For a discussion of other approaches to estimating the support of a probability distribution for the purposes of density estimation, see Section 1.4 of \citet{guan2016efficient} and the references therein.

Another issue to consider is the construction of the grid $\Tn$. One possibility is to use a fine, regular mesh. Data-based grids are also possible; one such approach is to construct $\Tn$ from the sample quantiles, so that $\tau_{n,j} = \widehat{Q}_n(j/k_n)$ for $j = 1,2,\ldots, k_n-1$ where $\widehat{Q}_n(q)$ is the $q$-quantile of the sample $\bx$. A further data-based option is to use the order statistics of the data as possible cut points. All of these approaches have their advantages and disadvantages, and the best choice of grid is ultimately dependent on the density we are trying to estimate; see e.g.~\citet{rozenholc2010irregular}. As part of our default procedure, we propose using a data-based mesh for $\Tn$, but note that in practical applications, one may also want to test data-based grids to check the sensitivity of the procedure to the choice of mesh.

\section{Asymptotic results}\label{sec:asymptotics}
To study the large-sample behavior of the Bayes histogram estimator \eqref{eq:bayes_histogram_estimator}, we use the theory developed by \citet{ghosal1999posterior,ghosal2000rates} to study posterior consistency and posterior convergence rates, but with some modifications to account for the fact that our estimator is based on the MAP partition $\wpa$. In the following, we assume that the data $\bx = (x_1, \ldots, x_n)$ is an i.i.d.~sample from a true density $f_0$.

\subsection{Consistency}
Our first result shows that the Bayes histogram estimator is Hellinger consistent under very general conditions on the prior and the true density $f_0$. The proof of Theorem~\ref{thm:consistency} is given in Section~\ref{subsec:consistency_proof}.
\begin{theorem}\label{thm:consistency}
    Suppose that $f_0$ is a probability density on the unit interval satisfying $\int_0^1 \big\{f_0(x)\big\}^r\mspace{2mu}\dd x < \infty$ for some $r \in (1, 2]$, and suppose that the random histogram prior sequence $\{P_n\}_{n\in \mathbb{N}}$ satisfies the following conditions:
        \begin{enumerate}
    \item[$(i)$] The sequence of prior probability mass functions for $k$, $p_n(k)$, is fully supported on $\{1,2,\ldots, k_n\}$, where $k_n\to\infty$ and $k_n = \smallO(n)$.
    Moreover, we have uniformly for all $k = 1,2,\ldots, k_n$ and $n$,
    \begin{equation*}
        \log p_n(k) = \smallO(n).
    \end{equation*}
    \item[$(ii)$] Conditional on the partition $\pa\in \mathcal{P}_{\Tn, k}$, the prior for $\btheta\ba \pa$ is $\mathrm{Dir}(\boldsymbol{a})$ where $0 < a_{j} \leq \Sigma$ for a constant $\Sigma > 0$.
    \item[$(iii)$] The mesh sequence $\Tn$ satisfies
    \begin{equation*}
         \max_{1\leq j\leq k_n} \{\tau_{n,j} - \tau_{n,j-1}\} = \smallO(1)
    \end{equation*}
    \end{enumerate}
    Then the Bayes histogram estimator $\widehat{f}_{\wpa}$ satisfies
    \begin{equation*}
        \E_{\bx\sim f_0}\Big\{\dhsq\big(f_0, \widehat{f}_{\wpa}\big)\Big\} = \smallO(1).
    \end{equation*}
\end{theorem}
The first and third condition ensure that the prior eventually assigns positive mass to an arbitrarily fine grid, which is needed to approximate $f_0$ in the Hellinger sense. The second condition is fulfilled if, for instance, $a_j = a\int_{\pa_j} g_0(x)\mspace{2mu}\dd x$ for a constant $a > 0$ and a strictly positive density $g_0$, or if the $a_j$ are all equal to some fixed, positive number. Theorem~\ref{thm:consistency} mirrors the result by \citet[Section~3.1]{barron1999consistency}, who studied posterior consistency for a regular random histogram prior, showing that the posterior distribution is consistent under the slightly stronger condition that $f_0$ is bounded.

\subsection{Convergence rate}
To establish a convergence rate result, we need to make some additional smoothness assumptions on the true density $f_0$. A function $f_0\colon [0,1]\to \mathbb{R}$ is said to be $\alpha$-H\"{o}lder continuous for $\alpha \in (0,1]$ if there exists a constant $L_0> 0$ such that $|f_0(x) - f_0(y)|\leq L_0|x-y|^{\alpha}$ for all $x,y\in [0,1]$. The parameter $\alpha$ determines the smoothness of the function $f_0$, with larger values of $\alpha$ corresponding to smoother functions. The H\"{o}lder exponent relates directly to the bias in the approximation of $f_0$ by a piecewise constant function on the partition $\pa$ (cf.~Lemma~\ref{lemma:histogram_approximation_rate} in the appendix), with larger exponents yielding better approximations for a given number of bins, leading to a faster convergence rates as a result.

\begin{theorem}\label{thm:convergence_rate}
    Suppose that $f_0\colon [0,1]\to \mathbb{R}$ is a strictly positive $\alpha$-H\"{o}lder continuous density and that the random irregular histogram prior sequence $\{P_n\}_{n\in \mathbb{N}}$ satisfies the following.
        \begin{enumerate}
    \item[$(i)$] The sequence of prior probability mass functions for $k$, $p_n(k)$, is fully supported on $\{1,2,\ldots, k_n\}$, where $n^{\gamma} = \mathcal{O}(k_n)$ for any $\gamma\in (0,1)$ and $k_n = \mathcal{O}(n)$.
    Moreover, there are positive constants $D_1,D_2, d_1, d_2$ such that for all $k\leq k_n$,
    \begin{equation*}
        D_1\exp\big(-d_1 k \log(k)\big) \leq p_n(k) \leq D_2 \exp\big(-d_2 k\log(k)\big).
    \end{equation*}
    \item[$(ii)$] Conditional on the partition $\pa\in \mathcal{P}_{\Tn, k}$, the prior for $\btheta\ba \pa$ is $\mathrm{Dir}(\boldsymbol{a})$ where $\sigma k^{-1} \leq a_{j} \leq \Sigma$ for all $j$ and two constants $\sigma, \Sigma > 0$.
    \item[$(iii)$] The mesh sequence $\Tn$ satisfies
    \begin{equation*}
         \max_{1\leq j\leq k_n} \{\tau_{n,j} - \tau_{n,j-1}\} = \mathcal{O}\big(k_n^{-1}\big),\quad \quad \min_{1\leq j\leq k_n} \{\tau_{n,j} - \tau_{n,j-1}\}\geq Bk_n^{-1},\ \mathrm{for\ some\ } B > 0.
    \end{equation*}
    \end{enumerate}
    Then for $\epsilon_n = \{n/\log(n)\}^{-\alpha/(2\alpha+1)}$, the Bayes histogram estimator satisfies
    \begin{equation*}
        \E_{\bx \sim f_0}\Big\{\dhsq\big(f_0, \widehat{f}_{\wpa}\big)\Big\} = \bigO\big(\epsilon_n^2\big).
    \end{equation*}
\end{theorem}
The proof of Theorem~\ref{thm:convergence_rate} can be found in Section~\ref{subsec:convergence_rate_proof}.
The first condition of the theorem is satisfied if, for instance, $p_n(k)$ is taken to be a Poisson distribution truncated to the interval $\{1,2,\ldots, k_n\}$. The third condition ensures that the mesh $\Tn$ does not deviate too much from a regular one, which guarantees that the deterministic component of the risk is at most of order $k^{-2\alpha}$ with respect to the squared Hellinger metric.

The rate obtained by the Bayes histogram estimator is equivalent to the minimax rate for $\alpha$-H\"{o}lder continuous densities with respect to the squared Hellinger metric of $n^{-2\alpha/(2\alpha+1)}$ up to a logarithmic factor \citep[Proposition 7.16]{massart2007concentration}. Moreover, the prior is rate-adaptive; it attains a near-optimal convergence rate without knowing the smoothness of the true density $f_0$. The convergence rate of the Bayes histogram estimator mirrors the posterior convergence rate obtained by \citet{scricciolo2007rates} for the corresponding regular histogram model and that of \citet{liu2023convergence} for Bayesian trees, with the minor difference that our rate is slightly better in terms of the logarithmic factor. The additional logarithmic factors that appear in the convergence rate are rather typical of Bayesian density estimators based on mixtures with Dirichlet-distributed weights; see, for example, the polygonally smoothed prior of \citet{scricciolo2007rates} and the extended Bernstein prior of \citet{kruijer2008beta}. \citet{hall1988stochastic} consider a regular random histogram model with a uniform prior on the interval probabilities. Their main result implies that the histogram estimator based on maximization of $p_n(k)$ converges at rate $\{n/\log(n)\}^{-1/3}$ if $f_0$ is bounded away from $0$ and has a uniformly continuous derivative $f_0'$, which is equal to the convergence rate of Theorem~\ref{thm:convergence_rate} under the weaker assumption that $\alpha = 1$. In light of this result, we conjecture that the rate obtained in Theorem~\ref{thm:convergence_rate} is sharp for any $\alpha\in (0,1]$.

\begin{remark}
The consistency and rate results presented here can also be extended to similar models for regular histograms, with minor modifications to the proofs. In particular, this means that the results of Theorems \ref{thm:consistency}~and~\ref{thm:convergence_rate} are also valid for the histogram estimator proposed by \citet{knuth2019optimal}.
\end{remark}

\section{Simulation study}\label{sec:simulation_study}
\begin{figure}[t]
    \centering
    \includegraphics[width=0.80\linewidth]{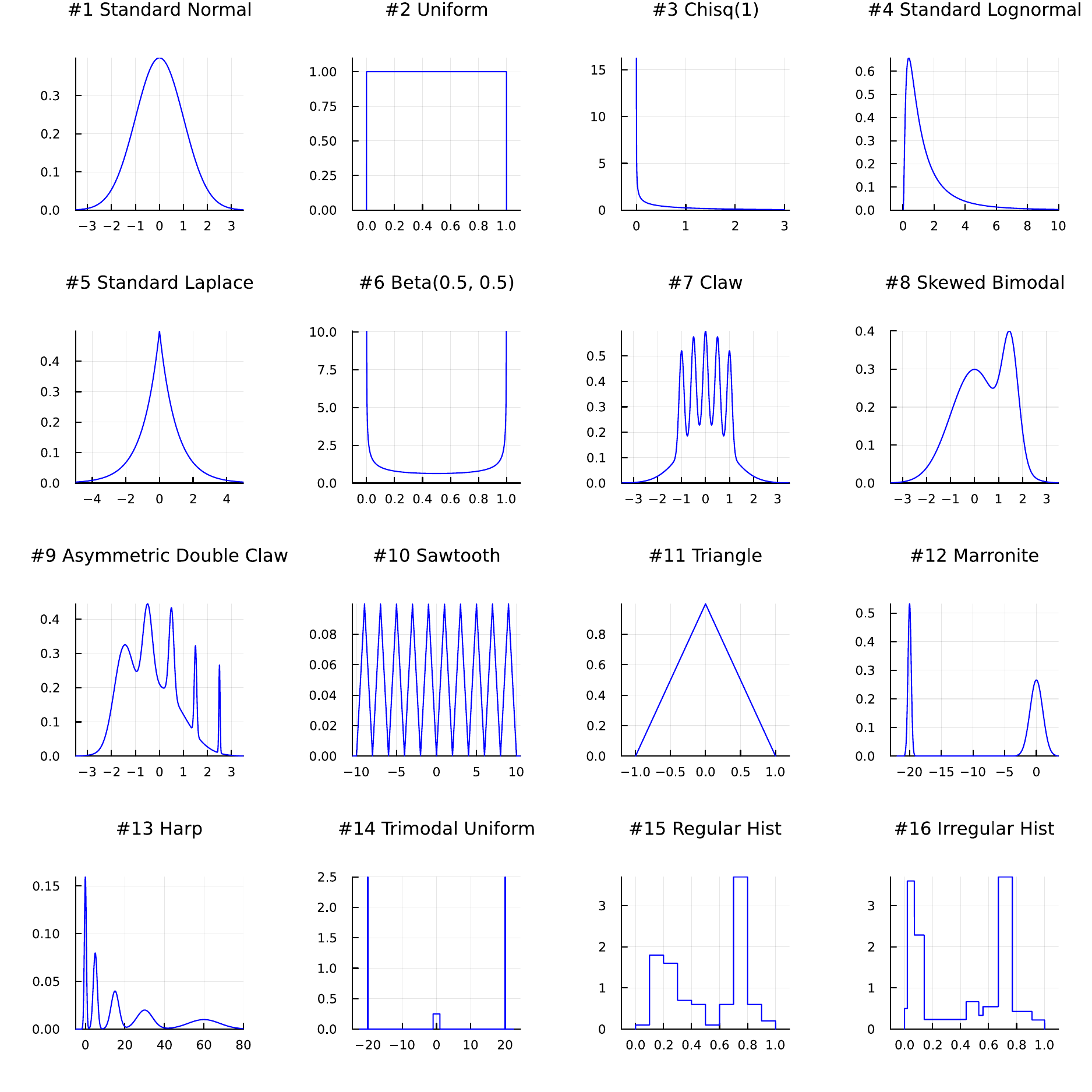}
    \vspace{-0.5cm}
    \caption{Test densities included in the simulation study}
    \label{fig:test_densities}
\end{figure}

To showcase the performance of our proposed irregular histogram method, we have conducted a simulation study in which we compare it to multiple state-of-the-art automatic histogram procedures on the set of test densities $\mathcal{D}$ given in Figure~\ref{fig:test_densities}.
These densities were chosen to reflect a variety of features with regard to skewness, tail behavior, number of modes and the sharpness of individual peaks. Densities 10, 12 and 14-16 are taken from \citet{rozenholc2010irregular}, densities 7-9 are from \citet{marron1992exact}, while density 13 was suggested by \citet{li2020essential}.

\subsection{Loss functions}
Although classical loss functions such as the $\mathbb{L}_2$- or Hellinger metrics provide a natural way of measuring the discrepancy between a density estimate and the true density in a simulation study, they do not adequately assess the quality of a histogram procedure in terms of identifying key features of a density, such as modes \citep{denby2009variations}. To also account for this important aspect of histogram estimation, we have included the peak identification loss of \citet{davies2009automatic} as a performance metric.
\begin{definition}[\citealp{davies2009automatic}]
\label{def:mode_peak}
    We say that a density $f$ has a \textit{finite mode} or \textit{peak} at $t$ if there is an interval\footnote{We include the case where $\mathcal{C}$ is a singleton set, which corresponds to the usual definition of a local maximum.} $\mathcal{C}$ such that $t$ is the midpoint of $\mathcal{C}$ and $f(s) = c$ for all $s\in \mathcal{C}$, and there exists a $\gamma > 0$ such that $f(s) < c$ for all $s\in \mathcal{C}_\gamma\setminus \mathcal{C}$, where $\mathcal{C}_\gamma = \cup_{s\in \mathcal{C}}\{x\colon |s-x|\leq \gamma\}$. The density $f$ is said to have an \textit{infinite peak} or \textit{mode} at $t$ if either one-sided limit at $t$ equals $\infty$.

    Suppose that $f_0$ has $l_0$ modes $t_1,t_2,\ldots,t_{l_0}$ and that the histogram estimate $\widehat{f}$ has modes at $s_1,\ldots, s_l$. Given a tolerance vector $\boldsymbol{\delta}\in (0,\infty)^{l_0}$, we say that a peak of the histogram $\widehat{f}$ \textit{matches the peak at $t_j$ of $f_0$ at tolerance $\delta_j$} if $\min_{1\leq i\leq l}|t_j - s_i|< \delta_j$. The peak $s_i$ is said to be \textit{spurious} if it does not match any peak of $f_0$. Letting $C = \sum_{j=1}^{l_0} \ind_{[0,\delta_j)}(\min_{i} |s_i - t_j|)$ denote the number of correctly identified modes, we define the \textit{peak identification} (PID) \textit{loss} as the sum of the number of spurious and unidentified peaks, $\ell\big(f_0, \widehat{f}\big) = (l_0 - C) + (l - C).$
\end{definition}
The behavior of the PID loss depends on the choice of the tolerance parameter $\boldsymbol{\delta}$. To determine the tolerance $\delta_j$ of a finite peak at $t_j$ of a density $f_0$, we have used the following approach: find the smallest $\gamma > 0$ such that
\begin{equation*}
    \frac{\int_{\mathcal{C}_{j,\gamma}} \big|f_0(x) - \bar{f}_{0,j,\gamma}\big|\mspace{2mu}\dd x}{\int_{\mathcal{C}_{j,\gamma}} f_0(x)\mspace{2mu}\dd x} > 0.2,\quad \mathcal{C}_{j,\gamma} = [t_j-\gamma, t_j + \gamma],
\end{equation*}
where we have defined $\bar{f}_{0,j,\gamma} = |\mathcal{C}_{j,\gamma}|^{-1}\int_{\mathcal{C}_{j,\gamma}} f_0(x)\mspace{2mu}\dd x$. We then take $\delta_j = \gamma/2$ as our tolerance for the peak at $t_j$. This criterion essentially measures how well $f_0$ is approximated by a constant function around any given mode, reflecting the fact that we should expect higher precision when trying to determine the location of a sharp peak. We visually confirmed that this criterion yielded sensible and non-overlapping tolerance regions for each of the test densities in Figure~\ref{fig:test_densities}. For the infinitely peaked $\mathrm{Chisq}(1)$ we set the tolerance region of the peak at $0$ to $[0, q_{0.1}]$, where $q_{\alpha}$ denotes the $\alpha$-quantile of this distribution, and for the $\mathrm{Beta}(0.5, 0.5)$, we used $[0, q_{0.1}]$ and $[q_{0.9}, 1]$ for the peaks at $0$ and $1$, respectively.

One objection to the use of the PID loss as a measure of mode detection is that it only captures a histogram procedure's ability to detect modes automatically and that no statistician would interpret any mode of a histogram as one corresponding to a mode in the true density. \citet{scott1979optimal} remarks that many data analysts prefer the use of relatively small bin widths when drawing a regular histogram, leaving the final smoothing to be done by eye.  Although we agree that this is a valid objection to the use of the peak identification loss in the regular case, the subjective manner in which the number and the location of modes have to be located is somewhat unsatisfactory. In contrast, a procedure that does well with respect to the PID loss can identify modes automatically, which is a major reason for the introduction of irregular histograms to begin with.

In addition to the PID loss, we have also included results for the Hellinger and $\mathbb{L}_2$ metrics in our simulations. Since closed-form expressions for these two loss functions are generally not available, Gauss quadrature was used to approximate the loss, performing the numeric integration piecewise over each interval where both the histogram estimate and the true density are continuous.

\subsection{Setup}
\label{subsec:setup}

In our study we have chosen to focus on histogram methods that have performed well in past simulation studies, see \citet{birge2006bins,davies2009automatic}. In addition, the method of \citet{knuth2019optimal} and the stochastic complexity approach of \citet{hall1988stochastic} were also included, as these effectively correspond to a regular equivalent of the random irregular histogram. The Essential Histogram of \citet{li2020essential} was excluded due to the high computational cost of its available software implementation, which made it impractical for the largest sample sizes considered in this study. We have also included an irregular histogram method based on Kullback--Leibler cross-validation, an approach that, to our knowledge, has so far not been explored for irregular histograms. This approach chooses the partition which maximizes $\mathrm{CV}_n(\pa) = \sum_{j=1}^k N_{j}\log(N_{j}-1) - \sum_{j=1}^k N_{j}\log |\pa_{j}|$
for all partitions with $N_j \geq 2$ for all $j$, and is an irregular counterpart to the cross-validation criterion proposed by \citet{hall1990akaike}. A detailed list of the included methods and their corresponding abbreviations can be found in Table~\ref{tab:simulations_methods}. Section~\ref{subsec:simulation_methods} provides further details on the implementations used for each procedure.
\begin{table}
    \centering
    \caption{Overview of the methods included in the simulation study}
    \label{tab:simulations_methods}
    \vspace{0.2cm}
    \begin{tabular}{c|c|c|c}
        \hline
        Type & Method & Abbreviation & Description \\ \hline
        &Wand's 2-step rule & Wand & \citet{wand1997data}\\ 
        & AIC & AIC & \citet{taylor1987aic} \\
        \multirow{2}{*}{Regular} & BIC & BIC & \citet{davies2009automatic}\\
        &Birg{é} and Rozenholc's rule & BR & \citet{birge2006bins}\\
        &Knuth's rule & Knuth & \citet{knuth2019optimal}\\
        &Stochastic Complexity & SC & \citet{hall1988stochastic}\\ \hline
        & Random irregular histogram & RIH & Section~\ref{seq:irreghist_model}\\
        & Penalty B in Rozenholc et al. & RMG-B & \citet{rozenholc2010irregular}\\
        \multirow{2}{*}{Irregular} & Penalty R in Rozenholc et al. & RMG-R & \citet{rozenholc2010irregular}\\
        & Taut string & TS & \citet{davies2004densities}\\
        & $\mathbb{L}_2$ cross-validation & L2CV & \citet{rudemo1982empirical}\\
        & Kullback--Leibler cross-validation & KLCV & Section~\ref{subsec:setup} \\
        \hline
    \end{tabular}
\end{table}

To judge the quality of the different histogram procedures, we used a Monte Carlo procedure to estimate the risks of each method for a set of test densities $f_0\in \mathcal{D}$ with respect to the Hellinger, $\mathbb{L}_2$ and PID losses, using the Monte Carlo average
\begin{equation}\label{eq:risk_estimate}
    \widehat{R}_n\big(f_0, \widehat{f}_m\big) = \frac{1}{B}\sum_{b=1}^B \ell\big(f_0, \widehat{f}_m\big), \quad (n, m, f_0)\in \mathcal{N}\times \mathcal{M}\times \mathcal{D},
\end{equation}
where $\mathcal{N} = \{50, 200, 1000, 5000, 25000\}$ denotes the set of sample sizes used in the study and $\mathcal{M}$ is the set of model indices. The Monte Carlo average in \eqref{eq:risk_estimate} was computed using $B = 500$ replications.

\subsection{Results}
\begin{figure}
    \centering
    {\includegraphics[width=0.470\linewidth]{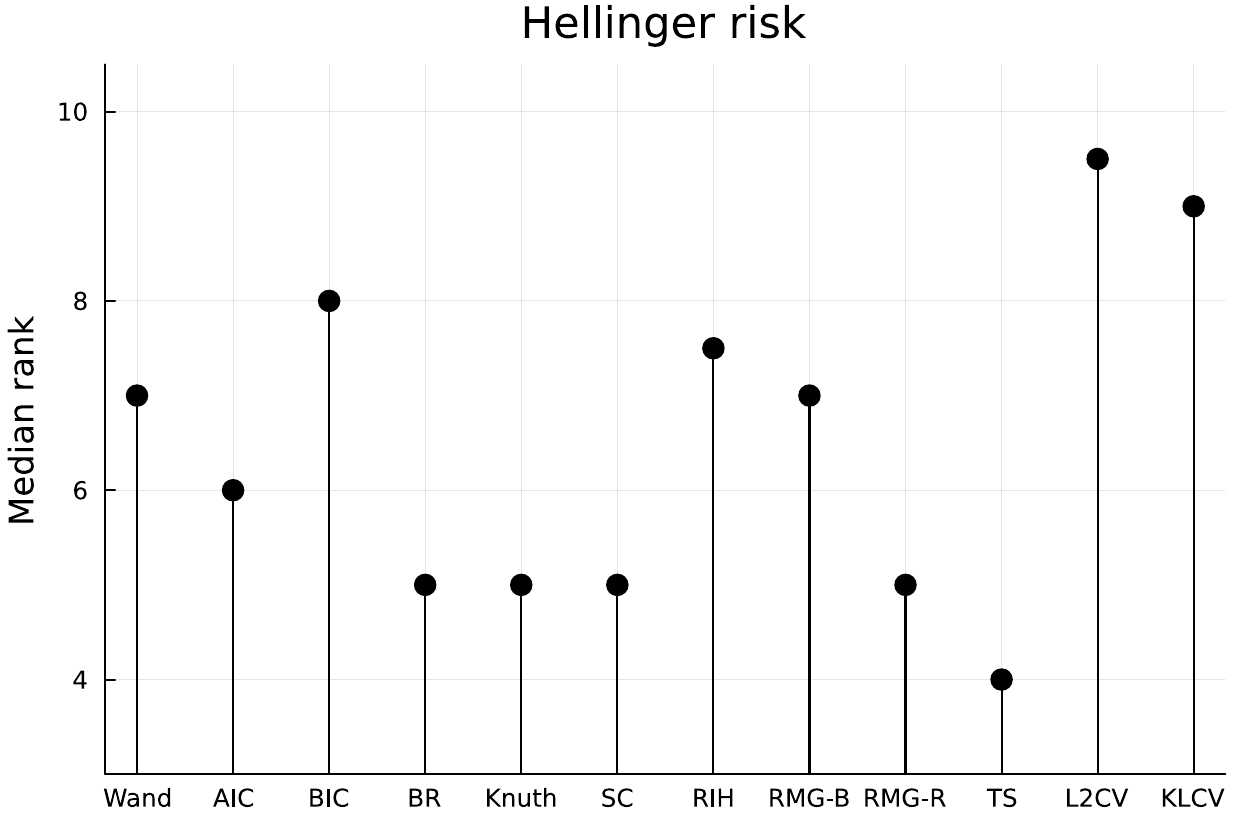} }
    \qquad
    {\includegraphics[width=0.470\linewidth]{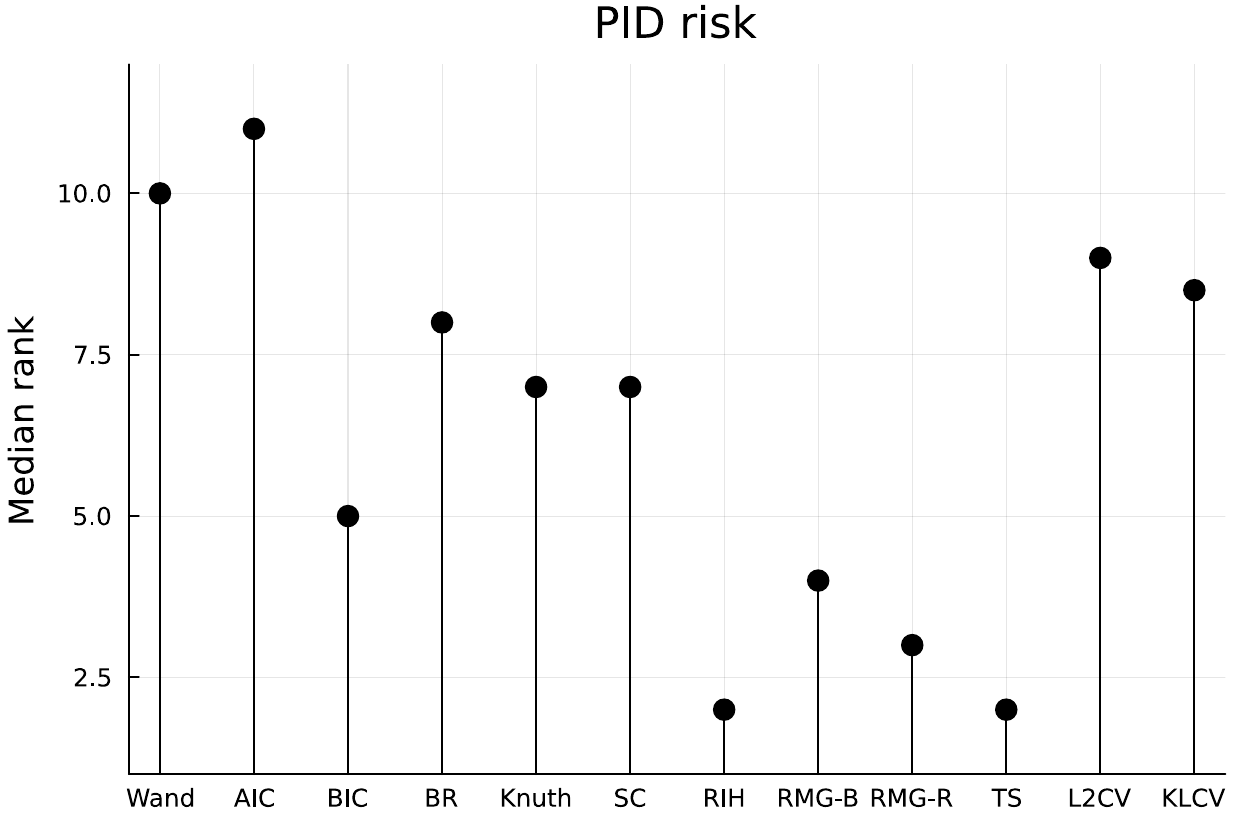} }
    \vspace{-0.8cm}
    \caption{Median ranks from the simulation study in terms of Hellinger risk (left) and PID risk (right).}
    \label{fig:average_ranks}
\end{figure}

Tables showing the estimated risks $\widehat{R}_n\big(f_0, \widehat{f}_m\big)$ for each of the three loss functions can be found in Section~\ref{subsubsec:tables_est_risk}, along with boxplots comparing the performances of the different methods included in the study.
To provide a more compact summary of the results, we ranked each method according to their risk for each combination of loss function, density and sample size, and computed the median rank by method over $\mathcal{D}$ and $\mathcal{N}$. The resulting medians for the Hellinger and PID losses are shown in Figure~\ref{fig:average_ranks}.

Although the relative performance of the different procedures depends on the true density at hand, our results indicate some broader trends among the regular and irregular histogram procedures.
\begin{itemize}
    \item Whether or not the Hellinger risk favors regular histogram methods varies based on the true density. For spatially homogeneous densities such as the standard normal or skewed bimodal densities, there is little benefit to using an irregular grid to begin with, resulting in generally smaller risks for regular histograms. For the heavy-tailed $\mathrm{Chisq}(1)$ and standard lognormal densities and the infinitely peaked $\mathrm{Beta}(1/2, 1/2)$, the irregular methods tend to perform better.
    \item With the exception of the L2CV and KLCV criteria, all the irregular histogram procedures included in the study outperformed their regular counterparts in terms of automatic mode identification for almost every combination of $f_0$ and $n$. This is apparent from the right panel of Figure~\ref{fig:average_ranks}; four of the irregular procedures produce much smaller median ranks than any of the other methods. In addition, we note that these four irregular methods exhibit lower PID risks as the sample size increases, a pattern that is often reversed for regular histograms.
\end{itemize}

Regarding the relative performance of the individual methods, our main findings were as follows:
\begin{itemize}
    \item The two cross-validation based procedures perform much worse in terms of mode identification than the other methods, and yielded the highest $\mathbb{L}_2$ and Hellinger losses for many densities. Plots of the histograms resulting from these procedures revealed that they still have a tendency to choose partitions with rather narrow bins even for smooth densities, despite the fact that a minimum bin width was enforced.
    \item Among the regular histogram procedures, the Bayesian criteria Knuth and SC are the overall best performers with respect to the Hellinger metric for the smallest sample sizes, while the criterion of \citet{birge2006bins} does better for larger samples. The BIC-based histogram is the clear winner in terms of PID risk, but does relatively poorly with respect to the two classical loss functions.
    \item The Taut String method of \citet{davies2004densities} is overall the best performer in terms of Hellinger and $\mathbb{L}_2$ risk out of the irregular histogram methods. Of the two criteria proposed by \citet{rozenholc2010irregular}, the performance is quite similar for the Hellinger loss, while penalty RMG-R tends to produce lower $\mathbb{L}_2$ risks, especially for smaller sample sizes. The random irregular histogram method performs comparably to the other non-cross-validation irregular procedures with respect to classical loss functions, and does particularly well at mode detection, as evidenced by its low PID risks across the board.
\end{itemize}

\begin{figure}[t]
    \centering
    \includegraphics[width=0.75\linewidth]{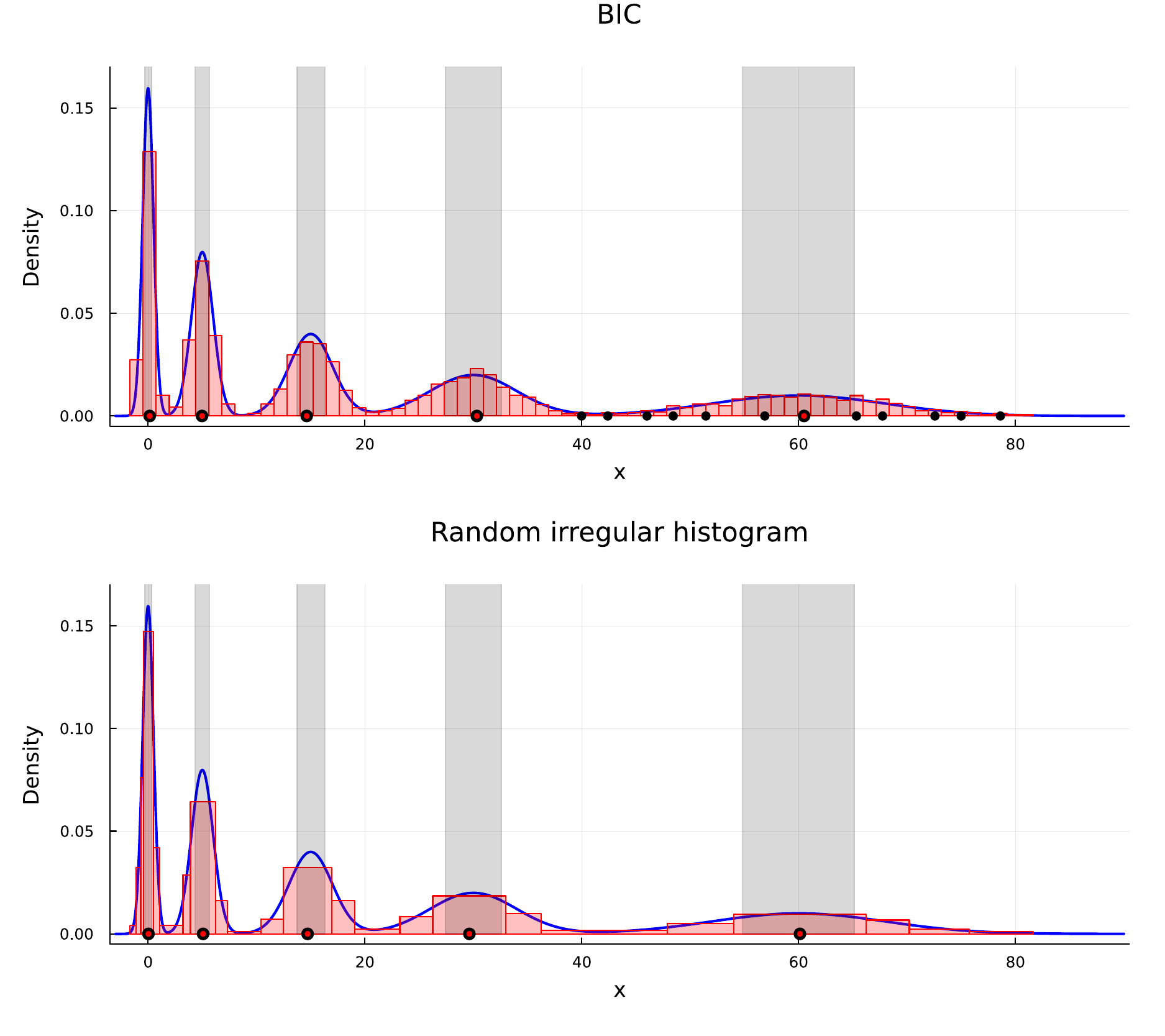}
    \vspace{-0.5cm}
    \caption{Illustration of automatic mode detection and the PID loss for a $n = 5000$ sample from the Harp density.}
    \label{fig:harp_mode_example}
\end{figure}

To illustrate the advantage of irregular histograms over their regular counterparts in automatic mode detection, we generated a sample of size $n = 5000$ from the Harp density in Figure~\ref{fig:test_densities}. The Harp density has multiple modes at different scales, making it difficult for regular histogram procedures to correctly identify the number of modes and the location of all individual peaks simultaneously, even for larger samples. Figure~\ref{fig:harp_mode_example} shows the resulting histograms from fitting the random irregular histogram and the regular BIC histogram to the simulated sample, which was chosen to yield a PID loss close to the corresponding estimated risks in Table~\ref{tab:complete_pid_risks_2}. The tolerance regions for each individual peaks is shown in gray, with the red dots indicating the positions of the histogram modes that match the peaks of the true density, whereas the spurious modes are drawn in black.
The disadvantages of using a fixed global bin width in this case are immediately apparent from Figure~\ref{fig:harp_mode_example}; the bin width chosen by the BIC histogram results in too much smoothing in the region surrounding the leftmost mode, while the density estimate is far too coarse near the rightmost mode, resulting in a high number of spurious peaks near this mode and in the right tail of the density. In contrast, the irregular histogram is able to adapt the bin width locally, providing more smoothing in flatter regions of the density, which results in high-precision estimates of each of the five true modes. On the other hand, the Hellinger loss for each of the two procedures are $0.176$ for the BIC and $0.272$ for the random irregular histogram, favoring the regular histogram. Ultimately, whether one prefers the regular or irregular estimate in this case comes down to the analysts preference for low estimation risk or automatic mode detection.

\section{Applications}\label{sec:examples}
In this section we apply the irregular histogram method proposed in this paper to two real-world datasets. For comparison, we also include results for the histogram method of \citet{knuth2019optimal}, as this method is a regular equivalent of the random irregular histogram.

\subsection{Old Faithful geyser data}
In our first application we consider the Old Faithful dataset from \citet{hardle2012smoothing}, which consists of measured waiting times in minutes between the consecutive eruptions of the Old Faithful geyser in Yellowstone National Park. Previous analyses of this dataset have typically shown a clear bimodal structure; see e.g.~\citet{venables1999modern} and \citet{kwasniok2021semiparametric} for analyses using several different density estimation procedures.

\begin{figure}
    \centering
    {\includegraphics[width=0.470\linewidth]{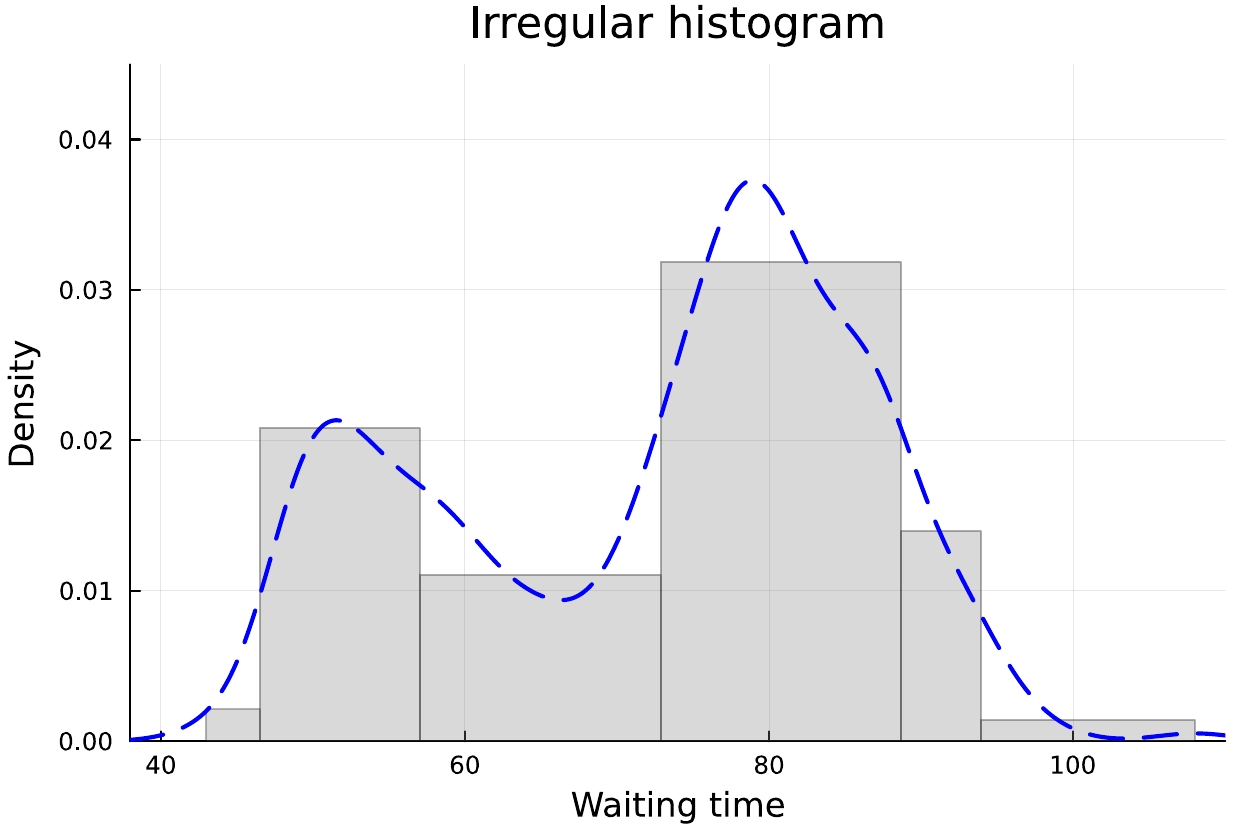} }
    \qquad
    {\includegraphics[width=0.470\linewidth]{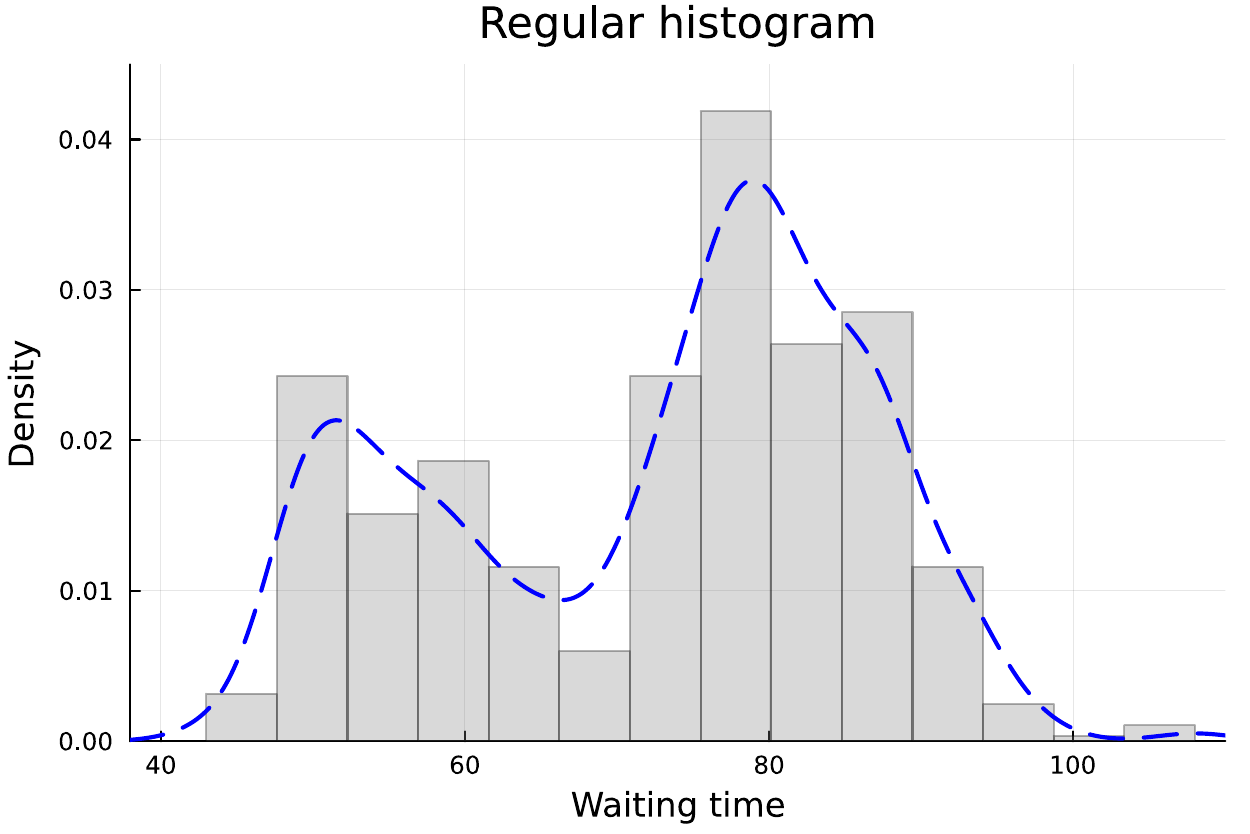} }
    \vspace{-0.8cm}
    \caption{Irregular (left) and regular (right) histograms fitted to the Old Faithful data. The dashed blue line shows the kernel estimate.}
    \label{fig:old_faithful}
\end{figure}

We fitted both the irregular random histogram and the regular histogram of \citet{knuth2019optimal} to the dataset, estimating the support of the histogram for both procedures. For comparison, we also computed a kernel density estimate using the \texttt{density()} function from base \proglang{R} with the bandwidth selector of \citet{sheather1991reliable}. The resulting density estimates are shown in Figure~\ref{fig:old_faithful}.
The irregular histogram method yields a smooth density estimate, using a relatively small number of bins, and displays a clear bimodal structure. In contrast, the regular histogram procedure chooses a fairly large number of bins in this case, resulting in a much rougher appearance. The bimodal structure is also less apparent in this case, as there are two peaks near each of the two modes in the kernel estimate. Overall, the irregular histogram displays a greater degree of agreement with the kernel estimate than Knuth's method.

\subsection{Multiple hypothesis testing}
In a multiple hypothesis testing scenario, statistical procedures often aim at controlling the false discovery rate (FDR), the expected number of true null hypotheses among the rejected ones \citep{benjamini1995controlling}. It is therefore of great interest to estimate the FDR based on the observed test statistics. This task is often carried out by using the $p$-values of the individual tests to construct an estimate of the proportion of true null hypotheses, from which the FDR can be estimated, for instance via the approach of \citet{storey2003statistical}. A model-based approach to estimating the true null proportion assumes that the $p$-values are distributed according to a mixture density of the form
\begin{equation}\label{eq:hypothesis_testing_model}
    f(p) = \pi_0\mspace{1mu} \ind_{[0,1]}(p) + (1-\pi_0)\mspace{1mu} h(p),
\end{equation}
where $\pi_0$ is the theoretical proportion of true null hypotheses and $h$ is the density of the false null hypotheses, assumed to have most of its mass near $0$ and $h(1) = 0$. A nonparametric approach to estimating $\pi_0$ is to use the observed $p$-values to compute a nonparametric density estimate $\widehat{f}$ and take $\widehat{\pi}_0 = \widehat{f}(1)$ as the estimate of $\pi_0$, which is reasonable in light of our assumptions that imply $\pi_0=f(1)$. For some previous proposed nonparametric approaches to estimating the density $f$ specifically for this purpose, see e.g.~\citet{langaas2005estimating,guan2008nonparametric} and the references therein. Our irregular histogram procedure seems like an attractive method for estimating $f$ in this case, as the support the distribution of the $p$-values is known. This avoids having to use boundary corrections, which are necessary for some classical density estimators such as kernel-based ones, since we are primarily concerned with estimating a value at the boundary of the support of the density.

To investigate the performance of our irregular histogram procedure in this setting, we conducted a small simulation study based on the model in \eqref{eq:hypothesis_testing_model}, before applying it to a real dataset. In our simulations, we took the probability density $h$ of the alternative hypotheses equal to the $\mathrm{Beta}(1,\beta)$ density for values of $\beta$ in the range $[2, 10]$. We varied the parameter $\pi_0\in \{0.5, 0.8, 0.95\}$, and considered samples of size $n\in \{200, 1000, 5000\}$. To measure the quality of the procedure, the root mean squared error (RMSE) of $\widehat{\pi}_0$ was estimated by $B = 500$ Monte Carlo samples.
For comparison, we included estimates of $\pi_0$ based on using the automatic histogram procedure of \citet{knuth2019optimal}. In a multiple testing scenario there can be a considerable proportion of observed $p$-values in the vicinity of $0$. The average RMSEs from the Monte Carlo simulations are shown in Tables~\ref{tab:rmse_beta5}~and~\ref{tab:rmse_beta10}.
\begin{table}
    \begin{minipage}{.5\linewidth}
        \centering
    \caption{Average RMSEs for $\beta = 4$}
    \label{tab:rmse_beta5}
    \vspace{0.2cm}
    \begin{tabular}{cc|cc}
    \toprule
    $\pi_0$ & $n$ & Irregular & Regular\\
    \midrule
    0.5 & 200 & 0.118 & 0.078\\
 & 1000 & 0.059 & 0.043\\
 & 5000 & 0.033 & 0.025\\ \hline
0.8 & 200 & 0.147 & 0.109\\
 & 1000 & 0.06 & 0.038\\
 & 5000 & 0.033 & 0.022\\ \hline
0.95 & 200 & 0.052 & 0.057\\
 & 1000 & 0.051 & 0.05\\
 & 5000 & 0.04 & 0.027\\
    \bottomrule
    \end{tabular}
\medskip
\end{minipage}\hfill
\begin{minipage}{.5\linewidth}
    \centering
    \caption{Average RMSEs for $\beta = 10$}
    \label{tab:rmse_beta10}
    \vspace{0.2cm}
    \begin{tabular}{cc|cc}
    \toprule
    $\pi_0$ & $n$ & Irregular & Regular\\
    \midrule
    0.5 & 200 & 0.065 & 0.117\\
 & 1000 & 0.031 & 0.063\\
 & 5000 & 0.015 & 0.034\\ \hline
0.8 & 200 & 0.059 & 0.099\\
 & 1000 & 0.038 & 0.062\\
 & 5000 & 0.017 & 0.034\\ \hline
0.95 & 200 & 0.053 & 0.057\\
 & 1000 & 0.044 & 0.051\\
 & 5000 & 0.015 & 0.023\\
    \bottomrule
\end{tabular}
\medskip
\end{minipage}
\end{table}

The relative performance of the two procedures depend heavily on the value of $\beta$ in the simulations, with Knuth's method performing better for $\beta = 4$ while the irregular procedure wins out for $\beta = 10$. This is related to the amount of smoothing introduced by the two procedures in the tail of the density. The tendency of the irregular procedure to select a large bin in regions where the density is very flat worsens the performance for the $\mathrm{Beta}(1,4)$, as this density still assigns a considerable amount of mass to $p > 0.5$, resulting in greater bias. In contrast, for $\beta = 10$ the right tail of $h$ decays very quickly, and the bias of the irregular histogram procedure becomes negligible. In this case, the variance reduction brought about by the greater amount of smoothing leads to better performance in terms of RMSE.

\begin{figure}[t]
    \centering
    {\includegraphics[width=0.470\linewidth]{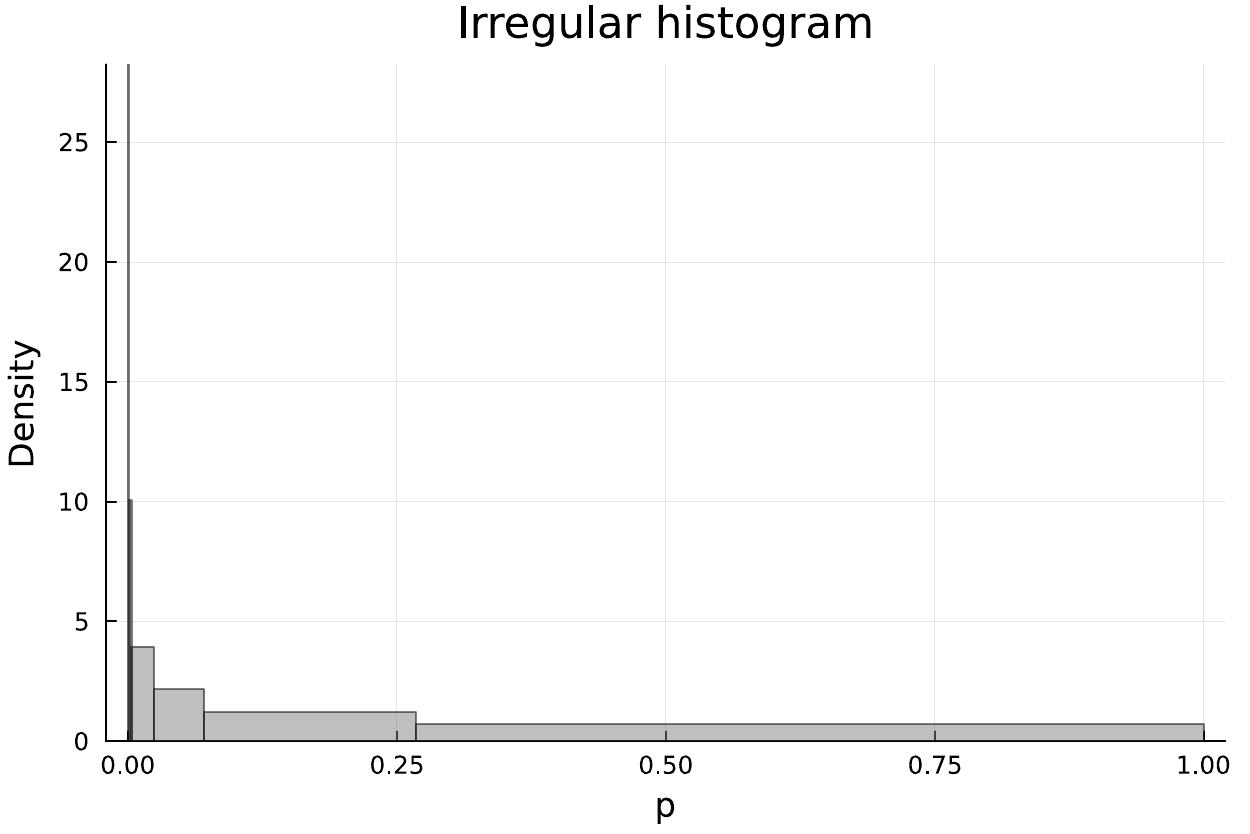} }
    \qquad
    {\includegraphics[width=0.470\linewidth]{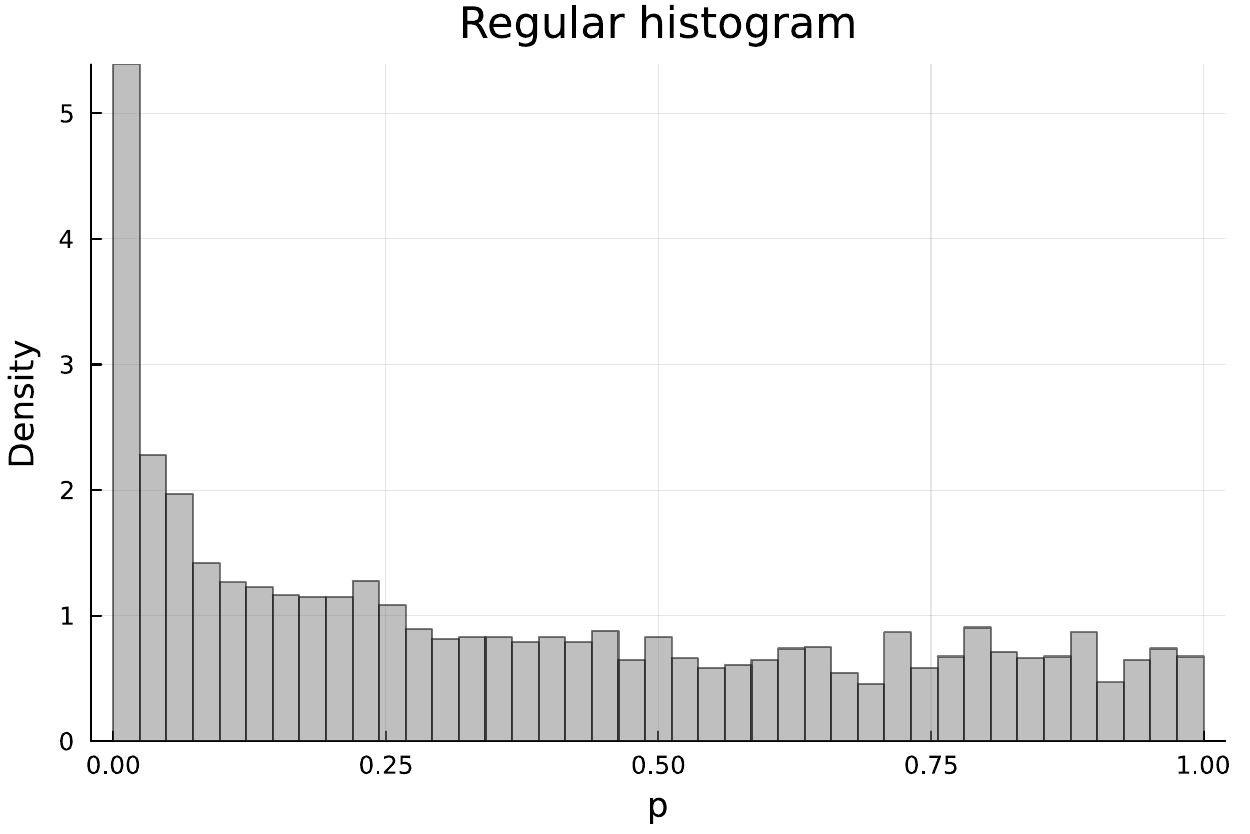} }
    \vspace{-0.5cm}
    \caption{Irregular (left) and regular (right) histograms fitted to the \citet{hedenfalk2001gene} data.}
    \label{fig:hedenfalk}
\end{figure}

To end the section, we consider an example from breast cancer research analyzed in \citet{storey2003statistical}. The study of \citet{hedenfalk2001gene} attempted to identify differences in gene expressivity in individuals with BRCA1- and BRCA2-mutation-positive tumors based on microarray measurements. \citet{storey2003statistical} use permutation tests to compute $p$-values for the measured differences in gene expressivity, leaving us with a sample of $n = 3170$ $p$-values. The resulting density estimates from applying the irregular histogram procedure and the regular histogram of Knuth are shown in Figure~\ref{fig:hedenfalk}.
The irregular histogram displays a very pronounced peak near $0$, owing to the fact that a non-negligible proportion of the $p$-values in this dataset concentrate in a narrow interval around the origin. Despite having a very narrow width, the leftmost interval contains a total of $73$ observations, accounting for more than $2 \%$ of the total sample. Although the regular histogram also shows a distinct mode near $0$, it is much less pronounced than that of the irregular histogram. The regular histogram also appears to undersmooth significantly in the right tail, as there are many sporadic bumps here that are unlikely to reflect any real structure in the data. The resulting estimates of $\pi_0$ are $\widehat{\pi}_{0,\mathrm{irr}} = 0.714$ and $\widehat{\pi}_{0,\mathrm{reg}} = 0.675$ for the irregular and regular histogram models, respectively. For comparison, we also estimated $\pi_0$ using the convex, decreasing density estimator in \citet{langaas2005estimating}, and the local false discovery rate method of \citet{phipson2013empirical}. This yielded estimates of $\widehat{\pi}_{0, \mathrm{cd}} = 0.671$ for the former method and $\widehat{\pi}_{0, \mathrm{lfdr}} = 0.744$ for the latter, indicating that the results from both of the histogram methods are quite reasonable in this case.

\section{Discussion}
We conclude the article with a brief discussion of possible extensions of the random irregular histogram, and how the ideas presented here can be used to construct computationally tractable Bayesian models for regression and hazard rate estimation.

The results of our simulation study show that the random irregular histogram model excels at mode detection for larger samples, a feat that cannot be achieved by regular histogram procedures designed for optimal performance with respect to classical estimation error criteria \citep{scott1992multivariate}. However, the improved performance of regular histograms in terms of classical loss functions for spatially homogeneous densities has led some authors to argue for their continued use, cf.~\citet{birge2006bins}. \citet{rozenholc2010irregular} propose constructing both a regular and an irregular histogram based on a penalized likelihood, and the resulting histogram is chosen to be the one with the lowest penalized likelihood of the two. Although this approach often succeeds in reducing the estimation error compared to an irregular procedure, the resulting density estimates still suffer from poor automatic mode detection whenever the regular histogram is chosen. Another possible approach in this regard would be to use a prior distribution (or penalty in the penalized likelihood case) such that, for a given value of $k$, the partitions which have more regular bin widths are assigned more prior probability or a smaller penalty. This could lead to an improved estimation error for spatially homogeneous densities, while keeping the advantage of an irregular procedure in terms of mode identification.

Although we have at present focused on density estimation in the univariate case, the random irregular histogram model can be extended to multivariate data as well. \citet{liu2023convergence} obtained a posterior convergence rate independent of the dimension of the data under a spatial sparsity constraint for a Bayesian tree method based on binary partitions. In light of these findings, an interesting avenue for further research would be to investigate whether similar theoretical results hold for the random irregular histogram in the multivariate case.

In this article, we have concerned ourselves with the task of finding the best piecewise constant density estimate, but many of the ideas presented here can be extended to other areas of statistics such as semiparametric regression and hazard rate estimation.
In the semiparametric regression setup, we consider the model $y_i = m(x_i) + \epsilon_i$ where $m\colon [0,1]\to \mathbb{R}$ is an unknown regression function and the $\epsilon_i$ are independent and $\mathrm{Normal}(0, \tau^{-2})$. A possible choice of nonparametric prior for $m$ is a piecewise constant model, where $m$ takes the fixed value $m_j$ on $\pa_j$. In this case, letting $m_j\ba \pa \sim \mathrm{Normal}(\mu, \beta^{-2})$ leads to a conjugate model for given $\pa$, which results in an analytical expression for the marginal likelihood. The computational techniques outlined in this paper can then be used to find the MAP partition $\wpa$ in this model, and the corresponding estimated regression function is a Bayesian variant of the regressogram; the regression equivalent of the histogram \citep{klemela2012bin}. In the hazard rate case, Gamma priors can be used to yield a similarly computationally tractable expression for the posterior of $\pa$.

\appendix
\renewcommand{\thesection}{Appendix \Alph{section}:}
\renewcommand{\thesubsection}{\Alph{section}.\arabic{subsection}}

\section{Sensitivity to choices of prior parameters}
To investigate the sensitivity of the irregular random histogram procedure to the choice of prior distribution, we conducted some small numerical experiments. In each experiment, we estimated the Hellinger risk for a given configuration of the prior parameters by generating random samples $\bx^{(b)}$ of size $n \in \mathcal{N} = \{100, 1000, 10\mspace{2mu}000\}$ from three test densities for $b = 1,2,\ldots, B$.
The risks were estimated by a simple Monte Carlo average,
\begin{equation}\label{eq:risk_estimate_prior}
    \widehat{R}_n(f_0, j) = \frac{1}{B}\sum_{i=1}^{B} d_H\big(f_0, \widehat{f}_{j}\big),\quad (n, m, f_0)\in \mathcal{N}\times \mathcal{M}\times \mathcal{D}',
\end{equation}
where $\mathcal{D}'$ denotes a set of test densities, $\mathcal{M}$ is the model index set for the different prior configurations and $\widehat{f}_j$ is the Bayes histogram estimate when prior $j$ is used. In our simulations, we took $B = 500$.
The chosen test densities were the $\mathrm{Gamma}(3,3)$, the $\mathrm{Beta}(3,3)$ and the $t$-distribution with $3$ degrees of freedom. For the $\mathrm{Beta}(3,3)$ we took $[0,1]$ to be the support of the density estimate, while for the $\mathrm{Gamma}(3,3)$ we computed the histogram estimate on $[0, x_{(n)}]$. For the $t$-distribution the left- and rightmost endpoints of the histogram were set equal to the minimum and maximum of the sample, respectively.

\subsection{Prior for the number of bins}
\label{subsubsec:numbin_prior}
\begin{figure}[t]
    \centering
    \includegraphics[width=0.7\linewidth]{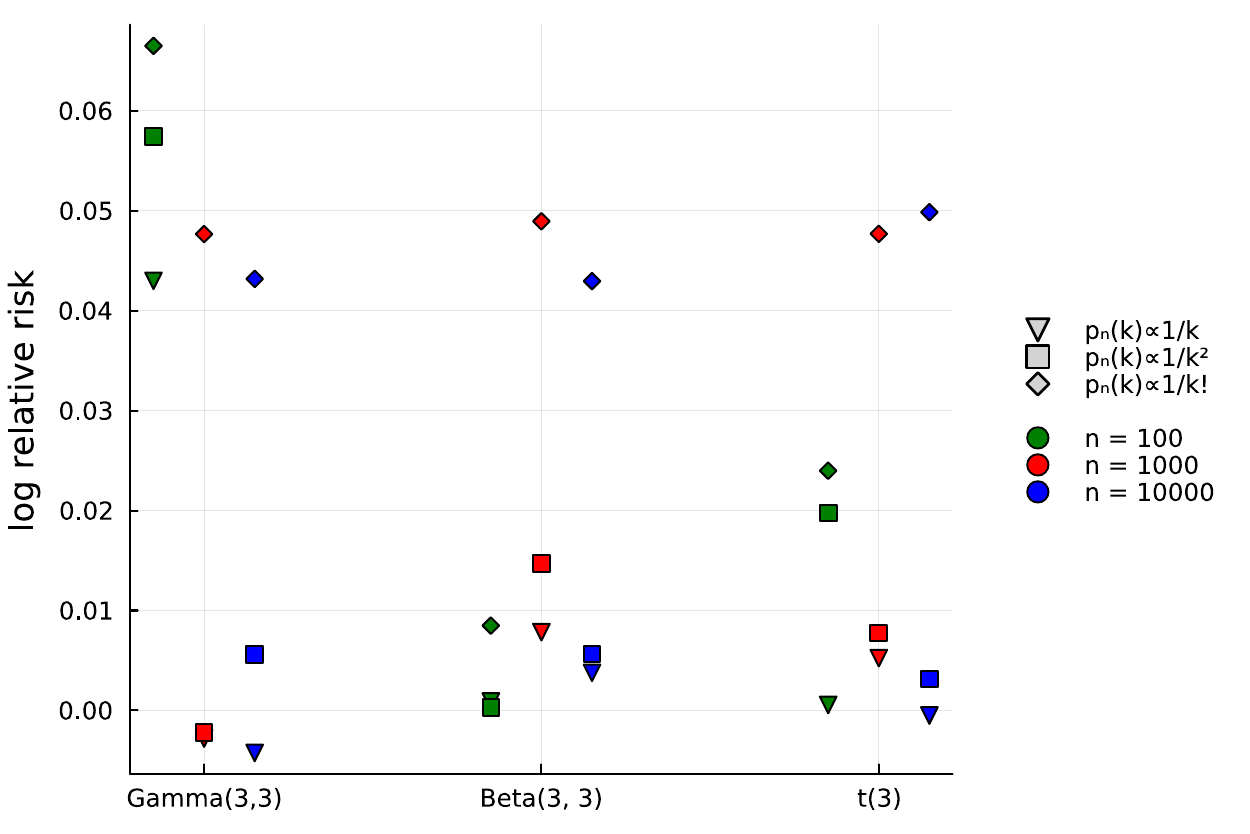}
    \caption{Logarithm of risk relative to the uniform prior $p_n(k)\propto 1$.}
    \label{fig:rel_risk_k}
\end{figure}

To start, we investigated the effect of different prior distributions for the number $k$ of bins. In the experiment, the value of the Dirichlet concentration parameter $a$ was kept fixed at $1$. We tested the prior distributions $p_n(k) \propto 1/k^{m}$ for $m\in \{0,1,2\}$ and the $\mathrm{Poiss}(1)$ distribution with $p_n(k) \propto 1/k!$ as this prior fulfills the conditions of Theorem~\ref{thm:convergence_rate}. In order to present the results, we computed the logarithm of the risk in \eqref{eq:risk_estimate_prior} relative to that of the uniform prior. The resulting log-relative risks are shown in Figure~\ref{fig:rel_risk_k}.
The uniform prior appears to yield the best performance for most combinations of the sample size and true density, with the exception of the $\mathrm{Gamma}(3,3)$ for $n = 10^4$. Both $p_n(k) \propto 1/k$ and $p_n(k)\propto 1/k^2$ generally perform similarly and not much worse than the uniform. The Poisson prior does the worst in terms of Hellinger risk, indicating that it puts too high a penalty on larger values of $k$ compared to what is optimal in this case.

\subsection{Choice of the Dirichlet concentration parameter}
\label{subsubsec:concentration_prior}
\begin{figure}[t]
    \centering
    {\includegraphics[width=0.470\linewidth]{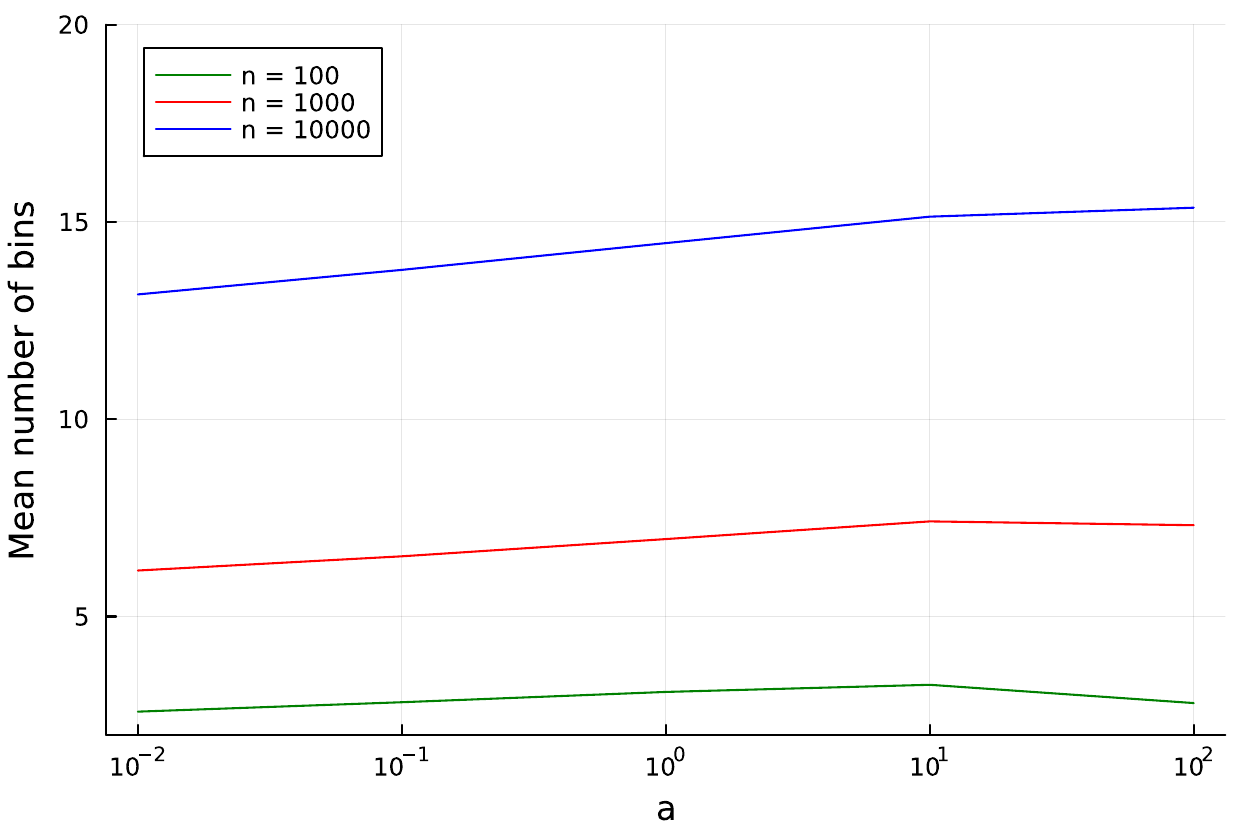} }
    \qquad
    {\includegraphics[width=0.470\linewidth]{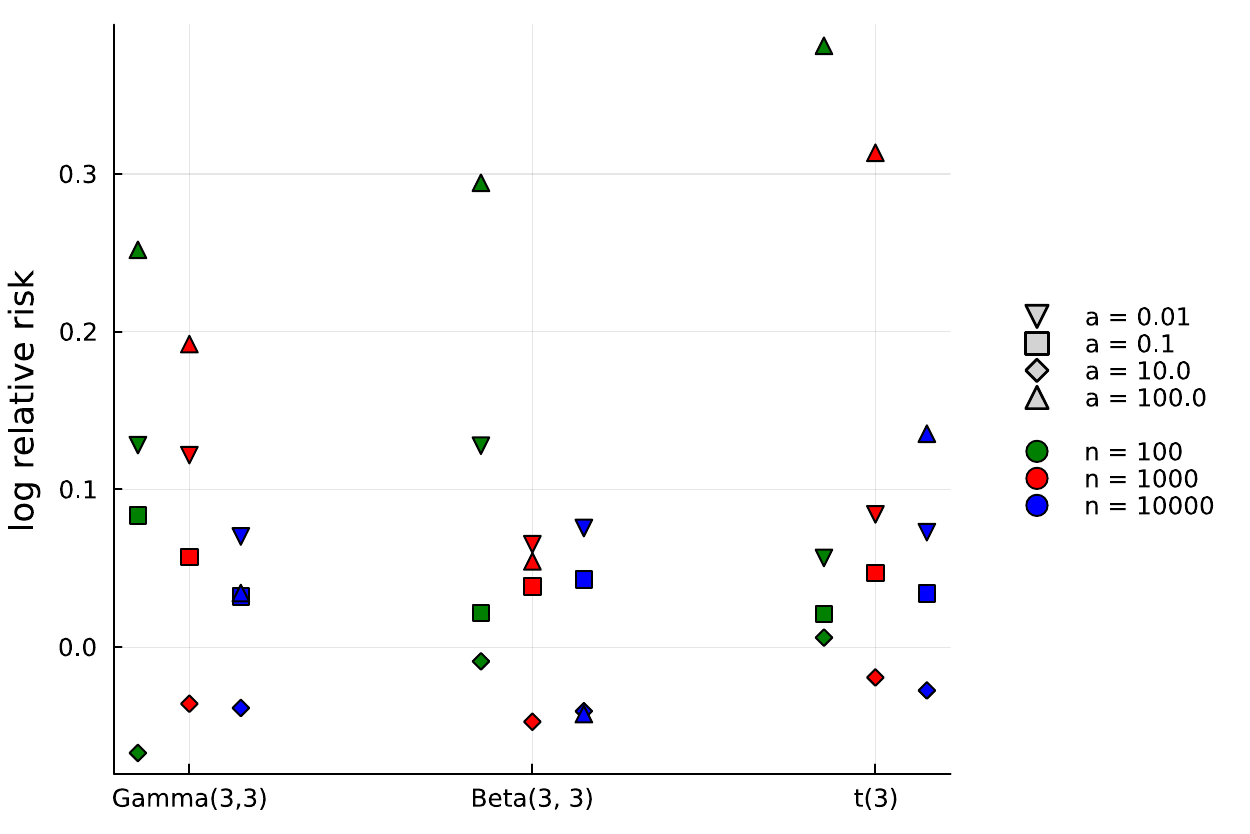} }
    \vspace{-0.5cm}
    \caption{Left: Mean value of $k$ aggregated over all test densities by sample size. Right: Logarithm of risk relative to $a = 1$.}%
    \label{fig:rel_risk_a}
\end{figure}

To study the effect of the concentration parameter $a$ on the quality of the resulting density estimates, a similar numerical experiment to that in the previous subsection was conducted. We tested the values $a \in \{0.01, 0.1, 1, 10, 100\}$. Across all values of $a$, we kept the prior for $k$ fixed to a uniform one. To investigate the effect of the concentration parameter on the number of bins in the MAP partition, we computed the mean number of bins chosen in the simulations for each choice of $a$. Furthermore, the relative Hellinger risk of the procedure was estimated for different values of $a$, using $a = 1$ as a reference.
The results are shown in Figure~\ref{fig:rel_risk_a}.
The mean number of bins chosen by the procedure tends to increase along with the concentration parameter $a$. This illustrates the point made in Section~\ref{subsec:prior_elicitation}; although smaller values of $a$ yield estimates of the bin probabilities that assign more weight to the data through \eqref{eq:posterior_mean_theta}, its effect on the posterior of $\pa$ cannot be neglected, and it can thus be beneficial to set the concentration parameter to a larger value in practice. We also plotted $a$ against the mean number of bins chosen separately for each density, but this showed a very similar picture to the aggregated results and is therefore omitted.

In terms of risk, the procedure with $a = 10$ performs the best in all cases. Its lower risk compared to $a = 1$ can likely be attributed to its tendency of choosing a greater number of bins. The smallest values of the concentration parameter $a = 0.01$ performs quite poorly, and the choice $a = 0.1$ is only slightly better. The choice $a = 100$ does very poorly for the smallest sample sizes. This is rather unsurprising in light of \eqref{eq:posterior_mean_theta}, as for $n = 100$ the prior mean ends up contributing significantly to the interval probability estimates. As the sample size increases, the influence of the prior mean weakens, and the relative risk of this procedure improves considerably. Overall, it seems that a choice of $a \in [1,10]$ should perform reasonably well for the sample sizes considered here.

\section{Proofs}\label{sec:proofs}
Our proofs are mainly based on the theory developed in \citet{ghosal1999posterior,ghosal2000rates}. These are based on the use of covering numbers, which provide a measure of the complexity of a given model. For a metric space $(\mathcal{X},d)$ the covering number $N\big(\epsilon,\mathcal{Y}, d)$ of a set $\mathcal{Y}\subseteq \mathcal{X}$ is defined as the minimal number of open balls in $\mathcal{X}$ needed to cover $\mathcal{Y}$. Covering numbers can be controlled using the slightly stronger concept of bracketing numbers. For a metric space $(\mathcal{F},d)$ consisting of functions from $\mathcal{A}\to \mathcal{B}$, we define the bracket $[f,g]$ for two functions $f,g\in \mathcal{F}$ to be the set of functions $h\colon \mathcal{A}\to\mathcal{B}$ satisfying $f(x)\leq h(x)\leq g(x)$ for all $x\in \mathcal{A}$. For two vectors in a Euclidian metric space, $\boldsymbol{a}, \boldsymbol{b}\in \mathbb{R}^k$, we define the bracket $[\boldsymbol{a}, \boldsymbol{b}]$ as the set of those $\boldsymbol{c}$ satisfying $a_j \leq c_j\leq b_j$ for all $j$. We define the bracketing number of a set $\mathcal{H}\subseteq \mathcal{F}$ to be the minimal number of brackets in $\mathcal{F}$ needed to cover $\mathcal{H}$.

Before proceeding with the proofs, we review some notation. Given an observed i.i.d.~sample $\bx = (x_1, \ldots, x_n)$, a likelihood function $f$ and a prior $P_n$ on $\mathcal{F}$, the space of probability densities on $[0,1]$, the posterior probability of a measurable set $\mathcal{A}$ is 
\begin{equation*}
    P_n(f\in\mathcal{A}\ba \bx) = \frac{\int_{\mathcal{A}} \prod_{i=1}^n f(x_i)\mspace{2mu}\dd P_n(f)}{\int_{\mathcal{F}} \prod_{i=1}^n f(x_i)\mspace{2mu}\dd P_n(f)}.
\end{equation*}
From the form of the random histogram prior given in \eqref{eq:prior_model_hierarchical}, this probability can be decomposed as
\begin{equation*}
    P_n(f\in\mathcal{A}\ba \bx) = \frac{\sum_{\pa\in \mathcal{P}_{\Tn}}p_n(\pa)\int_{\mathcal{A}_{\pa}} p(\btheta\ba \pa)\prod_{i=1}^n f_{\pa, \btheta}(x_i)\mspace{2mu}\dd \btheta}{\sum_{\pa\in \mathcal{P}_{\Tn}}p_n(\pa)\int_{\mathcal{S}_k} p(\btheta\ba \pa)\prod_{i=1}^n f_{\pa, \btheta}(x_i)\mspace{2mu}\dd \btheta},
\end{equation*}
where $\mathcal{A}_{\pa} = \{\btheta\in \mathcal{S}_k\colon f_{\pa, \btheta}\in \mathcal{A}\}$ and $f_{\pa, \btheta}$ is the density given in \eqref{eq:histogram_model}. Since our primary goal is to study the posterior mean of $f$ conditional on $\wpa$, the posterior distribution $f\ba \bx, \pa$ is of interest.  Conditional on the chosen partition $\pa\in \mathcal{P}_{\Tn,k}$, we have
\begin{equation*}
    P_n(f_{\pa, \btheta}\in\mathcal{A}\ba \bx, \pa) = \frac{\int_{\mathcal{A}_{\pa}} p(\btheta\ba \pa)\prod_{i=1}^n f_{\pa, \btheta}(x_i)\mspace{2mu}\dd \btheta}{\int_{\mathcal{S}_k} p(\btheta\ba \pa)\prod_{i=1}^n f_{\pa, \btheta}(x_i)\mspace{2mu}\dd \btheta}.
\end{equation*}
Moreover, the sample $x_1, \ldots, x_n$ is assumed to be i.i.d.~from a true density $f_0$ and we denote by $P_0^n$ the $n$-fold product measure induced by $f_0$.

\subsection{Convergence rate}\label{subsec:convergence_rate_proof}

For $M > 0$ and a sequence $\epsilon_n = \smallO(1)$, we define $\mathcal{A}_{n} = \big\{f\colon d_H\big(f_0, f\big) \geq M\epsilon_n\big\}$.
In line with the notation introduced in the beginning of the section, for $\pa\in \mathcal{P}_{\Tn, k}$ we denote by $\mathcal{A}_{n,\pa}$ the set given by $\big\{\btheta\in \Sk\colon d_H\big(f_0, f_{\pa,\btheta}\big) \geq M\epsilon_n\big\}$.

Our first result concerns the consistency of the conditional posterior mean based on a maximum a posteriori model selection procedure.
\begin{lemma}\label{lemma:bayes_est_concentration}
    Let $\epsilon_n, \bar{\epsilon}_n$ be sequences satisfying $\bar{\epsilon}_n \leq \epsilon_n$ and $n\bar{\epsilon}_n^2 \to \infty$.
    Suppose that for some $b > 0$ and a constant $c > 0$, we can find a sequence of interval partitions $\{\pa^{(n)}\}_{n\in \mathbb{N}}$ of $[0,1]$, where $\pa^{(n)}\in \mathcal{P}_{\Tn}$, such that the event \begin{equation}\label{eq:Bn_definition}
        e^{(2+c)n\bar{\epsilon}^2_n} p\big(\bx\ba \pa^{(n)}\big)\, p_n(\pa^{(n)}) / \prod_{i=1}^n f_0(x_i) > b,
    \end{equation}
    denoted $\mathcal{B}_n$, satisfies $P_0^n\big(\bx \in B_n^c\big) = \bigO\big(\epsilon_n^2\big)$. Suppose further that we can find sets $\big\{\mathcal{F}_n\big\}$ in $\mathcal{F}$ such that
    \begin{align}
         N\big(\epsilon_n/2, \mathcal{F}_n, d_H) &< \exp\big(n\epsilon^2_n\big),\label{eq:ghosal_condition_entropy} \\
         P_n\big(\mathcal{F}_n^c\big) &< \exp\big(-\{c+4\}n\bar{\epsilon}^2_n \big) \label{eq:ghosal_condition_prior_mass}, 
    \end{align}
    Then for all sufficiently large $M > 0$ we have $\E_{\bx\sim f_0}\big\{P_n\big(f_{\wpa,\btheta}\in \mathcal{A}_{n}\, \big |\, \bx,\wpa\big)\big\} = \smallO(1)$.
\end{lemma}
\begin{proof}
    Take $\{\mathcal{I}^{(n)}\}$ to satisfy \eqref{eq:Bn_definition}, and write $D_n(\pa) = p(\bx\ba \pa)\, p_n(\pa) / \prod_{i=1}^n f_0(x_i)$. We work with the following posterior probability:
    \begin{align*}
        P_n\Big(f_{\wpa, \btheta}\in \mathcal{A}_{n}\, \big |\, \bx,\wpa\Big) &= \frac{\int_{\mathcal{A}_{n,\wpa}} p(\btheta\ba \wpa)\prod_{i=1}^n f_{\wpa, \btheta}(x_i)\mspace{2mu}\dd \btheta}{\int_{\mathcal{S}_k} p(\btheta\ba \wpa)\prod_{i=1}^n f_{\wpa, \btheta}(x_i)\mspace{2mu}\dd \btheta} \\ &\leq \frac{\int_{\mathcal{A}_{n,\wpa}} p(\btheta\ba \wpa)\prod_{i=1}^n f_{\wpa, \btheta}(x_i)\mspace{2mu}\dd \btheta}{\int_{\mathcal{S}_k} p(\btheta\ba \wpa)\prod_{i=1}^n f_{\wpa, \btheta}(x_i)\mspace{2mu}\dd \btheta}\ind_{\mathcal{B}_n}(\bx) + \ind_{\mathcal{B}_n^c}(\bx).
    \end{align*}
    To get a useful bound on the above probability, we multiply by $p_n\big(\wpa\big)/\prod_{i=1}^n f_0(x_i)$ in both the numerator and the denominator in the first term of the above expression, resulting in the following inequalities
    \begin{align*}
    \begin{split}
        P_n\Big(f_{\wpa, \btheta}\in \mathcal{A}_{n}\, \big |\, \bx,\wpa\Big) &\leq D_n\big(\wpa\big)^{-1}p_n\big(\wpa\big)\int_{\mathcal{A}_{n,\wpa}} p\big(\btheta\bba \wpa\big)\prod_{i=1}^n \frac{f_{\wpa, \btheta}(x_i)}{f_0(x_i)} \, \dd \btheta\, \ind_{\mathcal{B}_n}(\bx) + \ind_{\mathcal{B}_n^c}(\bx) \\ &\leq D_n\big(\pa^{(n)}\big)^{-1}p_n\big(\wpa\big)\int_{\mathcal{A}_{n,\wpa}} p\big(\btheta\bba \wpa\big) \prod_{i=1}^n \frac{f_{\wpa, \btheta}(x_i)}{f_0(x_i)} \, \dd \btheta\,  \ind_{\mathcal{B}_n}(\bx)  + \ind_{\mathcal{B}_n^c}(\bx),
    \end{split}
    \end{align*}
    where we used the argmax property of $\wpa$, so that $p_n\big(\wpa\big)\, p\big(\bx\bba \wpa\big) \geq p_n\big(\pa^{(n)}\big)\, p\big(\bx\bba \pa^{(n)}\big)$ and hence $D_n\big(\wpa\big)\geq D_n\big(\pa^{(n)}\big)$ for the second inequality. The first term in the above expression can be further bounded by
    \begin{align*}
        &\, b^{-1}e^{(2+c)n\bar{\epsilon}_n^2} p_n\big(\wpa\big)\int_{\mathcal{A}_{n,\wpa}} p\big(\btheta\bba \wpa\big) \prod_{i=1}^n \frac{f_{\wpa, \btheta}(x_i)}{f_0(x_i)} \, \dd \btheta \\ \leq &\, 
        b^{-1}e^{(2+c)n\bar{\epsilon}_n^2} \sum_{\pa\in \mathcal{P}_{\Tn}} p_n(\pa)\int_{\mathcal{A}_{n,\pa}} p\big(\btheta\bba \pa\big) \prod_{i=1}^n \frac{f_{\pa, \btheta}(x_i)}{f_0(x_i)} \, \dd \btheta
        \\ = &\, b^{-1}e^{(2+c)n\bar{\epsilon}_n^2} \int_{\mathcal{A}_{n}} \prod_{i=1}^n\frac{f(x_i)}{f_0(x_i)}\dd P_n(f).
    \end{align*}
    Arguing as in the proof of Theorem 8.9 in \citet{ghosal2017fundamentals}, under the conditions \eqref{eq:ghosal_condition_entropy} and \eqref{eq:ghosal_condition_prior_mass} the expectation of the right hand side of the above under $P_0^n$ is $\mathcal{O}\big(e^{-\rho n\epsilon_n^2}\big)$ when it is multiplied by $e^{(2+c)n\bar{\epsilon}_n^2}$ for sufficiently large $M > 0$ and some $\rho > 0$ not depending on $n$. From this we deduce that
    \begin{equation}
    \label{eq:convegence_rate_posterior_condprob}
        \E_{\bx \sim f_0} \left\{P_n\Big(f_{\wpa,\btheta}\in \mathcal{A}_{n} \bba \bx,\wpa\Big) \right\} \leq \mathcal{O}\big(e^{-\rho n\epsilon_n^2}\big) + P_0^n\big(\bx \in \mathcal{B}_n^c\big).
    \end{equation}
    To conclude the proof, we note that by arguing as on page 507 of \citet{ghosal2000rates}, we have
    \begin{equation*}
        \dhsq\big(f_0, \widehat{f}_{\wpa}\big) \leq M^2\epsilon_n^2 + 2P_n\Big(f_{\wpa,\btheta}\in \mathcal{A}_n \bba \bx, \wpa\mspace{1mu}\Big),
    \end{equation*}
    which follows from the convexity of the map $g\mapsto \dhsq(f_0, g)$, the conditional version of Jensen's inequality and the fact that the Hellinger metric is bounded by $\sqrt{2}$. Since both terms on the right hand side of \eqref{eq:convegence_rate_posterior_condprob} are $\bigO(\epsilon_n^2)$, we obtain
    \begin{equation*}
        \E_{\bx \sim f_0}\big\{\dhsq\big(f_0, \widehat{f}_{\wpa}\big)\big\} = \mathcal{O}\big(\epsilon_n^2\big),
    \end{equation*}
    which proves the claim.
\end{proof}

We now turn our attention to finding useful bound on the covering number. Define
\begin{equation}\label{eq:histogram_densities_irr}
    \mathcal{H}_{\Tn,k} = \Big\{f\colon [0,1]\to \mathbb{R}\colon f(x) = \sum_{j=1}^{k} |\pa_j|^{-1}\theta_j\ind_{\pa_j}(x) \mathrm{\ for\ some\ } \pa\in \mathcal{P}_{\Tn,k}\ \mathrm{and}\ \btheta\in \mathcal{S}_{k}\Big\}.
\end{equation}
To show that the conditions of Lemma~\ref{lemma:bayes_est_concentration} hold, we need to bound on the bracketing number of $\mathcal{H}_{\Tn, k}$. This is achieved by relating it to the bracketing number of the simplex $\Sk$. Variants of the following lemmas appear in the proofs of \citet{ghosal2001bernstein,scricciolo2007rates}. We include the argument in its full length here for completeness.

\begin{lemma}\label{lemma:histogram_simplex_cover}
    Let $\pa$ be a partition of $[0,1]$.
    Let $l = \sum_{j=1}^k a_j \omega_j$ and $u = \sum_{j=1}^k b_j \omega_j$ with $\omega_j(x) = |\pa_j|^{-1}\ind_{\pa_j}(x)$.
    Suppose that $h(x) = \sum_{j=1}^k \theta_j\omega_j(x)$, where $\btheta\in \mathcal{S}_{k}$. If $[\mathbf{a}, \mathbf{b}]$ is an $\epsilon$-bracket containing $\btheta$, then $[l, u]$ is an $\epsilon$-bracket containing $h$. 
\end{lemma}
\begin{proof}
The claim that $h\in [l,u]$, follows from the fact that $|\pa_j|^{-1}a_j \leq |\pa_j|^{-1}\theta_j \leq |\pa_j|^{-1}b_j$ for all $j$ by assumption.

It remains to show that $d_H\big(l,u\big) < \epsilon$. We start off by noting that
\begin{equation*}
    \dhsq\big(a_j \omega_j, b_j \omega_j\big) = \int_{\pa_j} \left(\sqrt{|\pa_j|^{-1}a_j} - \sqrt{|\pa_j|^{-1}b_j}\,\right)^2 \dd x = \left(\sqrt{a_j} - \sqrt{b_j}\,\right)^2.
\end{equation*}
From Lemma 4 in \citet{genovese2000rates}, we have that $\dhsq\big(l, u \big) \leq \sum_{j=1}^k \dhsq\big(a_j \omega_j, b_j \omega_j \big)$ and hence,
\begin{equation*}
    \dhsq\big(l, u \big) \leq \sum_{j=1}^k \left(\sqrt{a_j} - \sqrt{b_j}\,\right)^2 = \dhsq\big(\mathbf{a},\mathbf{b}\big) < \epsilon^2,
\end{equation*}
which was to be shown.
\end{proof}

Define $\mathcal{H}_{\Tn, \pa}$ to be the set of piecewise constant densities based on the partition $\pa$.
We now give a bound on the covering number of $\mathcal{H}_{\Tn, \pa}$.
\begin{lemma}\label{lemma:bracket_num_bound}
    Let $\pa\in \mathcal{P}_{\Tn, k}$.
    Then the $\epsilon$-bracketing number of $\mathcal{H}_{\Tn, \pa}$ for $\epsilon \leq 1$ satisfies
    \begin{equation}\label{eq:bracketing_bound_histogram}
        N_{[\,]}\big(\epsilon, \mathcal{H}_{\Tn, \pa}, d_H\big) \leq \frac{k(2\pi e)^{k/2}}{\epsilon^{k-1}}
    \end{equation}
    Moreover, the $\epsilon$-convering number of $\mathcal{F}_n = \cup_{k=1}^{s_n} \mathcal{H}_{\Tn, k}$ satisfies
    \begin{equation}
        N_{[\,]}\big(\epsilon, \mathcal{F}_{n}, d_H\big) \leq \left(\frac{Lk_n}{\epsilon s_n}\right)^{s_n}
    \end{equation}
    for a universal constant $L > 0$.
\end{lemma}
\begin{proof}
    To show that the first claim holds, let $[\mathbf{a}_1, \mathbf{b}_1], \ldots, [\mathbf{a}_m, \mathbf{b}_m]$ be an $\epsilon$-bracketing of $\mathcal{S}_{k}$, and assume that $m$ is chosen so that the number of brackets is minimal. Define the associated functions $l_1, u_1, \ldots, l_m, u_m$ as in Lemma \ref{lemma:histogram_simplex_cover},
    \begin{equation*}
        l_i(x) = \sum_{j=1}^k a_{i,j}|\pa_j|^{-1}\ind_{\pa_j}(x),\quad u_i(x) = \sum_{j=1}^k b_{i,j}|\pa_j|^{-1}\ind_{\pa_j}(x).
    \end{equation*}
    Let $h\in \mathcal{H}_{\Tn, \pa}$ be given by $h(x) = \sum_{j=1}^k |\pa_j|^{-1}\theta_j \ind_{\pa_j}(x)$, for some $\btheta\in \mathcal{S}_{k}$. Since $\mathcal{S}_{k}\subseteq \cup_{l=1}^m [\mathbf{a}_l, \mathbf{b}_l]$, we have $\btheta\in [\mathbf{a}_i, \mathbf{b}_i]$ for some $i$. By Lemma~\ref{lemma:histogram_simplex_cover}, $[l_i, u_i]$ is an $\epsilon$-bracket containing $h$, and since $h$ was arbitrary, we conclude that $\mathcal{H}_{\Tn, \pa}\subseteq \cup_{i=1}^m [l_i, u_i]$. By the minimality of $m$, this shows that the Hellinger bracketing number of $\mathcal{H}_{\Tn, \pa}$ can be no greater than that of $\mathcal{S}_{k}$. By Lemma 2 in \citet{genovese2000rates}, the bracketing number of the $k$-simplex $\mathcal{S}_{k}$ satisfies
    \begin{equation*}
        N_{[\,]}\big(\epsilon, \mathcal{S}_{k}, d_H\big) \leq \frac{k(2\pi e)^{k/2}}{\epsilon^{k-1}}.
    \end{equation*}
    Using this fact we find that
    \begin{equation*}
        N_{[\,]}\big(\epsilon, \mathcal{H}_{\Tn, \pa}, d_H\big)\leq  N_{[\,]}\big(\epsilon, \mathcal{S}_{k}, d_H\big) \leq \frac{k(2\pi e)^{k/2}}{\epsilon^{k-1}},
    \end{equation*}
    which was to be shown.

    To prove the second part of the lemma, we note that for any $k\leq s_n$, a piecewise constant density based on the partition $\pa\in \mathcal{P}_{\Tn, k}$ can be written as 
    \begin{equation*}
    f_{\btheta, \pa}(x) = \sum_{j=1}^{k} |\pa_j|^{-1}\theta_j\ind_{\pa_j}(x), \text{ for } x\in [0,1],
    \end{equation*}
    for some $\btheta\in \Sk$.
    Now by choosing $\mathcal{J} \in \mathcal{P}_{\Tn, s_n}$ such that $\mathcal{J}$ is a refinement of $\pa$ and $\tilde{\btheta}\in \mathcal{S}_{s_n}$ appropriately, we have that
    \begin{equation*}
        f_{\mathcal{J}, \tilde{\btheta}}(x) = f_{\mathcal{I}, \btheta}(x),\ \mathrm{for\ each\ } x\in [0,1].
    \end{equation*}
    We conclude that $\mathcal{P}_{\Tn, k} \subseteq \mathcal{P}_{\Tn, s_n}$, and as such, $\mathcal{F}_n = \cup_{k=1}^{s_n}\mathcal{H}_{\Tn, k} \subseteq \mathcal{H}_{\Tn, s_n}$. Thus, it suffices to bound the covering number of $\mathcal{P}_{\Tn, s_n}$.
    Let $\mathcal{H}_{\Tn, \pa}$ be the set of piecewise constant densities based on the partition $\pa$. By construction, we have that 
    \begin{equation*}
    \mathcal{H}_{\Tn, s_n} = \bigcup_{\pa\in\, \mathcal{P}_{\Tn,s_n}}\mathcal{H}_{\Tn, \pa}.
    \end{equation*}
    By \eqref{eq:bracketing_bound_histogram}, the bracketing number of $\mathcal{H}_{\Tn, \pa}$ for $\pa\in \mathcal{P}_{\Tn, s_n}$ is bounded by the bracketing number of $\mathcal{S}_{s_n}$.
    We note that there are $\binom{k_n-1}{s_n-1}$ possible partitions of $[0,1]$ of size $s_n$. By a union bound, we have for any $\epsilon \leq 1$ that
    \begin{align*}
        N_{[\,]}\left(\epsilon, \mathcal{H}_{\Tn, s_n}, d_H\right) &\leq \sum_{\pa\in \mathcal{P}_{\Tn,s_n}} N_{[\,]}\left(\epsilon, \mathcal{H}_{\Tn, \pa}, d_H\right) \\ &\leq \sum_{\pa\in \mathcal{P}_{\Tn,s_n}} N_{[\,]}\left(\epsilon, \mathcal{S}_{s_n}, d_H\right) \\ & = 
        \binom{k_n-1}{s_n-1}N_{[\,]}\left(\epsilon, \mathcal{S}_{s_n}, d_H\right).
    \end{align*}
    Using the fact that the bracketing number bounds the covering number for the same $\epsilon$ together with previous estimates, we deduce the following chain of inequalities:
    \begin{align*}
        N\big(\epsilon, \mathcal{H}_{\Tn, s_n}, d_H\big) &\leq \sum_{\pa\in \mathcal{P}_{\Tn, s_n}}N_{[\,]}\big(\epsilon, \mathcal{H}_{\Tn, \pa}, d_H\big)  \\ &\leq \binom{k_n-1}{s_n-1}N_{[\,]}\big(\epsilon, \mathcal{S}_{s_n}, d_H\big)\\ &\leq \left(e k_n/s_n\right)^{s_n}\frac{s_n(2\pi e)^{s_n/2}}{\epsilon^{s_n-1}},
    \end{align*}
    where we used that $\binom{k_n-1}{s_n-1} \leq \binom{k_n}{s_n}\leq \left(ek_n/s_n\right)^{s_n}$. To conclude the proof, we note that as $s_n \leq 2^{s_n}$ for $s_n \geq 1$,
    \begin{equation*}
        N\big(\epsilon/2, \mathcal{H}_{\Tn, s_n}, d_H\big) \leq \left(\frac{e k_n}{s_n}\right)^{s_n}\frac{s_n(2\pi e)^{s_n/2}}{(\epsilon/2)^{s_n-1}} \leq  \left(\frac{Lk_n}{\epsilon s_n}\right)^{s_n},
    \end{equation*}
    for some constant $L > 0$, which was to be shown.
\end{proof}

In the sequel, for a given function $h\in \mathbb{L}_1\big([0,1]\big)$ we write $h_{\pa}$ for the piecewise constant version of $h$ based on the partition $\pa\in \mathcal{P}_{\Tn, k}$, defined by
\begin{equation*}
    h_\pa(x) = \sum_{j=1}^k h_j\ind_{\pa_j}(x),\quad \mathrm{for}\ x\in [0,1],
\end{equation*}
where $h_j = |\pa_j|^{-1}\int_{\pa_j} h(x)\, \dd x$.
The following lemma establishes an approximation rate result for H\"{o}lder continuous functions.

\begin{lemma}\label{lemma:histogram_approximation_rate}
    Let $h\colon [0,1]\to \mathbb{R}$ be $\alpha$-H\"{o}lder continuous. Let $\big\{\pa^{(k)}\big\}_{k=1}^{\infty}$ be a sequence of partitions of $[0,1]$ such that $\max_{j} \big|\pa^{(k)}_j\big| = \mathcal{O}\big(k^{-1}\big)$. Then the sequence of piecewise constant approximations $\{h_{\pa^{(k)}}\}_{k=1}^\infty$ of $h$ satisfies
    \begin{equation*}
        \big\lVert h - h_{\pa^{(k)}}\big\rVert_\infty = \mathcal{O}\big(k^{-\alpha}\big).
    \end{equation*}
    Moreover, if $h$ is strictly positive we have $\dhsq\big( h, h_{\pa^{(k)}}\big) = \mathcal{O}\big(k^{-2\alpha}\big)$
\end{lemma}
\begin{proof}
    Fix $k\in \mathbb{N}$ and let $x\in [0,1]$. Let $j$ be such that $x\in \pa^{(k)}_j$. Then,
    \begin{align*}
        |h(x) - h_{\pa^{(k)}}(x)| &= \Big\lvert h(x) - \big\lvert\pa^{(k)}_j\big\rvert^{-1}\int_{\pa^{(k)}_j} h(s)\, \dd s\Big\rvert \\ &\leq \big\lvert\pa^{(k)}_j\big\rvert^{-1}\int_{\pa^{(k)}_j} \lvert h(x) - h(s) \rvert\, \dd s \\ &\leq 
        \big\lvert\pa^{(k)}_j\big\rvert^{-1}\int_{\pa^{(k)}_j} L_0|x-s|^{\alpha}\, \dd s \\ &\leq
        |\pa^{(k)}_j|^{-1}L_0\big\lvert\pa^{(k)}_j\big\rvert^{1+\alpha}\\ &=
        L_0\big\lvert\pa^{(k)}_j\big\rvert^{\alpha}
    \end{align*}
    where we used that $|x-s| \leq |I_{k,j}|$ for $x,s\in I_{k,j}$. Since $\max_{j}\big\lvert\pa^{(k)}_j\big\rvert = \mathcal{O}(k^{-1})$, we find
    \begin{equation*}
        \big\lVert h - h_{\pa^{(k)}}\big\rVert_\infty = \sup_{x\in [0,1]} |h(x) - h_{\pa^{(k)}}(x)| \leq L_0\max_j\big\lvert \pa^{(k)}_j\big \rvert^{\alpha} = \mathcal{O}(k^{-\alpha}),
    \end{equation*}
    For the second part, we note that 
    \begin{equation*}
        \dhsq\big(h, h_{\pa^{(k)}}\big) \leq \int_0^1 \frac{\big\{h(x) - h_{\pa^{(k)}}(x)\big\}^2}{h(x)}\, \dd x,
    \end{equation*}
    cf.~Lemmas 2.5 and 2.7 in \citep{tsybakov2009nonparametric}. Since $h$ is continuous and strictly positive, the last expression is upper bounded by a constant multiple of $\lVert h - h_{\pa^{(k)}}\rVert_\infty^2$ as a consequence of the extreme value theorem.
\end{proof}

To show that the condition \eqref{eq:Bn_definition} holds, it suffices to show that the prior distribution assigns sufficient probability to the event that $f\in \mathcal{N}'_{\epsilon_n}(f_0)$, where
\begin{equation*}
    \mathcal{N}'_{\epsilon}(f_0) = \big\{f\colon \dhsq\big(f_0, f\big)\big\lVert f_0/f \big\rVert_\infty \leq \epsilon^2\big\}, \quad \epsilon > 0.
\end{equation*}
The following lemma gives a lower bound on the probability of $\mathcal{N}'_{\epsilon}(f_0)$.

\begin{lemma}
\label{lemma:prior_concentration_rate_irreg_hist}
    Suppose the hypotheses of Theorem~\ref{thm:convergence_rate} are met. Let $C_1, C_2$ be positive constants and let $k = k(\epsilon) \in \mathbb{N}$ satisfy
    \begin{equation}
        C_1 \epsilon^{-1/\alpha} \leq k \leq C_2 \epsilon^{-1/\alpha}.
    \end{equation}
    Then we can find a partition $\mathcal{I}^{(n)}\in\mathcal{P}_{\Tn,k}$ and two constants $c_1, c_2$ not depending on $\epsilon, k$ and $\pa^{(n)}$ such that for all sufficiently small $\epsilon > 0$,
    \begin{equation}
    \label{eq:prior_concentration_rate_hist}
        P_n\big(\mathcal{N}'_{\epsilon}(f_0)\big) \geq p_n\big(\pa^{(n)}\big)\, c_1\exp\big(-c_2k\log(1/\epsilon)\big),
    \end{equation}
    provided $n$ is sufficiently large.
\end{lemma}
\begin{proof}[\textbf{Proof}]
    We argue as in the beginning of the proof of Theorem 2 in \citet{scricciolo2007rates}.
    Let $\epsilon > 0$ to be determined later. If $n$ is such that $k(\epsilon) > k_n$, then the inequality \eqref{eq:prior_concentration_rate_hist} is trivially true as the right hand side is zero.
    Our strategy will be to relate the neighborhood $\mathcal{N}_\epsilon'(f_0)$ to neighborhoods in terms of the squared Hellinger distance, as the latter is more convenient to work with. To accomplish this, we first need an estimate of the squared Hellinger distance between the piecewise constant approximation and the true density $f_0$. We start by constructing the partition $\pa^{(n)}$ by including every $\lceil k_n/k \rceil$ index in the partition. The maximal size of a bin in this partition is then bounded by
    \begin{equation*}
        \max_{j} \big|\pa_j^{(n)}\big| \leq \lceil k_n/k \rceil \max_j \{\tau_{n,j} - \tau_{n,j-1}\} \leq A\lceil k_n/k \rceil k_n^{-1} \leq 2Ak^{-1},
    \end{equation*}
    provided $n$ is sufficiently large. This shows that $\max_j \big|\pa^{(n)}_j\big| = \mathcal{O}\big(k^{-1}\big)$. By a similar argument, we arrive at $\min_j \big|\pa^{(n)}_j\big| \geq Bk^{-1}/2$.
    Let $\gamma = c_0\epsilon$ for a positive constant $c_0$ to be determined later.
    To bound the squared Hellinger distance between $f_0$ and $f_{\pa^{(n)}, \btheta}$ we apply the triangle inequality together with the inequality $(a+b)^2 \leq 2a^2+2b^2$, yielding
    \begin{equation}\label{eq:convergence_rate_triangle}
        \dhsq\big(f_0, f_{\pa^{(n)},\btheta}\big) \leq 2\dhsq\big(f_0, f_{0,\pa^{(n)}}\big) + 2\dhsq\big(f_{0,\pa^{(n)}},f_{\pa^{(n)},\btheta}\big).
    \end{equation}
    To bound the first term of \eqref{eq:convergence_rate_triangle}, we leverage Lemma~\ref{lemma:histogram_approximation_rate} to find that
    \begin{equation*}
        \dhsq\big(f_0, f_{0,\pa^{(n)}}\big) \leq \tilde{A}k^{-2\alpha} \leq \tilde{A}C_1^{-2\alpha}\epsilon^{2}= \tilde{A}C_1^{-2\alpha}c_0^2\gamma^2,
    \end{equation*}
    for some constant $\tilde{A} > 0$.
    We now bound the last term on the right hand side of \eqref{eq:convergence_rate_triangle}. Denoting $\pi^{(n)}_j = \int_{\pa^{(n)}_j} f_0(x)\,\dd x$ and using the fact that $\dhsq\big(f_{\pa^{(n)},\btheta}, f_{0,\pa^{(n)}}\big)\leq \big\lVert f_{\pa^{(n)},\btheta} - f_{0,\pa^{(n)}}\big\rVert_1$, we obtain
    \begin{equation*}
        \dhsq\big(f_{\pa^{(n)},\btheta},f_{0,\pa^{(n)}}\big) \leq  \big\lVert f_{\pa^{(n)},\btheta} - f_{0,\pa^{(n)}}\big\rVert_1 =\sum_{j=1}^k \int_{\pa^{(n)}_j} \big|\pa^{(n)}_{j}\big|^{-1}\mspace{2mu}\big\lvert \theta_j - \pi_j^{(n)}\big\rvert \, \dd x =  \sum_{j=1}^{k} \big\lvert\theta_j-\pi_j^{(n)}\big\rvert = \big\lVert\btheta - \boldsymbol{\pi}^{(n)}\big\rVert_1 .
    \end{equation*}
   Now for any simplex vector $\btheta\in \mathcal{S}_{k}$ satisfying $\big\lVert\btheta - \boldsymbol{\pi}^{(n)}\big\rVert_1 \leq C_1^{-2\alpha}c_0^2\gamma^{1+1/\alpha}$ it follows from \eqref{eq:convergence_rate_triangle} that
    \begin{align*}
    \begin{split}
        \dhsq\big(f_0, f_{\pa^{(n)},\btheta}\big) &\leq 2 \tilde{A}k^{-2\alpha} + 2\big\lVert\btheta - \boldsymbol{\pi}^{(n)}\big\rVert_1 \\ &\leq 2\tilde{A}k^{-2\alpha} + 2C_1^{-2\alpha}c_0^2\gamma^{1+1/\alpha} \\ &\leq
        2(\tilde{A}C_1^{-2\alpha}c_0^2 + C_1^{-2\alpha}c_0^2)\gamma^2 \\ &= c_3 \gamma^2,
    \end{split}
    \end{align*}
    where we used that $\gamma^{1+1/\alpha} \leq \gamma^2$ for $\gamma \in (0,1)$ and $c_3 = 2c_0^2C_1^{-2\alpha}(\tilde{A} + 1)$ is a constant. Note that $c_3\downarrow 0$ as $c_0\downarrow 0$, so $c_3$ can be made arbitrarily small by choosing $c_0$ to be suitably small.

    Denote $\mathcal{A}_{\gamma} = \left\{\btheta\in \mathcal{S}_{k}\colon \big\lVert \btheta - \boldsymbol{\pi}^{(n)}\big\rVert_1 \leq C_1^{-2\alpha}c_0^2\gamma^{1+1/\alpha}\right\}$.
    To bound the likelihood ratio $\|f_0/f_{\pa^{(n)}, \btheta}\|_\infty$, we note that for $m = \inf_{x\in [0,1]} f_0(x)$, we have $\pi_j^{(n)} \geq m|\pa_j^{(n)}|$ for $j = 1,2,\ldots, k$. For sufficiently small $\gamma$, we thus have on $\mathcal{A}_{\gamma}$ that
    \begin{equation*}
        \theta_j \geq \pi_j^{(n)} - \big\lVert \btheta - \boldsymbol{\pi}^{(n)} \big\rVert_1 \geq m|\pa_j^{(n)}| - C_1^{-2\alpha}c_0^2\gamma^{1+1/\alpha}.
    \end{equation*}
    To ensure that $\theta_j\geq m\big|\pa^{(n)}_j\big|/2$, we note that $\gamma^{1+1/\alpha} \leq \mathcal{O}\big(k^{-1-\alpha}\big) = o\big(k^{-1}\big)$ and hence, any constant multiple of $\gamma^{1+1/\alpha}$ is upper bounded by $mBk^{-1}/4$ for large values of $k$. Consequently $C_1^{-2\alpha}c_0^2\gamma^{1+1/\alpha} \leq m\min_j |\pa_j^{(n)}|/2$ which further implies $\theta_j\geq m|\pa_j^{(n)}|/2$ by the above display. From the above arguments we arrive at $f_{\pa^{(n)},\btheta}(x) \geq m/2$ and hence
    \begin{equation*}
        \big\lVert f_0/f_{\pa^{(n)}, \btheta}\big\rVert_\infty \leq \frac{2\|f_0\|_\infty}{m}.
    \end{equation*}
    It follows that $\dhsq\big(f_0, f_{\pa^{(n)},\btheta}\big) \big\lVert f_0/f_{\pa^{(n)}, \btheta}\big\rVert_\infty \leq c_4\gamma^2$, for a constant $c_4 = c_4(c_0)\downarrow 0$ as $c_0\downarrow 0$. Applying the above estimates, we thus have that $\mathcal{A}_{\gamma} \subseteq \mathcal{N}_{\epsilon}'(f_0)$.
    As $\gamma^{1+1/\alpha} = o(k^{-1})$, any constant multiple of $\gamma^{1+1/\alpha}$ is less than $1/(\Sigma k)$ for all sufficiently large $k$.
    In addition, the lower bound $\sigma k^{-1} \leq a_{k,j}$ implies that the Dirichlet parameters are lower bounded by a constant multiple of $\gamma^{-b(1+1/\alpha)}$ for some $b > 0$, so the conditions of Lemma G.13 in \citet{ghosal2017fundamentals} are met, and we obtain the lower bound
    \begin{equation*}
        P_n\big(\mathcal{A}_{\gamma}\ba \pa^{(n)}\big) \geq c_1\exp\left(-c_5k\log(2C_1^{2\alpha}c_0^{-2}/\gamma^{1+1/\alpha})\right), 
    \end{equation*}
    for some $c_5 > 0$. It remains to show that we can write the probability on the right hand side in terms of $\epsilon$ only. Observe that for $\gamma = c_0\epsilon$ the expression $\log(2C_1^{2\alpha}c_0^{-2}/\gamma^{1+1/\alpha}) / \log(1/\epsilon)$ converges to a positive, finite limit as $\epsilon\downarrow 0$.
    From this we conclude that for a suitable constant $c_2 > 0$,
    \begin{equation*}
        P_n\big(\mathcal{N}_{\epsilon}'(f_0)\big) \geq p_n\big(\pa^{(n)}\big)\, c_1\exp\big(-c_2k\log(1/\epsilon)\big),
    \end{equation*}
    which was to be shown.
\end{proof}

We are finally in a position to prove Theorem~\ref{thm:convergence_rate}.
\begin{proof}[Proof of Theorem~\ref{thm:convergence_rate}]
    We verify the conditions of Lemma~\ref{lemma:bayes_est_concentration}, tending to the first condition with $\bar{\epsilon}_n = \epsilon_n = \{n/\log(n)\}^{-\alpha/(2\alpha+1)}$.
    
    Letting $\{s_n\}$ be a sequence of integers satisfying $C_1\epsilon_n^{-1/\alpha}\leq s_n \leq C_2\epsilon_n^{-1/\alpha}$ for positive constants $C_1, C_2$, the result of Lemma~\ref{lemma:prior_concentration_rate_irreg_hist} yields
    \begin{equation}\label{eq:lower_bound_probability}
        P_n\Big(f\colon \dhsq(f_0, f)\big\lVert f_0/f\big\rVert_\infty \leq \epsilon^2 \Big)\geq p_n\big(\pa^{(n)}\big)\, c_1\exp\big(-c_2s_n\log(1/\epsilon_n)\big),
    \end{equation}
    for some $\pa^{(n)}\in \mathcal{H}_{\Tn, s_n}$.
    To continue our argument, we need to lower bound the prior probability of the partition $\pa^{(n)}$. To this end, note that
    \begin{equation}\label{eq:s_n_prob_lower_bound}
        p_n\big(\pa^{(n)}\bba s_n\big) = \binom{k_n-1}{s_n-1}^{-1} \geq \binom{k_n}{s_n}^{-1} \geq \left(\frac{ek_n}{s_n}\right)^{-s_n} \geq \exp\big(-s_n\log(k_n)\big),
    \end{equation}
    where the last inequality holds for all sufficiently large $n$. Since $k_n = \mathcal{O}(n)$ by assumption, we have, for all sufficiently large $n$,
    \begin{equation*}
        p_n\big(\pa^{(n)}\ba s_n\big) \geq \exp\big(-s_n\log(A_1n)\big) \geq \exp\big(-A_2 s_n\log(n)\big),
    \end{equation*}
    for two constants $A_1, A_2 > 0$.
    \begin{equation*}
        p_n(s_n) \geq D_1\exp\big(-d_1s_n\log(s_n)\big) \geq D_1\exp\big(-d_1s_n\log(n)\big).
    \end{equation*}
    It is easily verified that $\log(1/\epsilon_n) / \log(n)$ converges to a positive, finite limit as $n\to\infty$. Moreover, a quick computation shows that $n\epsilon_n^2 = n^{1/(2\alpha+1)} \{\log(n)\}^{2\alpha/(2\alpha+1)}$ and hence, 
    \begin{equation*}
        s_n\log(n) \leq C_2 n^{1/(2\alpha+1)}\big\{\log(n)\big\}^{(2\alpha+1-1)/(2\alpha+1)} =  C_2 n\epsilon_n^2.
    \end{equation*}
    To proceed, note that since $\pa^{(n)}\in \mathcal{P}_{\Tn, s_n}$ we have $p_n\big(\pa^{(n)}\big) = p_n(s_n)\, p_n\big(\pa^{(n)}\ba s_n\big)$.
    We deduce that the right hand side of \eqref{eq:lower_bound_probability} is lower bounded by
    \begin{equation*}
        D_1\exp\big(-d_1s_n\log(n)\big)\exp\big(-A_2 s_n\log(n)\big)\,c_1\exp\big(-c_2C_2 n\epsilon_n^2\big),
    \end{equation*}
    where we used the assumption on $p_n(k)$. Since $\log(s_n)/\log(n)$ converges to a positive, finite limit as $n\to \infty$, we conclude that we can find constants $c_3, c_4 > 0$ such that
    \begin{equation*}
        P_n\Big(f\colon \dhsq\big(f_0, f\big)\big\lVert f_0/f\big\rVert_\infty\leq \epsilon_n^2 \Big) \geq c_3\exp\big(-c_4n\epsilon_n^2\big).
    \end{equation*}
    To verify that $P_0(\bx\in\mathcal{B}_n) = \bigO\big(\epsilon_n^2\big)$ for $\mathcal{B}_n$ the event that $e^{(2+c)n\bar{\epsilon}^2_n} p(\bx\ba \pa^{(n)})\, p_n(\pa^{(n)}) / \prod_{i=1}^n f_0(x_i) > b$, we appeal to Lemma 8.10 in \citet{ghosal2017fundamentals}, which together with equation $(8.8)$ in the same work implies that $\mathcal{B}_n$ has probability at least $d_6D^2(\sqrt{n}/\epsilon_n)^{-6} = \mathcal{O}\big(\epsilon_n^2\big)$ under $f_0$. The rate condition of Lemma~\ref{lemma:bayes_est_concentration} is thus seen to hold under the stated assumptions.

    To verify the prior condition \eqref{eq:ghosal_condition_prior_mass}, we let $\{t_n\}$ be a sequence of integers taken to satisfy 
    \begin{equation*}
        E_1 n^{1/(2\alpha+1)}\{\log(n)\}^{-1/(2\alpha+1)}\leq t_n \leq E_2n^{1/(2\alpha+1)}\{\log(n)\}^{-1/(2\alpha+1)},
    \end{equation*} 
    for two positive constants $E_1, E_2$ to be chosen later.
    From the definition of $t_n$ and previous calculations it is clear that $t_n\log(n) \geq E_1n\epsilon_n^2$.
    For $\mathcal{F}_n = \cup_{k=1}^{t_n} \mathcal{H}_{\Tn,k}$, where $\mathcal{H}_{\Tn,k}$ is as in \eqref{eq:histogram_densities_irr}, we then have that $P_n\big(\mathcal{F}_n^c\big) = \sum_{k=k_n+1}^{\infty} p_n(k)$. Using the assumption on $p_n(k)$ this is bounded above by $c_5\exp\big(-d_2t_n\log(t_n)\big)$ for $c_5 = D_2\sum_{k=0}^\infty \exp\big(-d_2k\log(k)\big) < \infty$. Furthermore, as $\log(t_n) / \log(n)$ converges to a positive limit as $n\to\infty$, we can find a new positive constant $d_2'$ such that 
    \begin{equation*}
        \sum_{k=k_n+1}^{\infty} p_n(k) < c_5 \exp\big(-d_2't_n\log(n)\big).
    \end{equation*}
    Taking $E_1 = (c_4+4)/d_2'$, we have $d_2't_n \geq (c_4+4)n\epsilon_n^2$ which combined with previous estimates implies that $P_n\big(\mathcal{F}_n^c\big) \geq \exp\{-(c_4+4)n\epsilon_n^2\}$, verifying \eqref{eq:ghosal_condition_prior_mass}.

    Finally, we show that the entropy condition \eqref{eq:histogram_densities_irr} holds. By Lemma~\ref{lemma:bracket_num_bound}, $N (\epsilon_n/2, \mathcal{F}_n, d_H)$ is upper bounded by $\exp\big(t_n\log(Lk_n/\{\epsilon_ns_n\})\big)$. Since $\log(k_n) + \log(1/\epsilon_n) = \bigO\big(\log(n)\big)$, the covering number can be further bounded by $\exp\big(c_5t_n\log(n)\big)$ for a suitable constant $c_5$. Since $t_n\log(n) = \mathcal{O}(n\epsilon_n^2)$ this implies that $N(\epsilon_n/2, \mathcal{F}_n, d_H) \leq \exp\big(c_6n\epsilon_n^2\big)$ for some $c_6 > 0$ as desired.
\end{proof}

\subsection{Consistency}\label{subsec:consistency_proof}
The proof of Theorem~\ref{thm:consistency} is in parts very similar to that of Theorem~\ref{thm:convergence_rate}, so we only fill in the details where needed.

Before proving Theorem~\ref{thm:consistency}, we need an approximation result.

\begin{lemma}\label{lemma:approx_step}
        Let $f_0\colon [0,1]\to\mathbb{R}$ be a probability density. Then for any $\epsilon > 0$ we can find a $\delta > 0$ such that if $\pa$ is any interval partition of $[0,1]$ with $\max_{j} |\pa_j| < \delta$, then there is a density $h$, possibly depending on $\epsilon$, that is piecewise constant on the partition $\pa$ satisfying $\big\lVert f_0 - h\big \lVert_1 < \epsilon$.
\end{lemma}
\begin{proof}
    We will first show that for any $\epsilon > 0$ and an arbitrary sequence of partitions $\mathcal{J}^{(n)}$ with $\max_j\big|\mathcal{J}_j^{(n)}\big|\to 0$, we can find an $N\in \mathbb{N}$ and a density $h_N$ that is piecewise constant on $\mathcal{J}^{(N)}$ and which satisfies $\big\lVert f_0 - h_N\big\rVert_1 < \epsilon$.
    To this end, we make use of the fact that step functions are dense in $\mathbb{L}_{\mspace{1mu}r}\big([0,1]\big)$ for $r\geq 1$ \citep[Corollary 12.11]{schilling2005measures}. Fix $\epsilon \in (0,1/2)$ and let $g_\epsilon$ be a strictly positive step function such that $\lVert f_0 - g_\epsilon \rVert_1 < \epsilon/5$.
    From the inverse triangle inequality, we have $1-\epsilon/5 < \lVert g_\epsilon \rVert_1 < 1+\epsilon/5$.
    Letting $m_\epsilon = g_\epsilon / \lVert g_\epsilon \rVert_1$, it follows by the triangle inequality that
    \begin{equation*}
        \big\lVert f_0 - m_\epsilon \big\rVert_1 \leq \big\lVert f_0 - g_\epsilon \big\rVert_1 + \big\lVert g_\epsilon - m_\epsilon \big\rVert_1 < \frac{\epsilon}{5} + \big\lVert g_\epsilon\big\rVert_1 \left(1-\frac{1}{\lVert g_\epsilon\rVert_1}\right) < \frac{\epsilon}{5} + \frac{2\epsilon}{5}(1+\epsilon/5) < \frac{4\epsilon}{5}.
    \end{equation*}
    where we used $|1-1/x| < 2x$ for $x> 1/2$. Since $m_\epsilon$ is a step function, it is almost everywhere continuous on $[0,1]$. Let $\{\mathcal{J}^{(n)}\}$ be a sequence of interval partitions of $[0,1]$ such that $\max_j \big|\mathcal{J}_j^{(n)}\big|\to 0$. Define $h_{n} = m_{\epsilon, \mathcal{J}^{(n)}}$ to be the piecewise constant version of $m_\epsilon$ based on the partition $\mathcal{J}^{(n)}$.
    To show convergence $\mathbb{L}_1$, we first show that $h_n$ converges to $m_\epsilon$ almost everywhere. To this end, fix $\mu > 0$ and let $x\in [0,1]$ be a continuity point of $m_\epsilon$ so that $|m_\epsilon(x)-m_\epsilon(y)| < \mu$ whenever $|x-y| < \nu$ for all sufficiently small $\nu$. For all large $n$ we have $\max_j\big|\mathcal{J}^{(n)}_j\big| \leq \nu$, so for $x, y\in \mathcal{J}_j^{(n)}$, integrating the inequalitites $-\mu < m_\epsilon(x)-m_\epsilon(y) < \mu$ over this interval and multiplying by $1/\big|\mathcal{J}^{(n)}_j\big|$ shows that 
    \begin{equation*}
        -\mu < m_\epsilon(x) - m_{\epsilon, \mathcal{J}^{(n)}}(x) < \mu.
    \end{equation*}
    Since $\mu$ was arbitrary we conclude that $h_{n}(x) \to m_{\epsilon}(x)$.
    As $h_{n}\to m_\epsilon$ almost everywhere and $h_{n}, m_{\epsilon}$ are densities for all $n$, Scheff{é}'s Lemma guarantees that $\big\lVert m_{\epsilon} - h_{n}\big\rVert_1\to 0$. In particular, this result implies that $\lVert m_{\epsilon} - h_N\rVert_1 < \epsilon/5$ for some $N\in \mathbb{N}$, so that
    \begin{equation*}
        \big\lVert f_0 - h_N \big\rVert_1 \leq \big\lVert f_0 - m_\epsilon \big\rVert_1 + \big\lVert m_\epsilon - h_N \big\rVert_1 < \epsilon.
    \end{equation*}

    Suppose now that the hypothesis of the lemma is false for the sake of contradiction. Then there exists an $\epsilon > 0$ such that for every $n \in \mathbb{N}$, we can find a partition $\pa^{(n)}$ with $\max_{j} \big|\pa^{(n)}_j\big| < 1/n$ such that $\big\lVert f_0 - h_n \big\rVert_1 \geq \epsilon$ for all densities $h_n$ which are piecewise constant on $\pa^{(n)}$. Since $\max_{j} \big|\pa^{(n)}_j\big|\to 0$, this contradicts our previous result.
\end{proof}

We are now ready to prove Theorem~\ref{thm:consistency}.

\begin{proof}[Proof of Theorem~\ref{thm:consistency}]
    By an argument similar to that of Lemma~\ref{lemma:bayes_est_concentration}, it can be established the Bayes histogram estimator is consistent if for a constant $c > 0$ and a sequence of sets $\mathcal{F}_n\subseteq \mathcal{F}$,
    \begin{align*}
        N\big(\epsilon, \mathcal{F}_n, d_H\big) &< n\epsilon^2,\\
        P_n(\mathcal{F}_n^c\big) &< \exp(-cn),
    \end{align*}
    in addition to the condition that for all $\delta > 0$, we can find a sequence of partitions $\pa^{(n)}\in \mathcal{P}_{\Tn}$ such that the event $e^{n\delta}p\big(\pa^{(n)}\big)\mspace{1mu}p\big(\bx\ba \pa^{(n)}\big) / \prod_{i=1}^n f_0(x_i) > b$ for some $b > 0$, denoted $\mathcal{B}_n$, has probability tending to $1$. The first two conditions can be handled as in the proof of Theorem~\ref{thm:convergence_rate} by taking $\mathcal{F}_n = \cup_{k=1}^{k_n} \mathcal{H}_{\Tn,k}$, as $N\big(\epsilon, \mathcal{F}_n, d_H\big) < n\epsilon^2$ for all sufficiently large $n$ and $P_n\big(\mathcal{F}_n^c\big) = 0$ in this case.

    We first show that $f_0$ can be approximated to arbitrary precision by a piecewise constant density in the squared Hellinger metric. Since $\dhsq (f, g) \leq \lVert f - g\rVert_1$ for all densities $f,g$, we work with the $\mathbb{L}_1$ metric instead as it is more convenient in this case.
    Let now $\mu\in (0,1)$ to be decided later. We will now show that we can always construct a partition $\pa^{(n)}\in \mathcal{P}_{\Tn, k}$ and find a density $h_n$ piecewise constant on $\pa^{(n)}$ with $k\in \mathbb{N}$ fixed such that $\big\lVert f_0 - h_n \big\rVert_1 < \mu$. By Lemma~\ref{lemma:approx_step}, we can find such an $h_n$ whenever $\max_j\big|\pa_j^{(n)}\big|< \nu$ for some $\nu > 0$. Let $k > 2/\nu$, and let $N\in \mathbb{N}$ be such that $\max_{1\leq j\leq k} \{\tau_{n,j} - \tau_{n,j-1}\} < \nu/3$ for all $n \geq N$. For each $l\in \{1,2,\ldots, k\}$, we can then find at least one $\tau_{n,j}\in \big((l-1)/k, l/k\big]$. We now construct the partition $\pa^{(n)}$ by including exactly one $\tau_{n,j}$ for each regular interval $\big((l-1)/k, l/k\big]$. Since the length of every such regular interval is less than $\nu/2$, it follows that the adjacent endpoints in the interval partition $\pa^{(n)}$ constructed this way are at most $\nu$ apart and consequently we must have $\big\lVert f_0 - h_n \big\rVert_1 < \mu$.
    
    Denote $V(f_0, f) = \int_0^1 f_0(x) \log^2\big\{f_0(x)/f(x)\big\}\mspace{2mu}\dd x$ and put $\mathcal{N}''_\epsilon(f_0) = \big\{f\colon K(f_0, f) \leq \epsilon^2,\, V(f_0, f) \leq \epsilon^2\big\}$. To relate Hellinger neighbourhoods to $\mathcal{N}_\epsilon''(f_0)$, we appeal to
    Theorem 5 in \citet{wong1995inequalities}, which implies that
    \begin{equation*}
        \max\big\{K(f_0, f_{\pa^{(n)}, \btheta}), V(f_0, f_{\pa^{(n)}, \btheta}) \big\}< \epsilon^2 \mathrm{\ whenever\ } \dhsq\big(f_0, f_{\pa^{(n)}, \btheta}\big) < 3\mu
    \end{equation*}
    for all sufficiently small $\mu > 0$ provided that for some $\gamma \in (0,1]$ and the set $\mathcal{D}_\gamma = \big\{f_0 / f_{\pa^{(n)}, \btheta}\geq e^{1/\gamma}\big\}$
    \begin{equation*}
        L_\gamma^2 = \int_{\mathcal{D}_\gamma} f_0(x)\left\{\frac{f_0(x)}{f_{\pa^{(n)}, \btheta}(x)}\right\}^\gamma\dd x  < \infty.
    \end{equation*}
    This holds true for $\gamma = r-1$ if $\min_j\theta_j > \mu^2/2$ due to the fact that $f_0\in \mathbb{L}_{\mspace{1mu}r}\big([0,1]\big)$ for some $r \in (0,1]$ by assumption. To show that the the prior assigns sufficient mass to Hellinger neighbourhoods of $f_0$, note that by the triangle inequality,
    \begin{equation*}
        \dhsq\big(f_0, f_{\pa^{(n)}, \btheta}\big) \leq \big\lVert f_0 - h_n\big\rVert_1 + \big\lVert h_n - f_{\pa^{(n)}, \theta}\big\rVert_1 = \big\lVert f_0 - h_n\big\rVert_1 + \big\lVert \boldsymbol{\eta}^{(n)} - \btheta \big\rVert_1,
    \end{equation*}
    where $\eta_j^{(n)} = \int_{\pa_j^{(n)}} h_n(x)\mspace{2mu}\dd x$.
    The first term in the above expression is less than $\mu$ for all sufficiently large $n$ by the previous calculation, and we conclude that $\{f_{\pa^{(n)}, \btheta}\colon \dhsq(f_0, f_{\pa^{(n)}, \btheta}) < 3\mu\}\subseteq \mathcal{N}_\epsilon''(f_0)$    
    for $\btheta$ belonging to the set
    \begin{equation*}
        \mathcal{A}_\mu = \big\{\btheta\in \mathcal{S}_{s_n}\colon \big\lVert \btheta - \boldsymbol{\eta}^{(n)} \big\rVert_1 \leq 2\mu,\, \min_{j} \theta_j >  \mu^2/2\big\}.
    \end{equation*}
    Since the prior for $\btheta\ba \pa^{(n)}$ is a $k$-dimensional $\mathrm{Dir}(\boldsymbol{a})$ distribution and $\boldsymbol{a} \in (0,\Sigma)^k$ by assumption, we can find $z > 0$ such that $a_j$ is lower bounded by a constant multiple of $\mu^z$ for all $j$. Appealing to Lemma G.13 in \citet{ghosal2017fundamentals}, we find that $P\big(\mathcal{A}_\mu\ba \pa^{(n)}\big) > c_1\exp\big(-c_2k\log(1/\mu)\big)$ for two positive constants $c_1, c_2$. From \eqref{eq:s_n_prob_lower_bound} we have $p_n\big(\pa^{(n)} \ba k\big) \geq \exp\big(
    -k\log(k_n)\big)$ and hence,
    \begin{equation*}
       P_n\big(f\in \mathcal{N}_\epsilon''(f_0)\big)\geq p_n(k)\, p_n\big(\pa^{(n)}\ba k\big) \,P\big(\mathcal{A}_\mu\ba \pa^{(n)}\big) \geq c_3\exp\big(\log p_n(k) - c_4k\log(n)\big),
    \end{equation*}
    for constants $c_3, c_4 > 0$. By Lemma 8.10 in \citet{ghosal2017fundamentals}, we obtain the lower bound
    \begin{equation*}
        e^{n\delta}p\big(\pa^{(n)}\big)\mspace{1mu}p\big(\bx\ba \pa^{(n)}\big) / \prod_{i=1}^n f_0(x_i) \geq c_3\exp\big(n\delta + \log p_n(k) - c_4k\log(n)-(1+D)n\epsilon^2\big),
    \end{equation*}
    where $D > 0$, except for on a set of probability less than $(\sqrt{n}\epsilon^2)^{-2}$.
    By our assumptions the terms in the exponent on the right hand side converge to $0$ when divided by $n$, expect for $n\delta - (1+D)n\epsilon^2$. Hence, the right hand side of the expression diverges to $\infty$ provided $\epsilon$ is taken to be sufficiently small, so that $\mathcal{B}_n^c$ is contained in a set with probability tending to $0$, which concludes the proof.
\end{proof}

\section{Simulation study}
The following section includes some further details on the setup of the simulation study in Section~\ref{sec:simulation_study} and presents the results in more detail.

\subsection{Methods used and implementation}
\label{subsec:simulation_methods}
An overview of all the methods included in the simulation study is given in Table~\ref{tab:simulations_methods}. The sowftware implementations used for each method are as follows:
For the Wand method we used the \texttt{dpih()} function in the \texttt{KernSmooth} library \citep{wand2021kernsmooth}. The \texttt{ftnonpar} package \citep{davies2012ftnonpar} was used to compute the Taut String histogram. Finally, for the two approaches of \citet{rozenholc2010irregular} we used the \texttt{histogram} \proglang{R} library \citep{mildenberger2019histogram}. For all the other methods we have used our own software implementation.

For all the densities in Figure~\ref{fig:test_densities}, we estimated the support in the way described in the beginning of Section~\ref{subsec:prior_elicitation}. This was done even for those densities that have known compact supports to ensure that all density estimates were comparable, as some of the software implementations used in the simulation study did not allow us to manually specify the support of the histogram estimate.

With the exception of the Wand method, we computed the optimal number $k$ of bins for each regular histogram procedure among all regular histograms consisting of less than $\lceil 4n/\log^2(n)\rceil$. Although \citet{birge2006bins} specifically recommend maximizing the penalized likelihood up to $\lceil n/\log(n)\rceil$ for their method, we have found that for the test densities under consideration here, the maximum always occurs at a much lower value of $k$ for the largest sample sizes, so we can save a considerable amount of computation by restricting our search to a smaller set.

For the RIH, L2CV and KLCV methods we used the greedy search heuristic of \citet{rozenholc2010irregular} to reduce the computational burden of computing the optimal partition according to these criteria. Based on some preliminary simulations, we decided to use a data-driven grid for both cross-validation methods and restricted our search to only consider partitions to a minimum bin width greater than $\log^{1.5}(n)/n$, similar to the approach used by \citet{rozenholc2010irregular} for the L2CV method. This restriction was put in place because these criteria would frequently yield density estimates with sharp and narrow spikes even for smooth densities in the unrestricted case. For the random irregular histogram method, we used a fine regular grid consisting of $\lceil 4n/\log^2(n)\rceil$ bins.

\subsection{Results from the simulation study}
\label{subsubsec:tables_est_risk}
This subsection contains tables showing the complete output from the simulation study for the Hellinger, PID and $\mathbb{L}_2$ losses, respectively. Complete tables of the results with respect to each loss function are shown in Tables \ref{tab:complete_hellinger_risks_1} to \ref{tab:complete_l2_risks_2}.

\begin{figure}
    \centering
    {\includegraphics[width=0.470\linewidth]{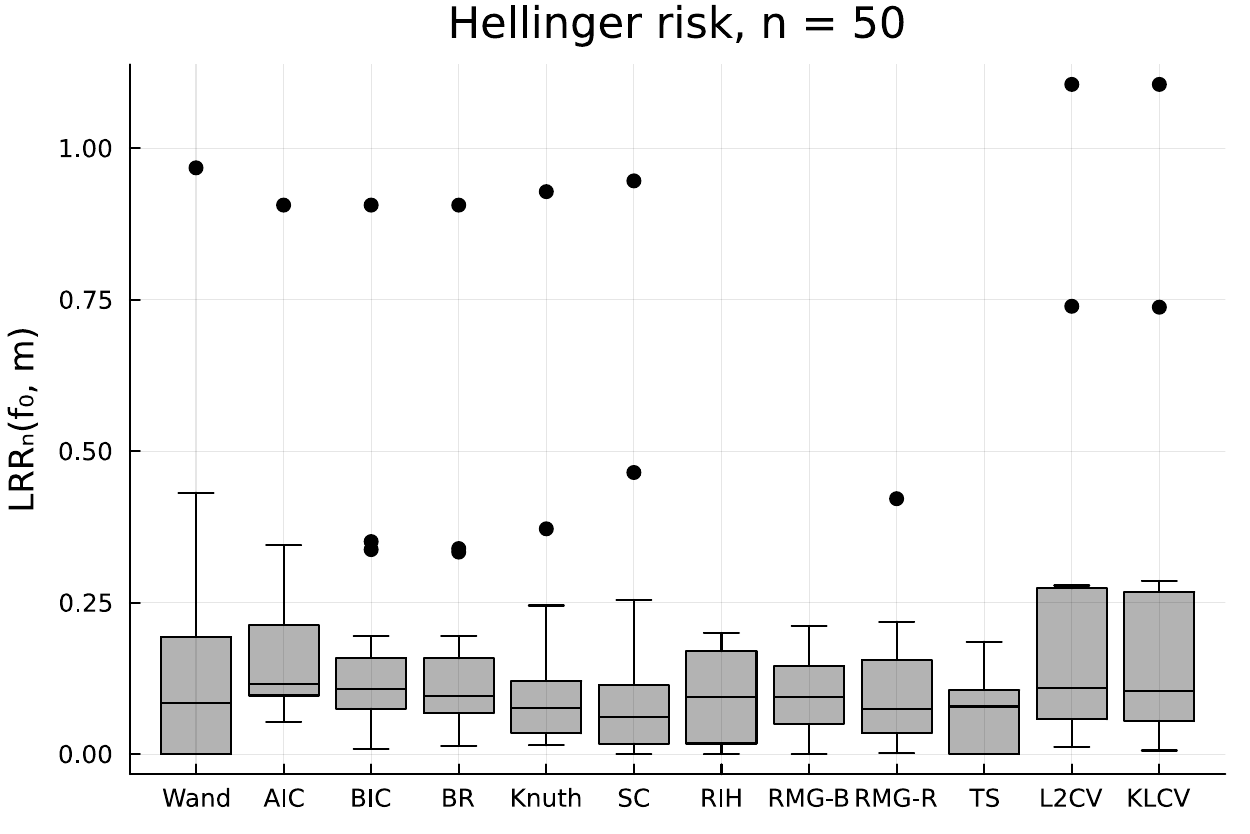} }
    \qquad
    {\includegraphics[width=0.470\linewidth]{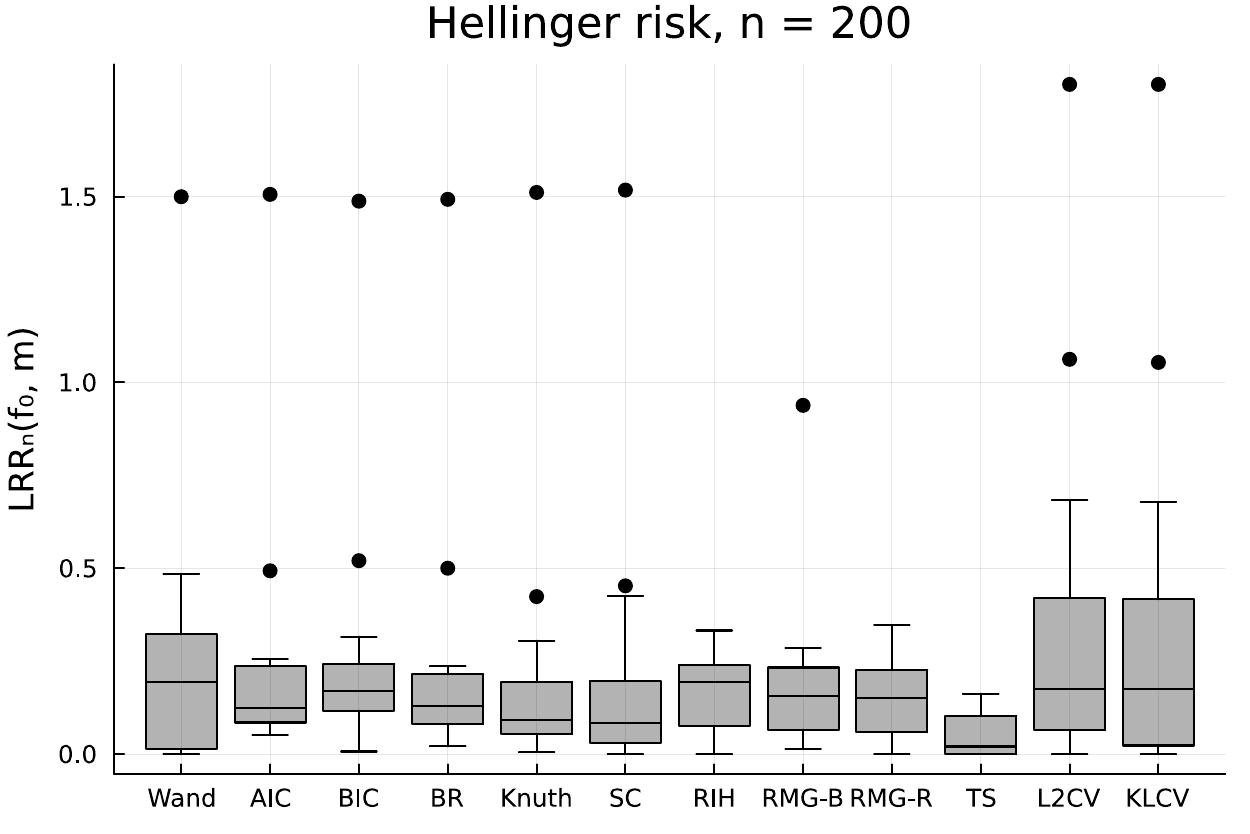} }
    \par\vspace{1cm}
    {\includegraphics[width=0.470\linewidth]{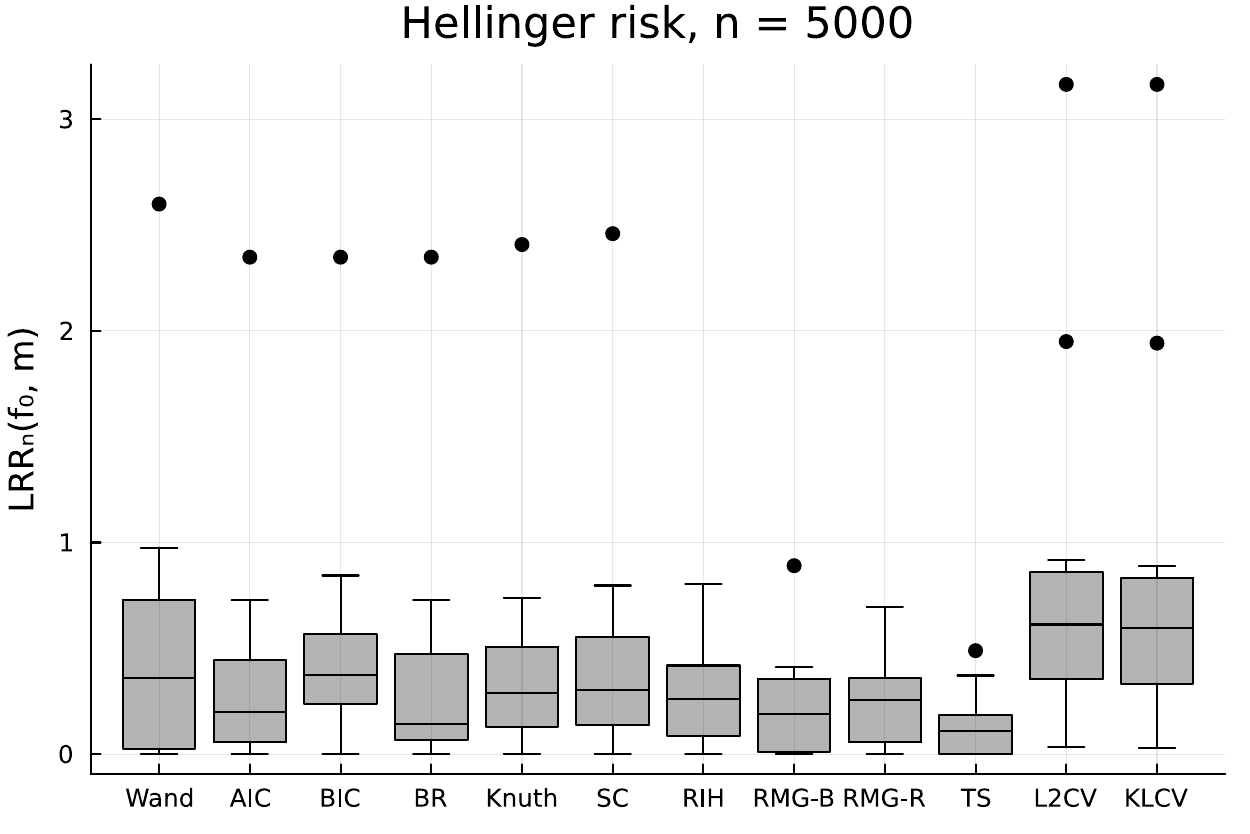} }
    \qquad
    {\includegraphics[width=0.470\linewidth]{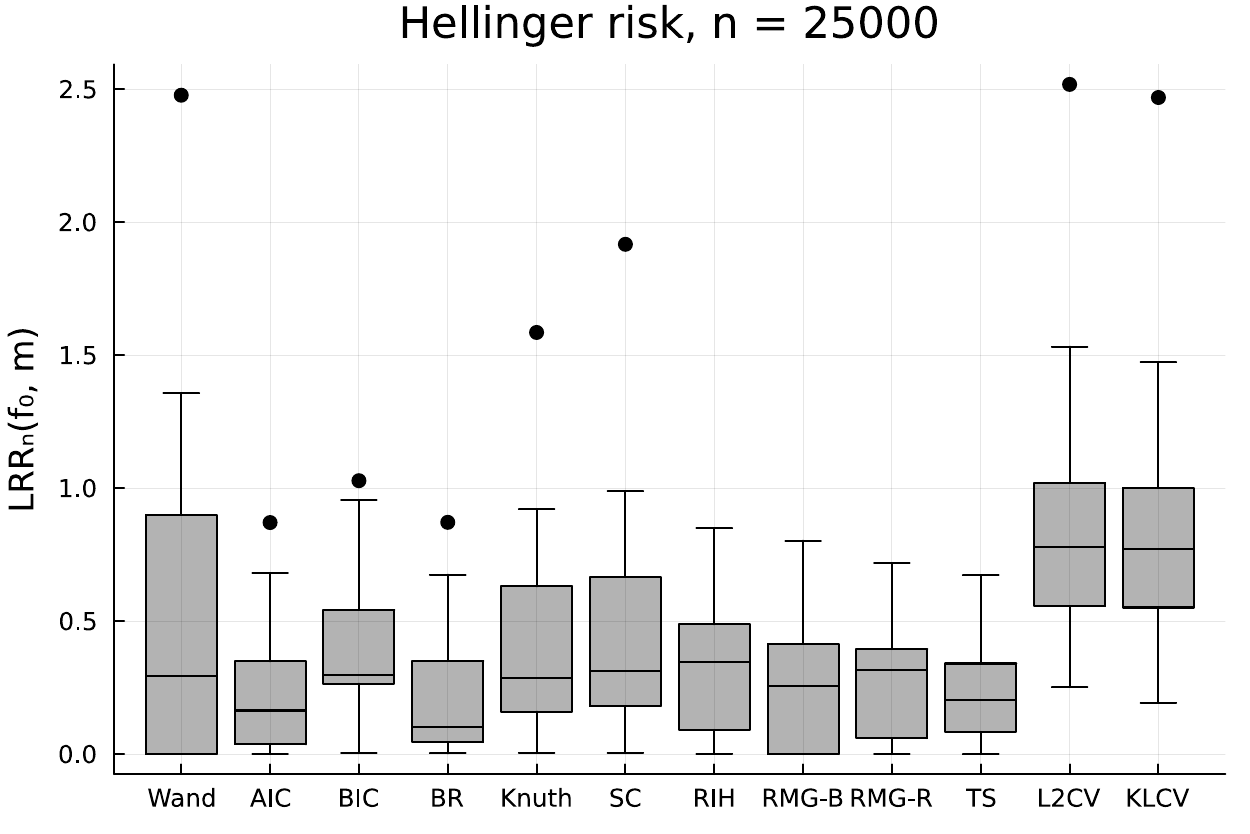} }
    \vspace{-0.5cm}
    \caption{Boxplots of $\mathrm{LRR}_n(f_0, m)$ for $d_H$.}%
    \label{fig:lrr_hell}
\end{figure}

\begin{figure}
    \centering
    {\includegraphics[width=0.470\linewidth]{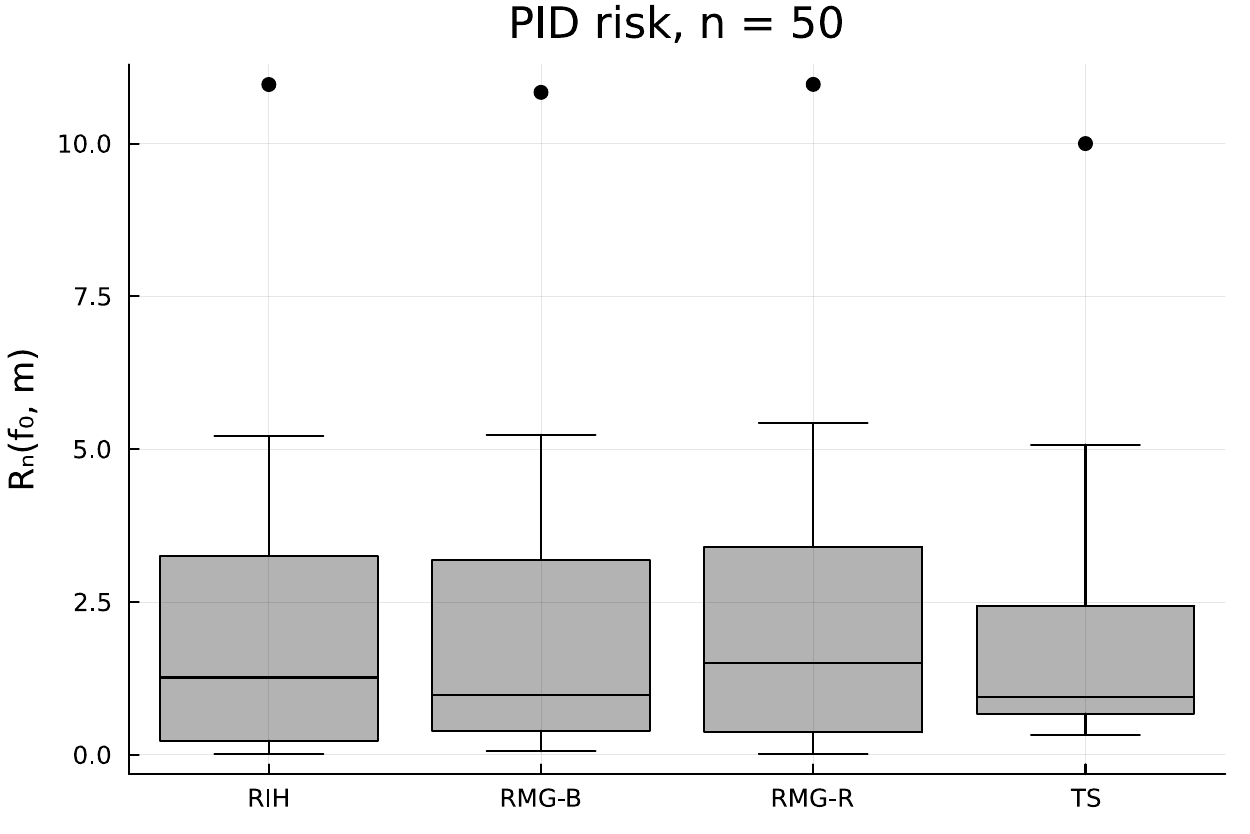} }
    \qquad
    {\includegraphics[width=0.470\linewidth]{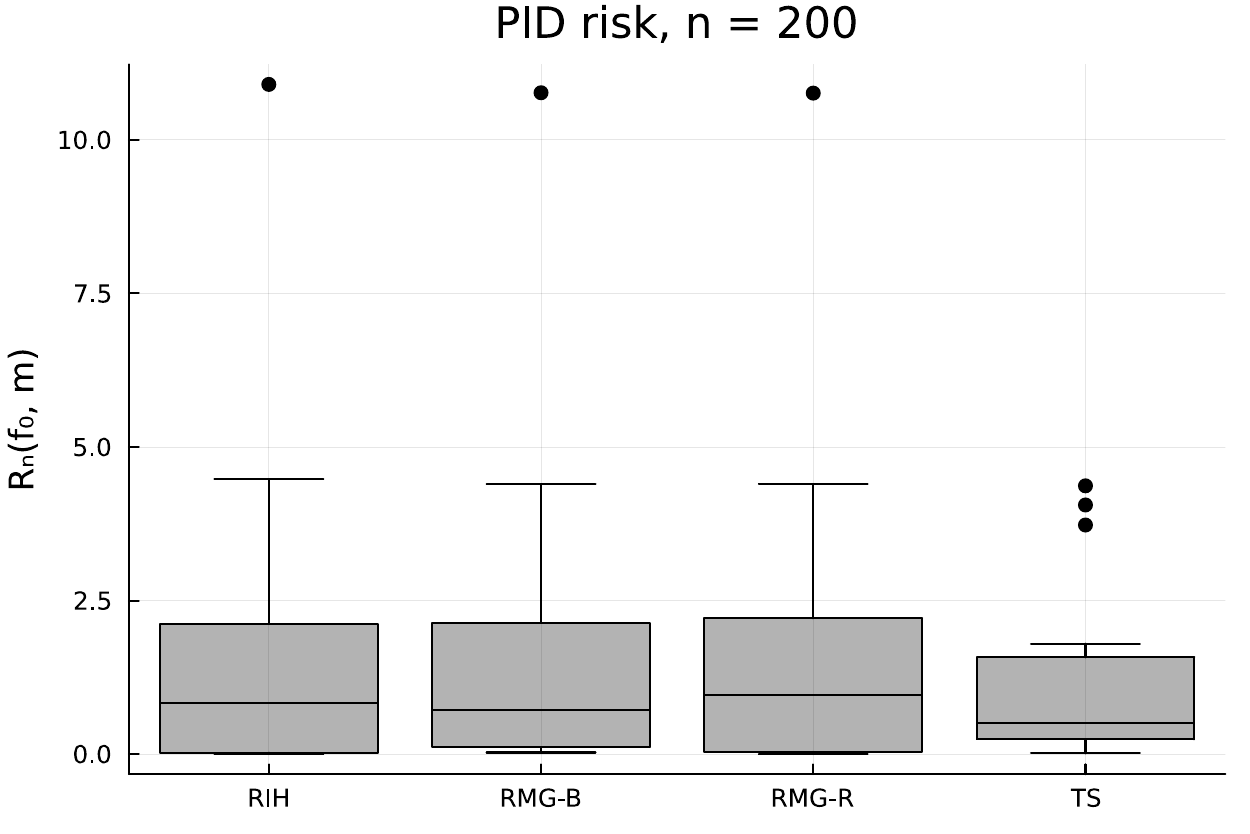} }
    \vspace{-0.5cm}
    \caption{Boxplots of PID risks for four of the irregular histogram methods for sample sizes $n = 50$ and $n = 200$.}
    \label{fig:r_pid_small}
\end{figure}

\begin{figure}
    \centering
    {\includegraphics[width=0.470\linewidth]{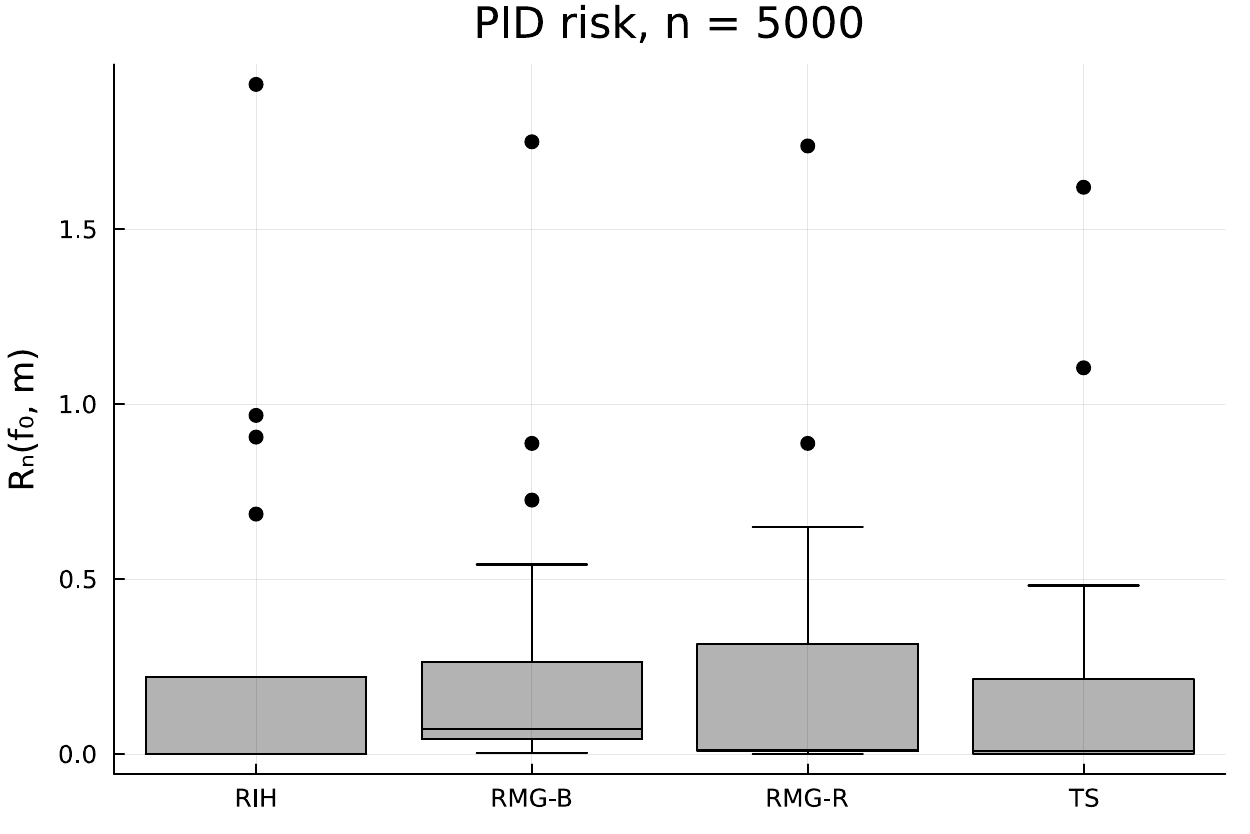} }
    \qquad
    {\includegraphics[width=0.470\linewidth]{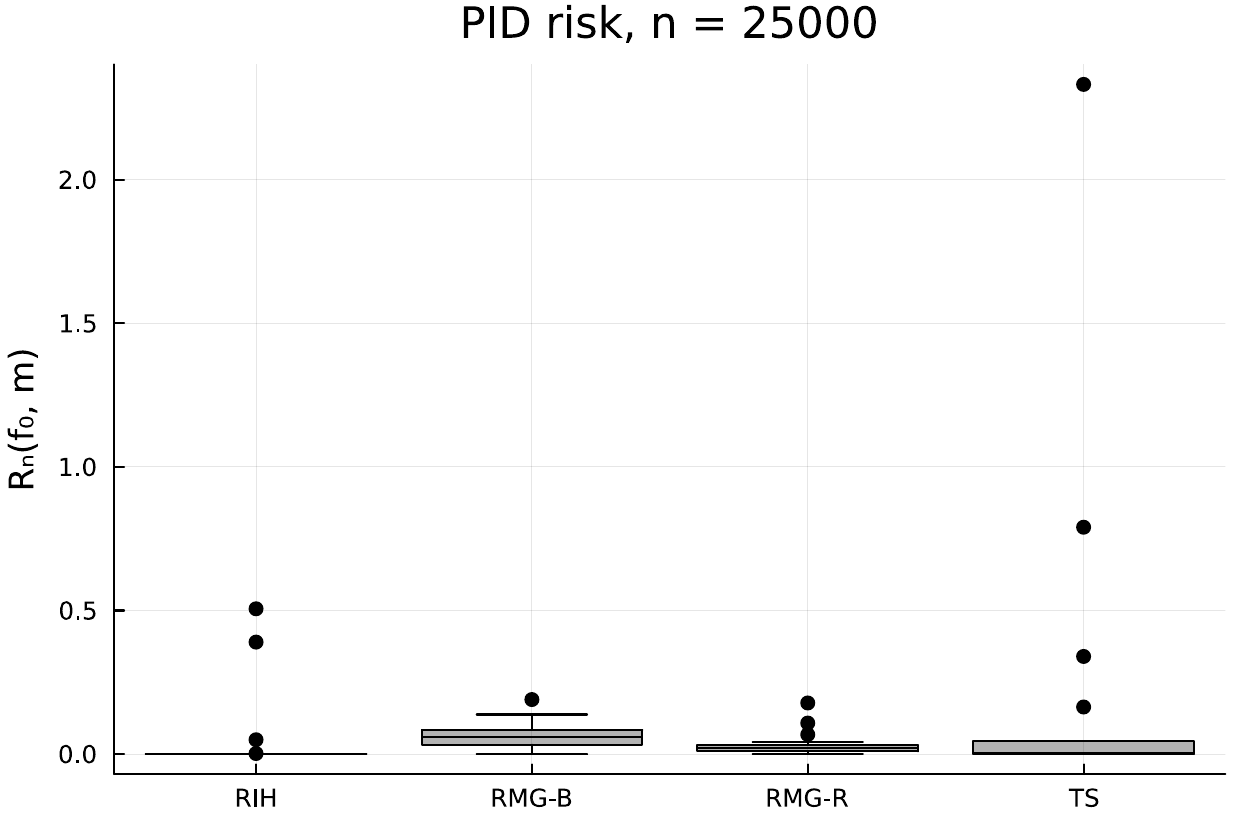} }
    \vspace{-0.5cm}
    \caption{Boxplots of PID risks for four of the irregular histogram methods for samples of size $n = 5000$ and $n = 25000$}
    \label{fig:r_pid_large}
\end{figure}

\begin{figure}
    \centering
    {\includegraphics[width=0.470\linewidth]{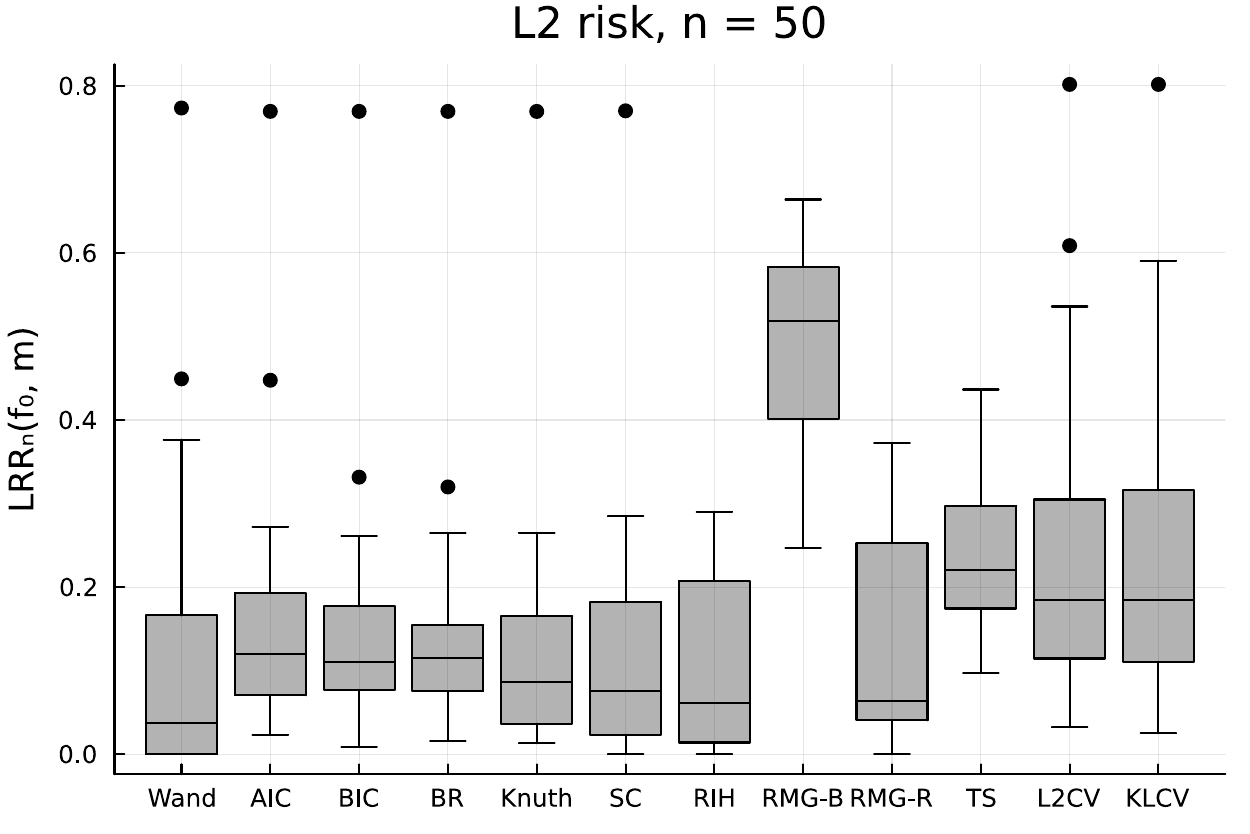} }
    \qquad
    {\includegraphics[width=0.470\linewidth]{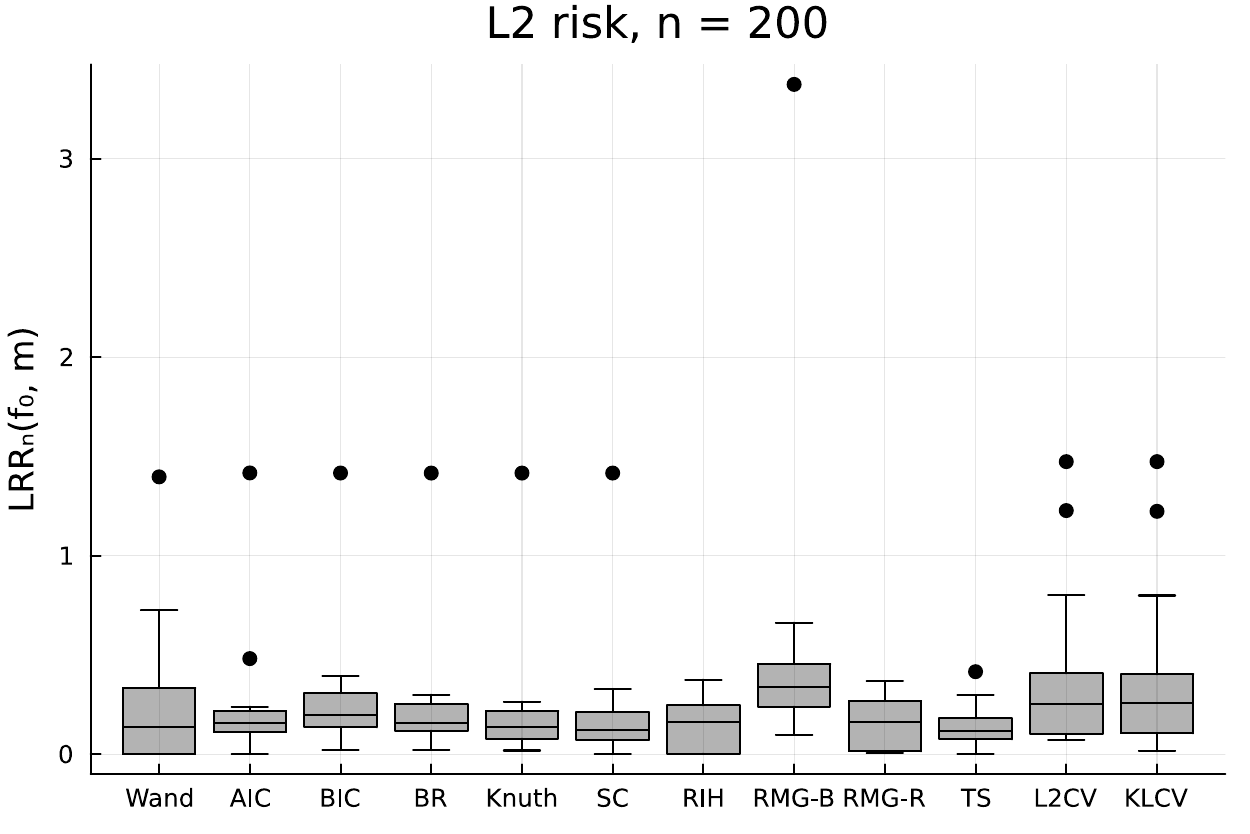} }
    \par\vspace{1cm}
    {\includegraphics[width=0.470\linewidth]{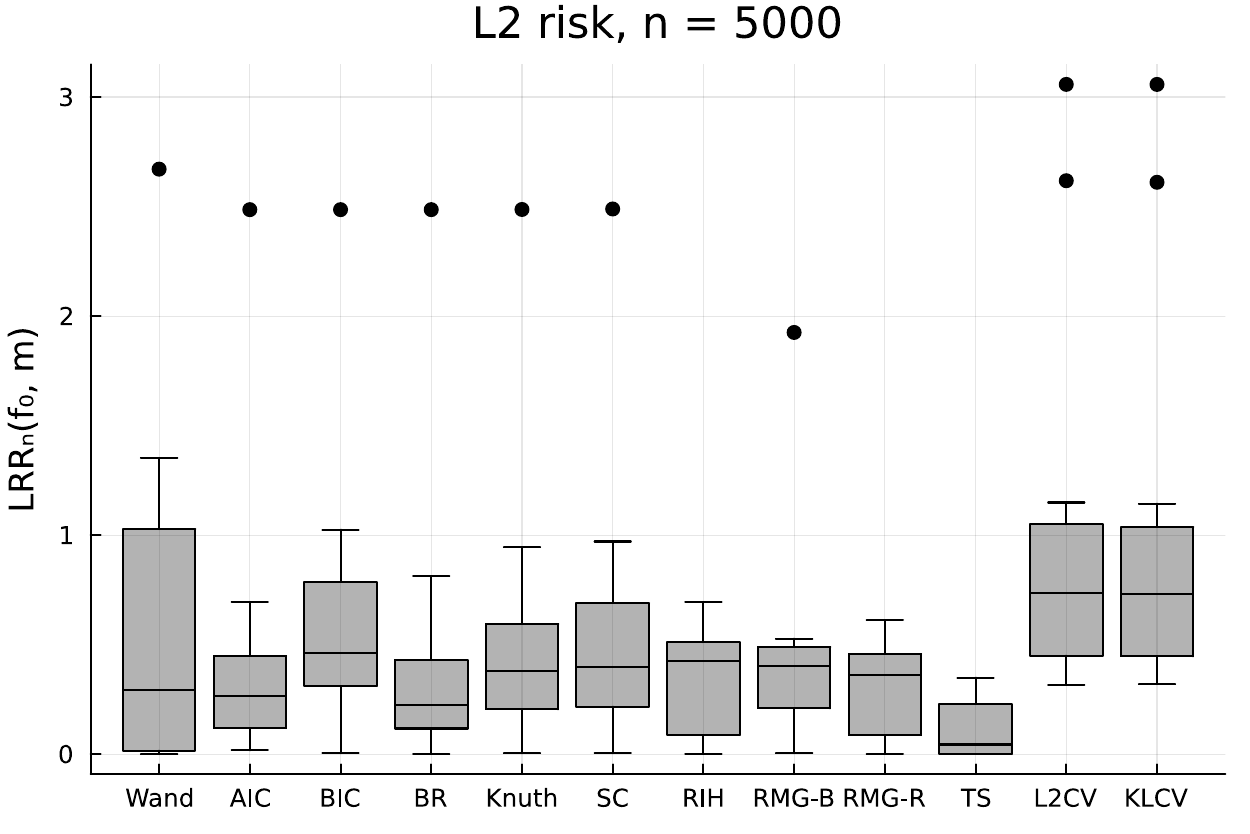} }
    \qquad
    {\includegraphics[width=0.470\linewidth]{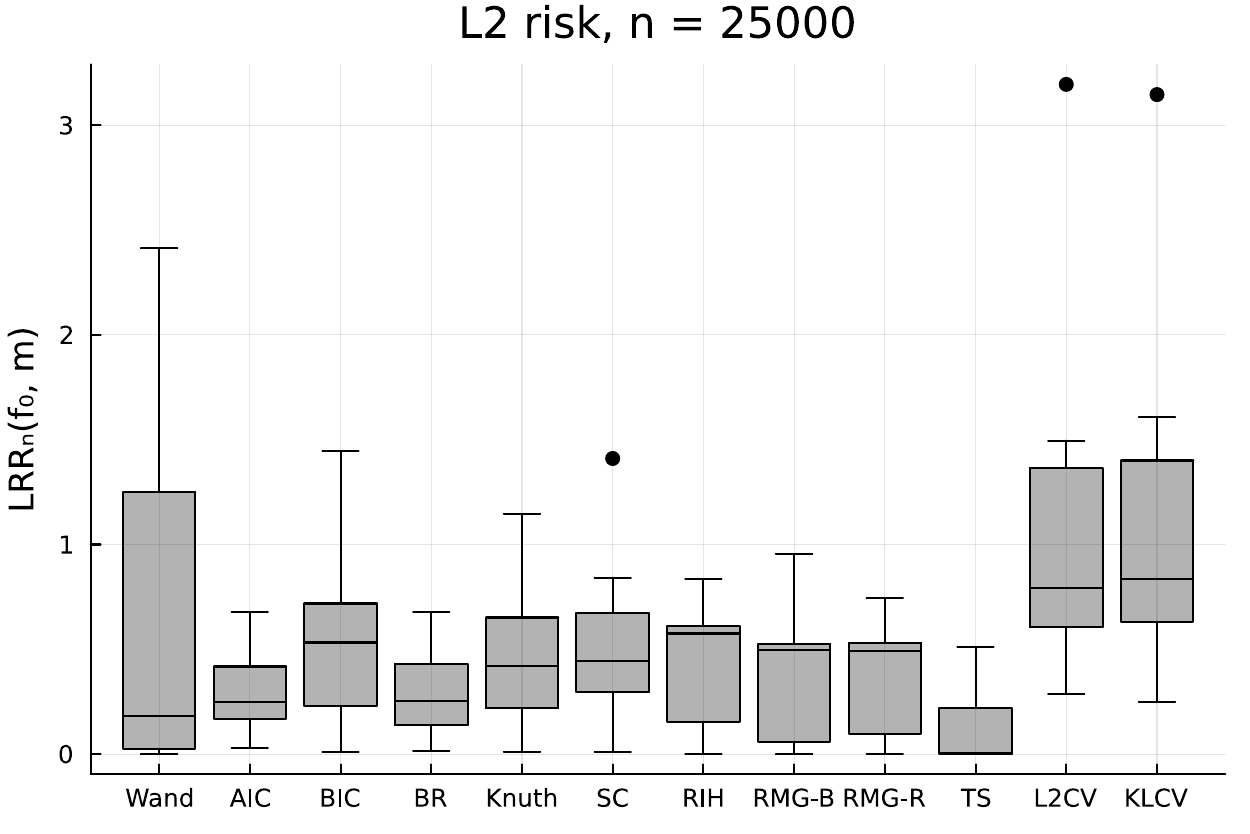} }
    \vspace{-0.5cm}
    \caption{Boxplots of $\mathrm{LRR}_n(f_0, m)$ for the $\mathbb{L}_2$ metric.}%
    \label{fig:lrr_l2}
\end{figure}

To provide a more compact visual summary of the results with respect to the Hellinger and $\mathbb{L}_2$ losses, we computed the logarithm of the risk relative to the best-performing method for each density $f_0\in \mathcal{D}$ and sample size $n\in \mathcal{N}$,
\begin{equation*}
    \mathrm{LRR}_n(f_0, m) = \log \widehat{R}_n\big(f_0, \widehat{f}_{m}\big) - \log \min_{m'\in \mathcal{M}} \widehat{R}_n\big(f_0, \widehat{f}_{m'}\big).
\end{equation*}
Boxplots of $\mathrm{LRR}_n(f_0, m)$ for sample sizes of $\{50, 200, 5000, 25000\}$ are shown in Figure~\ref{fig:lrr_hell}~and~\ref{fig:lrr_l2} for the Hellinger and $\mathbb{L}_2$ losses, respectively. Figure~\ref{fig:r_pid_small}~and~\ref{fig:r_pid_large} shows the PID risks for the four irregular methods RIH, RMG-B, RMG-R and TS. The irregular L2CV and KLCV histograms were excluded from this comparison, as they tended to perform much worse than the other irregular procedures in this regard.

\begin{table}
    \centering
    \caption{Estimated Hellinger risks for densities $1$-$8$.}
    \vspace{0.2cm}
    \begin{adjustbox}{width=0.9\textwidth, totalheight=0.95\textheight}
    \begin{tabular}{cccccccccccccc}
\toprule
Density & n & Wand & AIC & BIC & BR & Knuth & SC & RIH & RMG-B & RMG-R & TS & L2CV & KLCV\\
\midrule
1 & 50 & 0.27 & 0.298 & 0.3 & 0.297 & 0.298 & 0.293 & 0.325 & 0.333 & 0.321 & 0.325 & 0.292 & 0.294\\
 & 200 & 0.167 & 0.185 & 0.194 & 0.187 & 0.185 & 0.183 & 0.22 & 0.222 & 0.218 & 0.189 & 0.194 & 0.195\\
 & 1000 & 0.096 & 0.104 & 0.117 & 0.106 & 0.109 & 0.11 & 0.14 & 0.139 & 0.136 & 0.127 & 0.139 & 0.137\\
 & 5000 & 0.057 & 0.06 & 0.071 & 0.061 & 0.066 & 0.067 & 0.089 & 0.085 & 0.084 & 0.092 & 0.112 & 0.11\\
 & 25000 & 0.034 & 0.035 & 0.044 & 0.035 & 0.04 & 0.041 & 0.055 & 0.052 & 0.051 & 0.066 & 0.088 & 0.086\\ \hline
2 & 50 & 0.226 & 0.239 & 0.194 & 0.197 & 0.197 & 0.207 & 0.189 & 0.194 & 0.189 & 0.204 & 0.249 & 0.246\\
 & 200 & 0.13 & 0.117 & 0.095 & 0.099 & 0.096 & 0.097 & 0.095 & 0.096 & 0.095 & 0.102 & 0.188 & 0.187\\
 & 1000 & 0.07 & 0.052 & 0.043 & 0.045 & 0.043 & 0.043 & 0.043 & 0.043 & 0.043 & 0.047 & 0.152 & 0.151\\
 & 5000 & 0.04 & 0.022 & 0.019 & 0.019 & 0.019 & 0.019 & 0.018 & 0.019 & 0.018 & 0.02 & 0.13 & 0.129\\
 & 25000 & 0.023 & 0.01 & 0.008 & 0.009 & 0.008 & 0.008 & 0.008 & 0.008 & 0.008 & 0.009 & 0.104 & 0.099\\ \hline
3 & 50 & 0.433 & 0.446 & 0.443 & 0.444 & 0.404 & 0.408 & 0.384 & 0.391 & 0.393 & 0.316 & 0.417 & 0.421\\
 & 200 & 0.336 & 0.339 & 0.348 & 0.341 & 0.316 & 0.326 & 0.268 & 0.261 & 0.286 & 0.207 & 0.32 & 0.321\\
 & 1000 & 0.253 & 0.254 & 0.274 & 0.255 & 0.247 & 0.259 & 0.173 & 0.161 & 0.195 & 0.136 & 0.239 & 0.239\\
 & 5000 & 0.192 & 0.194 & 0.218 & 0.194 & 0.196 & 0.208 & 0.109 & 0.1 & 0.133 & 0.094 & 0.183 & 0.181\\
 & 25000 & 0.146 & 0.148 & 0.173 & 0.148 & 0.156 & 0.167 & 0.068 & 0.062 & 0.09 & 0.065 & 0.141 & 0.136\\ \hline
4 & 50 & 0.361 & 0.349 & 0.345 & 0.346 & 0.323 & 0.334 & 0.347 & 0.346 & 0.338 & 0.284 & 0.33 & 0.34\\
 & 200 & 0.266 & 0.231 & 0.24 & 0.232 & 0.222 & 0.236 & 0.237 & 0.232 & 0.232 & 0.183 & 0.211 & 0.213\\
 & 1000 & 0.188 & 0.166 & 0.178 & 0.167 & 0.166 & 0.178 & 0.153 & 0.144 & 0.15 & 0.13 & 0.148 & 0.147\\
 & 5000 & 0.129 & 0.136 & 0.156 & 0.142 & 0.151 & 0.157 & 0.097 & 0.09 & 0.095 & 0.089 & 0.11 & 0.109\\
 & 25000 & 0.095 & 0.091 & 0.151 & 0.093 & 0.122 & 0.148 & 0.063 & 0.058 & 0.063 & 0.076 & 0.075 & 0.07\\ \hline
5 & 50 & 0.32 & 0.334 & 0.327 & 0.327 & 0.322 & 0.321 & 0.357 & 0.363 & 0.356 & 0.316 & 0.325 & 0.327\\
 & 200 & 0.209 & 0.218 & 0.222 & 0.214 & 0.208 & 0.212 & 0.249 & 0.249 & 0.245 & 0.205 & 0.205 & 0.206\\
 & 1000 & 0.129 & 0.129 & 0.142 & 0.128 & 0.129 & 0.134 & 0.16 & 0.156 & 0.154 & 0.139 & 0.128 & 0.127\\
 & 5000 & 0.08 & 0.077 & 0.091 & 0.078 & 0.081 & 0.086 & 0.101 & 0.096 & 0.096 & 0.098 & 0.098 & 0.096\\
 & 25000 & 0.049 & 0.046 & 0.059 & 0.047 & 0.051 & 0.056 & 0.063 & 0.058 & 0.059 & 0.069 & 0.079 & 0.074\\ \hline
6 & 50 & 0.322 & 0.384 & 0.345 & 0.347 & 0.346 & 0.331 & 0.32 & 0.332 & 0.325 & 0.317 & 0.334 & 0.331\\
 & 200 & 0.248 & 0.271 & 0.267 & 0.263 & 0.26 & 0.253 & 0.228 & 0.229 & 0.229 & 0.211 & 0.259 & 0.258\\
 & 1000 & 0.194 & 0.197 & 0.204 & 0.193 & 0.193 & 0.189 & 0.151 & 0.144 & 0.153 & 0.14 & 0.202 & 0.202\\
 & 5000 & 0.154 & 0.147 & 0.159 & 0.144 & 0.147 & 0.144 & 0.098 & 0.089 & 0.1 & 0.093 & 0.165 & 0.162\\
 & 25000 & 0.122 & 0.11 & 0.124 & 0.109 & 0.113 & 0.111 & 0.061 & 0.056 & 0.065 & 0.062 & 0.13 & 0.123\\ \hline
7 & 50 & 0.337 & 0.356 & 0.355 & 0.353 & 0.349 & 0.343 & 0.357 & 0.363 & 0.35 & 0.37 & 0.341 & 0.339\\
 & 200 & 0.25 & 0.255 & 0.251 & 0.248 & 0.243 & 0.242 & 0.246 & 0.249 & 0.246 & 0.245 & 0.244 & 0.245\\
 & 1000 & 0.169 & 0.157 & 0.172 & 0.151 & 0.147 & 0.15 & 0.19 & 0.179 & 0.177 & 0.131 & 0.152 & 0.15\\
 & 5000 & 0.091 & 0.094 & 0.113 & 0.094 & 0.099 & 0.103 & 0.111 & 0.109 & 0.108 & 0.073 & 0.108 & 0.106\\
 & 25000 & 0.055 & 0.055 & 0.071 & 0.055 & 0.061 & 0.064 & 0.072 & 0.066 & 0.066 & 0.043 & 0.087 & 0.083\\ \hline
8 & 50 & 0.285 & 0.312 & 0.309 & 0.307 & 0.306 & 0.302 & 0.312 & 0.314 & 0.305 & 0.318 & 0.302 & 0.301\\
 & 200 & 0.174 & 0.191 & 0.199 & 0.191 & 0.189 & 0.189 & 0.206 & 0.208 & 0.204 & 0.193 & 0.2 & 0.2\\
 & 1000 & 0.102 & 0.111 & 0.127 & 0.113 & 0.116 & 0.117 & 0.138 & 0.136 & 0.133 & 0.125 & 0.141 & 0.14\\
 & 5000 & 0.061 & 0.064 & 0.077 & 0.065 & 0.07 & 0.071 & 0.086 & 0.083 & 0.082 & 0.068 & 0.112 & 0.11\\
 & 25000 & 0.036 & 0.038 & 0.048 & 0.038 & 0.043 & 0.044 & 0.055 & 0.051 & 0.051 & 0.049 & 0.088 & 0.086\\
\bottomrule
\end{tabular}

    \end{adjustbox}
    \label{tab:complete_hellinger_risks_1}
\end{table}

\begin{table}
    \centering
    \caption{Estimated Hellinger risks for densities $9$-$16$.}
    \vspace{0.2cm}
    \begin{adjustbox}{width=0.9\textwidth, totalheight=0.95\textheight}
    \begin{tabular}{cccccccccccccc}
\toprule
Density & n & Wand & AIC & BIC & BR & Knuth & SC & RIH & RMG-B & RMG-R & TS & L2CV & KLCV\\
\midrule
9 & 50 & 0.309 & 0.34 & 0.336 & 0.334 & 0.335 & 0.329 & 0.338 & 0.343 & 0.334 & 0.347 & 0.32 & 0.322\\
 & 200 & 0.229 & 0.242 & 0.24 & 0.234 & 0.232 & 0.231 & 0.251 & 0.252 & 0.247 & 0.241 & 0.235 & 0.235\\
 & 1000 & 0.159 & 0.152 & 0.17 & 0.152 & 0.154 & 0.154 & 0.165 & 0.16 & 0.157 & 0.148 & 0.16 & 0.16\\
 & 5000 & 0.104 & 0.09 & 0.111 & 0.09 & 0.095 & 0.098 & 0.103 & 0.1 & 0.099 & 0.073 & 0.117 & 0.115\\
 & 25000 & 0.062 & 0.059 & 0.069 & 0.059 & 0.061 & 0.063 & 0.067 & 0.062 & 0.061 & 0.043 & 0.09 & 0.088\\ \hline
10 & 50 & 0.404 & 0.417 & 0.389 & 0.39 & 0.391 & 0.397 & 0.385 & 0.39 & 0.386 & 0.417 & 0.416 & 0.414\\
 & 200 & 0.349 & 0.305 & 0.345 & 0.33 & 0.31 & 0.285 & 0.345 & 0.347 & 0.346 & 0.335 & 0.355 & 0.355\\
 & 1000 & 0.322 & 0.183 & 0.216 & 0.185 & 0.183 & 0.177 & 0.23 & 0.23 & 0.226 & 0.191 & 0.201 & 0.2\\
 & 5000 & 0.189 & 0.107 & 0.134 & 0.107 & 0.114 & 0.112 & 0.162 & 0.151 & 0.147 & 0.112 & 0.133 & 0.133\\
 & 25000 & 0.079 & 0.063 & 0.083 & 0.063 & 0.071 & 0.07 & 0.103 & 0.095 & 0.093 & 0.07 & 0.105 & 0.104\\ \hline
11 & 50 & 0.256 & 0.291 & 0.285 & 0.282 & 0.288 & 0.282 & 0.295 & 0.301 & 0.297 & 0.301 & 0.277 & 0.277\\
 & 200 & 0.155 & 0.171 & 0.178 & 0.172 & 0.171 & 0.168 & 0.193 & 0.195 & 0.193 & 0.178 & 0.194 & 0.193\\
 & 1000 & 0.087 & 0.094 & 0.108 & 0.097 & 0.101 & 0.098 & 0.127 & 0.126 & 0.123 & 0.11 & 0.148 & 0.147\\
 & 5000 & 0.052 & 0.054 & 0.066 & 0.056 & 0.061 & 0.059 & 0.08 & 0.078 & 0.076 & 0.075 & 0.126 & 0.124\\
 & 25000 & 0.031 & 0.031 & 0.041 & 0.032 & 0.037 & 0.036 & 0.05 & 0.047 & 0.046 & 0.053 & 0.101 & 0.099\\ \hline
12 & 50 & 0.586 & 0.491 & 0.54 & 0.531 & 0.552 & 0.606 & 0.46 & 0.416 & 0.408 & 0.381 & 0.797 & 0.796\\
 & 200 & 0.294 & 0.282 & 0.299 & 0.285 & 0.331 & 0.373 & 0.308 & 0.279 & 0.282 & 0.244 & 0.705 & 0.7\\
 & 1000 & 0.183 & 0.163 & 0.241 & 0.174 & 0.221 & 0.263 & 0.187 & 0.171 & 0.172 & 0.152 & 0.214 & 0.19\\
 & 5000 & 0.09 & 0.093 & 0.135 & 0.098 & 0.132 & 0.156 & 0.115 & 0.105 & 0.107 & 0.104 & 0.093 & 0.093\\
 & 25000 & 0.048 & 0.054 & 0.081 & 0.056 & 0.08 & 0.093 & 0.071 & 0.064 & 0.065 & 0.072 & 0.065 & 0.064\\ \hline
13 & 50 & 0.586 & 0.563 & 0.589 & 0.588 & 0.523 & 0.508 & 0.518 & 0.542 & 0.537 & 0.559 & 0.586 & 0.575\\
 & 200 & 0.489 & 0.395 & 0.477 & 0.417 & 0.377 & 0.371 & 0.485 & 0.889 & 0.492 & 0.348 & 0.526 & 0.524\\
 & 1000 & 0.275 & 0.231 & 0.273 & 0.235 & 0.229 & 0.229 & 0.42 & 0.588 & 0.391 & 0.2 & 0.323 & 0.32\\
 & 5000 & 0.155 & 0.13 & 0.171 & 0.131 & 0.134 & 0.136 & 0.271 & 0.295 & 0.243 & 0.121 & 0.183 & 0.182\\
 & 25000 & 0.074 & 0.075 & 0.101 & 0.075 & 0.084 & 0.085 & 0.173 & 0.165 & 0.152 & 0.074 & 0.13 & 0.13\\ \hline
14 & 50 & 1.079 & 1.015 & 1.015 & 1.015 & 1.038 & 1.056 & 0.484 & 0.41 & 0.625 & 0.428 & 1.239 & 1.239\\
 & 200 & 0.913 & 0.919 & 0.903 & 0.907 & 0.924 & 0.93 & 0.25 & 0.209 & 0.204 & 0.227 & 1.236 & 1.236\\
 & 1000 & 0.699 & 0.747 & 0.747 & 0.747 & 0.764 & 0.78 & 0.113 & 0.092 & 0.091 & 0.105 & 0.965 & 0.964\\
 & 5000 & 0.548 & 0.426 & 0.426 & 0.426 & 0.452 & 0.476 & 0.051 & 0.041 & 0.041 & 0.048 & 0.964 & 0.964\\
 & 25000 & 0.217 & 0.019 & 0.019 & 0.019 & 0.089 & 0.124 & 0.023 & 0.018 & 0.018 & 0.023 & 0.036 & 0.039\\ \hline
15 & 50 & 0.398 & 0.39 & 0.398 & 0.397 & 0.363 & 0.352 & 0.388 & 0.405 & 0.39 & 0.391 & 0.404 & 0.407\\
 & 200 & 0.273 & 0.249 & 0.251 & 0.244 & 0.236 & 0.234 & 0.238 & 0.242 & 0.238 & 0.224 & 0.228 & 0.221\\
 & 1000 & 0.184 & 0.141 & 0.14 & 0.134 & 0.134 & 0.133 & 0.103 & 0.104 & 0.103 & 0.129 & 0.152 & 0.146\\
 & 5000 & 0.128 & 0.07 & 0.068 & 0.068 & 0.068 & 0.068 & 0.049 & 0.05 & 0.05 & 0.063 & 0.123 & 0.12\\
 & 25000 & 0.086 & 0.03 & 0.03 & 0.03 & 0.03 & 0.03 & 0.025 & 0.024 & 0.023 & 0.027 & 0.098 & 0.094\\ \hline
16 & 50 & 0.415 & 0.416 & 0.41 & 0.41 & 0.368 & 0.359 & 0.355 & 0.374 & 0.358 & 0.356 & 0.467 & 0.468\\
 & 200 & 0.297 & 0.29 & 0.281 & 0.278 & 0.268 & 0.264 & 0.224 & 0.228 & 0.225 & 0.23 & 0.242 & 0.229\\
 & 1000 & 0.2 & 0.171 & 0.226 & 0.172 & 0.17 & 0.166 & 0.111 & 0.111 & 0.111 & 0.116 & 0.155 & 0.148\\
 & 5000 & 0.141 & 0.089 & 0.119 & 0.09 & 0.107 & 0.103 & 0.055 & 0.054 & 0.053 & 0.063 & 0.124 & 0.12\\
 & 25000 & 0.084 & 0.04 & 0.04 & 0.04 & 0.04 & 0.039 & 0.022 & 0.022 & 0.022 & 0.03 & 0.1 & 0.094\\
\bottomrule
\end{tabular}

    \end{adjustbox}
    \label{tab:complete_hellinger_risks_2}
\end{table}

\begin{table}
    \centering
    \caption{Estimated PID risks for densities $1$-$8$.}
    \vspace{0.2cm}
    \begin{adjustbox}{width=0.9\textwidth, totalheight=0.95\textheight}
    \begin{tabular}{cccccccccccccc}
\toprule
Density & n & Wand & AIC & BIC & BR & Knuth & SC & RIH & RMG-B & RMG-R & TS & L2CV & KLCV\\
\midrule
1 & 50 & 0.492 & 0.696 & 0.25 & 0.262 & 0.428 & 0.63 & 0.046 & 0.108 & 0.052 & 0.338 & 0.112 & 0.126\\
 & 200 & 0.514 & 1.09 & 0.14 & 0.19 & 0.246 & 0.28 & 0.0 & 0.052 & 0.002 & 0.012 & 0.426 & 0.388\\
 & 1000 & 1.124 & 1.842 & 0.066 & 0.356 & 0.19 & 0.176 & 0.0 & 0.044 & 0.006 & 0.0 & 4.45 & 4.136\\
 & 5000 & 3.166 & 3.582 & 0.216 & 1.538 & 0.486 & 0.426 & 0.0 & 0.034 & 0.012 & 0.002 & 24.022 & 22.716\\
 & 25000 & 7.394 & 6.938 & 0.8 & 4.326 & 1.34 & 1.112 & 0.0 & 0.034 & 0.028 & 0.004 & 80.834 & 75.788\\ \hline
2 & 50 & 1.368 & 0.654 & 0.052 & 0.072 & 0.108 & 0.34 & 0.024 & 0.168 & 0.02 & 0.368 & 0.794 & 0.864\\
 & 200 & 1.814 & 0.616 & 0.012 & 0.108 & 0.014 & 0.05 & 0.0 & 0.062 & 0.004 & 0.252 & 2.232 & 2.178\\
 & 1000 & 3.648 & 0.626 & 0.0 & 0.06 & 0.0 & 0.0 & 0.0 & 0.036 & 0.012 & 0.27 & 9.586 & 9.356\\
 & 5000 & 7.708 & 0.558 & 0.0 & 0.038 & 0.0 & 0.0 & 0.0 & 0.024 & 0.0 & 0.47 & 38.31 & 37.486\\
 & 25000 & 15.572 & 0.662 & 0.0 & 0.026 & 0.0 & 0.0 & 0.0 & 0.0 & 0.0 & 0.34 & 118.566 & 104.676\\ \hline
3 & 50 & 4.932 & 4.516 & 3.452 & 3.474 & 4.308 & 4.194 & 1.37 & 0.738 & 2.0 & 0.84 & 2.008 & 2.004\\
 & 200 & 10.596 & 7.696 & 4.824 & 6.352 & 6.968 & 6.36 & 0.6 & 0.142 & 1.988 & 0.308 & 2.242 & 2.152\\
 & 1000 & 27.918 & 19.048 & 8.706 & 16.974 & 14.674 & 11.84 & 0.0 & 0.044 & 1.132 & 0.0 & 4.298 & 3.908\\
 & 5000 & 75.49 & 54.402 & 19.5 & 50.702 & 33.506 & 26.298 & 0.0 & 0.068 & 0.004 & 0.0 & 15.022 & 13.738\\
 & 25000 & 196.452 & 154.122 & 48.028 & 147.712 & 83.588 & 63.848 & 0.0 & 0.064 & 0.008 & 0.0 & 52.39 & 40.55\\ \hline
4 & 50 & 3.082 & 3.256 & 2.9 & 2.894 & 3.132 & 3.022 & 1.592 & 1.554 & 1.8 & 1.062 & 1.94 & 1.938\\
 & 200 & 8.214 & 4.148 & 3.702 & 3.918 & 3.958 & 3.846 & 1.454 & 1.39 & 1.644 & 0.432 & 1.578 & 1.54\\
 & 1000 & 20.592 & 6.088 & 5.116 & 5.474 & 5.424 & 5.2 & 0.904 & 0.74 & 1.106 & 0.092 & 1.53 & 1.344\\
 & 5000 & 52.7 & 19.38 & 7.302 & 14.198 & 7.708 & 7.294 & 0.066 & 0.062 & 0.276 & 0.028 & 6.004 & 5.14\\
 & 25000 & 168.888 & 59.098 & 13.02 & 56.224 & 30.608 & 15.322 & 0.0 & 0.054 & 0.01 & 0.0 & 26.102 & 19.264\\ \hline
5 & 50 & 1.578 & 1.38 & 0.874 & 0.874 & 1.202 & 1.212 & 0.384 & 0.442 & 0.402 & 0.514 & 0.388 & 0.452\\
 & 200 & 2.978 & 2.382 & 0.704 & 0.936 & 1.156 & 1.018 & 0.03 & 0.114 & 0.046 & 0.026 & 0.114 & 0.088\\
 & 1000 & 7.386 & 4.472 & 1.162 & 2.536 & 1.978 & 1.676 & 0.0 & 0.034 & 0.002 & 0.0 & 1.566 & 1.298\\
 & 5000 & 17.622 & 9.266 & 2.696 & 6.878 & 4.092 & 3.446 & 0.0 & 0.05 & 0.014 & 0.0 & 14.264 & 12.85\\
 & 25000 & 38.554 & 18.642 & 5.962 & 15.826 & 8.36 & 6.824 & 0.0 & 0.076 & 0.024 & 0.0 & 60.984 & 52.118\\ \hline
6 & 50 & 2.072 & 3.29 & 1.658 & 1.754 & 2.676 & 3.314 & 0.942 & 0.752 & 1.138 & 0.83 & 1.926 & 1.892\\
 & 200 & 3.206 & 6.048 & 2.872 & 3.854 & 4.702 & 5.812 & 0.418 & 0.178 & 0.592 & 0.284 & 3.222 & 3.132\\
 & 1000 & 5.602 & 17.178 & 5.074 & 10.554 & 8.73 & 12.602 & 0.002 & 0.034 & 0.004 & 0.0 & 7.648 & 7.402\\
 & 5000 & 11.966 & 52.07 & 10.422 & 40.59 & 21.156 & 27.398 & 0.0 & 0.046 & 0.01 & 0.0 & 34.92 & 33.774\\
 & 25000 & 31.296 & 153.122 & 30.11 & 138.168 & 57.666 & 69.232 & 0.0 & 0.06 & 0.026 & 0.0 & 118.792 & 98.536\\ \hline
7 & 50 & 5.832 & 5.664 & 5.486 & 5.488 & 5.568 & 5.56 & 5.218 & 5.24 & 5.216 & 5.076 & 5.388 & 5.364\\
 & 200 & 5.996 & 5.32 & 5.516 & 5.516 & 5.306 & 5.254 & 4.26 & 4.26 & 4.28 & 3.73 & 4.818 & 4.862\\
 & 1000 & 5.548 & 7.412 & 4.338 & 4.776 & 3.988 & 3.858 & 3.276 & 1.864 & 2.0 & 0.05 & 1.202 & 0.946\\
 & 5000 & 12.228 & 16.064 & 3.204 & 12.736 & 6.936 & 5.874 & 0.0 & 0.074 & 0.01 & 0.0 & 11.762 & 10.34\\
 & 25000 & 38.996 & 30.568 & 9.886 & 26.936 & 15.29 & 13.61 & 0.0 & 0.066 & 0.028 & 0.002 & 66.768 & 57.108\\ \hline
8 & 50 & 1.376 & 1.408 & 1.356 & 1.38 & 1.37 & 1.422 & 1.162 & 1.21 & 1.206 & 1.342 & 1.31 & 1.33\\
 & 200 & 0.826 & 1.408 & 0.854 & 0.834 & 0.912 & 0.958 & 1.112 & 1.138 & 1.13 & 1.502 & 0.638 & 0.644\\
 & 1000 & 1.698 & 2.658 & 0.37 & 0.818 & 0.586 & 0.598 & 1.136 & 1.102 & 1.09 & 0.988 & 3.934 & 3.676\\
 & 5000 & 5.262 & 5.212 & 0.58 & 2.94 & 1.184 & 1.07 & 0.968 & 0.888 & 0.888 & 0.104 & 22.972 & 21.732\\
 & 25000 & 12.188 & 10.952 & 1.422 & 7.308 & 2.392 & 2.084 & 0.39 & 0.114 & 0.108 & 0.002 & 79.588 & 74.554\\
\bottomrule
\end{tabular}

    \end{adjustbox}
    \label{tab:complete_pid_risks_1}
\end{table}

\begin{table}
    \centering
    \caption{Estimated PID risks for densities $9$-$16$.}
    \vspace{0.2cm}
    \begin{adjustbox}{width=0.9\textwidth, totalheight=0.95\textheight}
    \begin{tabular}{cccccccccccccc}
\toprule
Density & n & Wand & AIC & BIC & BR & Knuth & SC & RIH & RMG-B & RMG-R & TS & L2CV & KLCV\\
\midrule
9 & 50 & 5.182 & 5.034 & 5.186 & 5.174 & 5.106 & 5.026 & 4.99 & 4.968 & 4.886 & 4.948 & 5.042 & 4.962\\
 & 200 & 4.574 & 5.034 & 4.83 & 4.682 & 4.742 & 4.692 & 4.476 & 4.394 & 4.4 & 4.368 & 3.6 & 3.52\\
 & 1000 & 4.326 & 7.814 & 3.864 & 4.642 & 4.006 & 4.066 & 2.928 & 2.482 & 2.454 & 1.698 & 3.42 & 3.448\\
 & 5000 & 7.61 & 14.008 & 3.572 & 10.462 & 5.726 & 5.338 & 0.906 & 0.726 & 0.648 & 0.13 & 18.086 & 17.208\\
 & 25000 & 20.534 & 36.982 & 6.222 & 25.948 & 9.582 & 8.848 & 0.05 & 0.056 & 0.016 & 0.0 & 73.482 & 67.678\\ \hline
10 & 50 & 10.182 & 10.856 & 10.892 & 10.846 & 10.93 & 10.864 & 10.966 & 10.838 & 10.968 & 10.0 & 10.476 & 10.522\\
 & 200 & 12.484 & 5.586 & 10.912 & 7.986 & 7.216 & 5.838 & 10.904 & 10.768 & 10.762 & 4.056 & 10.074 & 9.948\\
 & 1000 & 6.906 & 3.606 & 3.426 & 2.18 & 1.974 & 2.434 & 0.628 & 0.472 & 0.16 & 0.532 & 0.38 & 0.43\\
 & 5000 & 0.964 & 7.982 & 0.104 & 3.924 & 0.49 & 0.874 & 0.0 & 0.172 & 0.05 & 0.012 & 10.444 & 9.786\\
 & 25000 & 0.012 & 17.92 & 0.018 & 11.796 & 0.624 & 1.012 & 0.0 & 0.138 & 0.068 & 0.002 & 84.522 & 80.274\\ \hline
11 & 50 & 0.484 & 0.74 & 0.132 & 0.218 & 0.318 & 0.654 & 0.02 & 0.056 & 0.044 & 0.332 & 0.162 & 0.152\\
 & 200 & 0.378 & 0.88 & 0.082 & 0.144 & 0.17 & 0.382 & 0.002 & 0.026 & 0.0 & 0.036 & 0.976 & 0.9\\
 & 1000 & 0.462 & 1.594 & 0.026 & 0.294 & 0.148 & 0.228 & 0.0 & 0.034 & 0.008 & 0.014 & 7.152 & 6.78\\
 & 5000 & 0.812 & 3.098 & 0.04 & 1.058 & 0.238 & 0.31 & 0.0 & 0.034 & 0.01 & 0.002 & 33.902 & 32.454\\
 & 25000 & 1.498 & 6.306 & 0.04 & 2.596 & 0.258 & 0.35 & 0.002 & 0.028 & 0.02 & 0.002 & 108.578 & 102.754\\ \hline
12 & 50 & 2.846 & 2.534 & 2.582 & 2.586 & 2.53 & 2.49 & 0.296 & 0.396 & 0.272 & 0.82 & 1.508 & 1.444\\
 & 200 & 0.526 & 0.384 & 0.392 & 0.36 & 0.36 & 0.362 & 0.044 & 0.112 & 0.06 & 0.212 & 0.978 & 1.024\\
 & 1000 & 0.524 & 1.72 & 0.198 & 1.08 & 0.684 & 0.268 & 0.0 & 0.044 & 0.002 & 0.002 & 0.034 & 0.19\\
 & 5000 & 3.652 & 3.784 & 0.866 & 2.86 & 1.35 & 0.98 & 0.0 & 0.106 & 0.002 & 0.002 & 3.544 & 3.274\\
 & 25000 & 16.588 & 8.876 & 1.502 & 7.492 & 2.96 & 1.934 & 0.0 & 0.054 & 0.014 & 0.002 & 29.402 & 26.248\\ \hline
13 & 50 & 4.708 & 5.412 & 5.526 & 5.524 & 5.428 & 5.378 & 4.738 & 4.492 & 5.43 & 3.356 & 5.464 & 5.608\\
 & 200 & 5.01 & 5.584 & 4.762 & 5.136 & 5.414 & 5.38 & 1.058 & 1.232 & 0.802 & 0.588 & 3.264 & 3.322\\
 & 1000 & 4.33 & 15.188 & 3.942 & 12.916 & 11.226 & 10.878 & 0.0 & 0.138 & 0.008 & 0.044 & 2.706 & 2.2\\
 & 5000 & 16.696 & 42.412 & 12.32 & 36.552 & 25.872 & 25.232 & 0.0 & 0.12 & 0.016 & 0.006 & 17.222 & 15.536\\
 & 25000 & 64.008 & 77.988 & 26.232 & 71.332 & 41.472 & 40.33 & 0.0 & 0.116 & 0.042 & 0.006 & 83.18 & 67.84\\ \hline
14 & 50 & 4.886 & 4.0 & 4.0 & 4.0 & 4.0 & 4.0 & 0.012 & 0.386 & 0.786 & 0.728 & 4.0 & 4.0\\
 & 200 & 4.962 & 4.568 & 4.0 & 4.148 & 4.26 & 4.032 & 0.008 & 0.288 & 0.0 & 1.062 & 4.0 & 4.0\\
 & 1000 & 4.884 & 5.194 & 5.182 & 5.194 & 5.194 & 5.194 & 0.0 & 0.034 & 0.0 & 1.378 & 2.384 & 2.372\\
 & 5000 & 7.358 & 3.388 & 3.388 & 3.388 & 3.388 & 3.388 & 0.0 & 0.004 & 0.0 & 1.62 & 3.61 & 3.602\\
 & 25000 & 11.128 & 5.914 & 5.914 & 5.914 & 5.914 & 5.914 & 0.0 & 0.004 & 0.0 & 2.332 & 8.258 & 7.356\\ \hline
15 & 50 & 2.142 & 2.07 & 2.114 & 2.124 & 2.108 & 2.084 & 1.804 & 1.85 & 1.844 & 1.824 & 2.794 & 2.818\\
 & 200 & 2.372 & 3.648 & 1.37 & 1.86 & 2.528 & 2.756 & 1.858 & 1.89 & 1.842 & 1.484 & 1.532 & 1.482\\
 & 1000 & 7.056 & 7.612 & 0.896 & 3.216 & 2.34 & 2.616 & 1.98 & 2.008 & 1.96 & 1.324 & 6.514 & 6.224\\
 & 5000 & 17.35 & 7.138 & 0.052 & 1.78 & 0.38 & 0.384 & 1.914 & 1.75 & 1.738 & 0.482 & 30.226 & 28.682\\
 & 25000 & 38.954 & 1.576 & 0.0 & 0.222 & 0.0 & 0.0 & 0.506 & 0.19 & 0.178 & 0.164 & 102.66 & 91.492\\ \hline
16 & 50 & 4.218 & 3.294 & 3.502 & 3.52 & 3.314 & 3.29 & 2.758 & 2.762 & 2.912 & 2.138 & 4.448 & 4.472\\
 & 200 & 2.68 & 4.886 & 2.436 & 2.84 & 3.61 & 3.56 & 2.896 & 2.882 & 2.874 & 1.798 & 2.114 & 2.044\\
 & 1000 & 7.692 & 14.472 & 3.706 & 12.306 & 11.216 & 11.548 & 1.738 & 1.522 & 1.524 & 1.192 & 5.562 & 5.292\\
 & 5000 & 19.014 & 28.098 & 12.53 & 25.878 & 17.404 & 18.646 & 0.686 & 0.542 & 0.436 & 1.104 & 29.646 & 27.908\\
 & 25000 & 44.898 & 28.788 & 28.602 & 28.694 & 28.602 & 28.602 & 0.0 & 0.014 & 0.008 & 0.79 & 106.614 & 89.904\\
\bottomrule
\end{tabular}

    \end{adjustbox}
    \label{tab:complete_pid_risks_2}
\end{table}

\begin{table}
    \centering
    \caption{Estimated $\mathbb{L}_2$ risks for densities $1$-$8$ (where applicable).}
    \vspace{0.2cm}
    \begin{adjustbox}{width=0.9\textwidth, totalheight=0.8\textheight}
    \begin{tabular}{cccccccccccccc}
\toprule
Density & n & Wand & AIC & BIC & BR & Knuth & SC & RIH & RMG-B & RMG-R & TS & L2CV & KLCV\\
\midrule
1 & 50 & 0.165 & 0.194 & 0.195 & 0.19 & 0.195 & 0.189 & 0.22 & 0.32 & 0.217 & 0.234 & 0.197 & 0.197\\
 & 200 & 0.106 & 0.122 & 0.128 & 0.123 & 0.121 & 0.119 & 0.15 & 0.172 & 0.15 & 0.119 & 0.145 & 0.145\\
 & 1000 & 0.062 & 0.07 & 0.081 & 0.071 & 0.074 & 0.074 & 0.099 & 0.103 & 0.096 & 0.068 & 0.115 & 0.115\\
 & 5000 & 0.038 & 0.04 & 0.051 & 0.042 & 0.046 & 0.047 & 0.063 & 0.063 & 0.06 & 0.039 & 0.097 & 0.097\\
 & 25000 & 0.022 & 0.024 & 0.032 & 0.024 & 0.028 & 0.029 & 0.04 & 0.037 & 0.037 & 0.022 & 0.071 & 0.076\\ \hline
2 & 50 & 0.309 & 0.308 & 0.209 & 0.218 & 0.214 & 0.243 & 0.197 & 0.33 & 0.198 & 0.259 & 0.362 & 0.356\\
 & 200 & 0.197 & 0.154 & 0.097 & 0.108 & 0.098 & 0.102 & 0.095 & 0.134 & 0.096 & 0.117 & 0.325 & 0.323\\
 & 1000 & 0.117 & 0.069 & 0.043 & 0.05 & 0.043 & 0.044 & 0.043 & 0.052 & 0.044 & 0.053 & 0.288 & 0.287\\
 & 5000 & 0.071 & 0.029 & 0.019 & 0.021 & 0.019 & 0.019 & 0.018 & 0.024 & 0.018 & 0.024 & 0.253 & 0.251\\
 & 25000 & 0.044 & 0.014 & 0.009 & 0.01 & 0.009 & 0.009 & 0.008 & 0.008 & 0.008 & 0.01 & 0.205 & 0.196\\ \hline
4 & 50 & 0.195 & 0.224 & 0.253 & 0.254 & 0.227 & 0.233 & 0.248 & 0.363 & 0.252 & 0.216 & 0.255 & 0.262\\
 & 200 & 0.147 & 0.165 & 0.202 & 0.184 & 0.178 & 0.189 & 0.179 & 0.196 & 0.184 & 0.136 & 0.174 & 0.177\\
 & 1000 & 0.101 & 0.145 & 0.173 & 0.152 & 0.158 & 0.168 & 0.125 & 0.122 & 0.128 & 0.11 & 0.137 & 0.137\\
 & 5000 & 0.06 & 0.12 & 0.16 & 0.136 & 0.155 & 0.159 & 0.08 & 0.073 & 0.084 & 0.063 & 0.095 & 0.094\\
 & 25000 & 0.037 & 0.064 & 0.157 & 0.067 & 0.116 & 0.151 & 0.051 & 0.045 & 0.054 & 0.037 & 0.049 & 0.047\\ \hline
5 & 50 & 0.176 & 0.196 & 0.2 & 0.2 & 0.195 & 0.19 & 0.225 & 0.313 & 0.229 & 0.218 & 0.198 & 0.201\\
 & 200 & 0.114 & 0.127 & 0.152 & 0.142 & 0.136 & 0.138 & 0.166 & 0.222 & 0.165 & 0.129 & 0.134 & 0.134\\
 & 1000 & 0.069 & 0.08 & 0.108 & 0.089 & 0.094 & 0.098 & 0.111 & 0.11 & 0.107 & 0.075 & 0.084 & 0.084\\
 & 5000 & 0.041 & 0.049 & 0.075 & 0.054 & 0.064 & 0.069 & 0.072 & 0.069 & 0.069 & 0.044 & 0.064 & 0.064\\
 & 25000 & 0.024 & 0.03 & 0.051 & 0.032 & 0.043 & 0.047 & 0.045 & 0.041 & 0.043 & 0.025 & 0.048 & 0.051\\ \hline
7 & 50 & 0.277 & 0.291 & 0.289 & 0.287 & 0.287 & 0.281 & 0.286 & 0.384 & 0.28 & 0.329 & 0.286 & 0.284\\
 & 200 & 0.237 & 0.216 & 0.233 & 0.229 & 0.224 & 0.221 & 0.219 & 0.239 & 0.22 & 0.243 & 0.235 & 0.236\\
 & 1000 & 0.171 & 0.134 & 0.174 & 0.139 & 0.142 & 0.143 & 0.19 & 0.188 & 0.175 & 0.135 & 0.146 & 0.146\\
 & 5000 & 0.082 & 0.08 & 0.119 & 0.085 & 0.102 & 0.106 & 0.109 & 0.109 & 0.107 & 0.071 & 0.099 & 0.099\\
 & 25000 & 0.043 & 0.047 & 0.076 & 0.049 & 0.063 & 0.067 & 0.071 & 0.066 & 0.066 & 0.039 & 0.076 & 0.079\\ \hline
8 & 50 & 0.184 & 0.209 & 0.199 & 0.197 & 0.2 & 0.198 & 0.2 & 0.311 & 0.194 & 0.219 & 0.207 & 0.206\\
 & 200 & 0.117 & 0.132 & 0.133 & 0.129 & 0.128 & 0.127 & 0.138 & 0.168 & 0.137 & 0.13 & 0.152 & 0.151\\
 & 1000 & 0.071 & 0.077 & 0.09 & 0.08 & 0.082 & 0.082 & 0.099 & 0.101 & 0.095 & 0.082 & 0.117 & 0.117\\
 & 5000 & 0.042 & 0.045 & 0.058 & 0.046 & 0.051 & 0.052 & 0.06 & 0.059 & 0.058 & 0.039 & 0.097 & 0.097\\
 & 25000 & 0.025 & 0.026 & 0.036 & 0.027 & 0.032 & 0.033 & 0.04 & 0.037 & 0.037 & 0.022 & 0.073 & 0.076\\
\bottomrule
\end{tabular}

    \end{adjustbox}
    \label{tab:complete_l2_risks_1}
\end{table}

\begin{table}
    \centering
    \caption{Estimated $\mathbb{L}_2$ risks for densities $9$-$16$.}
    \vspace{0.2cm}
    \begin{adjustbox}{width=0.9\textwidth, totalheight=0.95\textheight}
    \begin{tabular}{cccccccccccccc}
\toprule
Density & n & Wand & AIC & BIC & BR & Knuth & SC & RIH & RMG-B & RMG-R & TS & L2CV & KLCV\\
\midrule
9 & 50 & 0.215 & 0.241 & 0.233 & 0.231 & 0.235 & 0.231 & 0.234 & 0.32 & 0.231 & 0.256 & 0.23 & 0.231\\
 & 200 & 0.169 & 0.174 & 0.177 & 0.172 & 0.172 & 0.169 & 0.188 & 0.233 & 0.185 & 0.19 & 0.183 & 0.182\\
 & 1000 & 0.117 & 0.11 & 0.127 & 0.11 & 0.114 & 0.112 & 0.134 & 0.139 & 0.127 & 0.114 & 0.129 & 0.129\\
 & 5000 & 0.074 & 0.065 & 0.081 & 0.066 & 0.071 & 0.072 & 0.081 & 0.081 & 0.079 & 0.057 & 0.101 & 0.1\\
 & 25000 & 0.044 & 0.041 & 0.053 & 0.041 & 0.048 & 0.049 & 0.054 & 0.051 & 0.05 & 0.031 & 0.076 & 0.078\\ \hline
10 & 50 & 0.145 & 0.149 & 0.136 & 0.137 & 0.137 & 0.142 & 0.136 & 0.173 & 0.135 & 0.209 & 0.152 & 0.151\\
 & 200 & 0.131 & 0.115 & 0.128 & 0.123 & 0.118 & 0.11 & 0.129 & 0.136 & 0.129 & 0.166 & 0.136 & 0.136\\
 & 1000 & 0.123 & 0.071 & 0.082 & 0.071 & 0.07 & 0.069 & 0.09 & 0.105 & 0.089 & 0.082 & 0.08 & 0.08\\
 & 5000 & 0.069 & 0.042 & 0.049 & 0.041 & 0.043 & 0.042 & 0.064 & 0.061 & 0.058 & 0.043 & 0.058 & 0.058\\
 & 25000 & 0.027 & 0.025 & 0.029 & 0.024 & 0.025 & 0.025 & 0.041 & 0.038 & 0.037 & 0.024 & 0.047 & 0.047\\ \hline
11 & 50 & 0.256 & 0.313 & 0.306 & 0.299 & 0.312 & 0.307 & 0.323 & 0.364 & 0.326 & 0.347 & 0.309 & 0.308\\
 & 200 & 0.164 & 0.191 & 0.19 & 0.185 & 0.184 & 0.183 & 0.229 & 0.296 & 0.226 & 0.2 & 0.247 & 0.246\\
 & 1000 & 0.097 & 0.111 & 0.116 & 0.108 & 0.11 & 0.108 & 0.156 & 0.16 & 0.151 & 0.117 & 0.208 & 0.207\\
 & 5000 & 0.058 & 0.064 & 0.069 & 0.062 & 0.065 & 0.064 & 0.1 & 0.098 & 0.095 & 0.071 & 0.182 & 0.181\\
 & 25000 & 0.034 & 0.038 & 0.041 & 0.036 & 0.038 & 0.038 & 0.063 & 0.06 & 0.058 & 0.043 & 0.141 & 0.146\\ \hline
12 & 50 & 0.335 & 0.302 & 0.32 & 0.317 & 0.3 & 0.306 & 0.23 & 0.348 & 0.239 & 0.253 & 0.393 & 0.393\\
 & 200 & 0.216 & 0.203 & 0.219 & 0.207 & 0.205 & 0.209 & 0.188 & 0.216 & 0.188 & 0.16 & 0.355 & 0.355\\
 & 1000 & 0.129 & 0.119 & 0.178 & 0.125 & 0.131 & 0.156 & 0.122 & 0.119 & 0.12 & 0.084 & 0.128 & 0.132\\
 & 5000 & 0.067 & 0.07 & 0.107 & 0.074 & 0.088 & 0.099 & 0.079 & 0.074 & 0.076 & 0.048 & 0.066 & 0.066\\
 & 25000 & 0.034 & 0.041 & 0.066 & 0.043 & 0.055 & 0.062 & 0.049 & 0.044 & 0.047 & 0.027 & 0.046 & 0.047\\ \hline
13 & 50 & 0.162 & 0.15 & 0.16 & 0.16 & 0.148 & 0.146 & 0.15 & 0.224 & 0.155 & 0.205 & 0.162 & 0.162\\
 & 200 & 0.145 & 0.121 & 0.138 & 0.126 & 0.121 & 0.12 & 0.115 & 2.822 & 0.116 & 0.097 & 0.143 & 0.142\\
 & 1000 & 0.097 & 0.071 & 0.096 & 0.074 & 0.076 & 0.077 & 0.085 & 1.185 & 0.08 & 0.049 & 0.094 & 0.095\\
 & 5000 & 0.058 & 0.038 & 0.057 & 0.04 & 0.047 & 0.048 & 0.054 & 0.184 & 0.049 & 0.027 & 0.044 & 0.044\\
 & 25000 & 0.024 & 0.022 & 0.039 & 0.023 & 0.031 & 0.031 & 0.034 & 0.038 & 0.031 & 0.015 & 0.027 & 0.027\\ \hline
14 & 50 & 1.129 & 1.124 & 1.124 & 1.124 & 1.124 & 1.125 & 0.521 & 0.88 & 0.756 & 0.676 & 1.161 & 1.161\\
 & 200 & 1.074 & 1.096 & 1.095 & 1.095 & 1.095 & 1.095 & 0.266 & 0.371 & 0.269 & 0.358 & 1.16 & 1.16\\
 & 1000 & 0.904 & 1.003 & 1.003 & 1.003 & 1.003 & 1.004 & 0.119 & 0.122 & 0.119 & 0.162 & 1.116 & 1.116\\
 & 5000 & 0.758 & 0.63 & 0.63 & 0.63 & 0.63 & 0.632 & 0.053 & 0.053 & 0.052 & 0.074 & 1.116 & 1.116\\
 & 25000 & 0.255 & 0.027 & 0.027 & 0.027 & 0.028 & 0.032 & 0.024 & 0.023 & 0.023 & 0.038 & 0.04 & 0.04\\ \hline
15 & 50 & 0.783 & 0.704 & 0.766 & 0.765 & 0.69 & 0.662 & 0.753 & 1.199 & 0.75 & 0.829 & 0.809 & 0.816\\
 & 200 & 0.556 & 0.465 & 0.484 & 0.465 & 0.455 & 0.451 & 0.395 & 0.487 & 0.401 & 0.424 & 0.425 & 0.402\\
 & 1000 & 0.393 & 0.271 & 0.278 & 0.263 & 0.264 & 0.263 & 0.172 & 0.189 & 0.175 & 0.224 & 0.314 & 0.304\\
 & 5000 & 0.274 & 0.139 & 0.135 & 0.134 & 0.134 & 0.134 & 0.09 & 0.093 & 0.091 & 0.114 & 0.263 & 0.26\\
 & 25000 & 0.189 & 0.061 & 0.06 & 0.06 & 0.06 & 0.06 & 0.046 & 0.042 & 0.042 & 0.055 & 0.187 & 0.209\\ \hline
16 & 50 & 0.857 & 0.771 & 0.795 & 0.795 & 0.745 & 0.729 & 0.761 & 1.309 & 0.763 & 0.887 & 1.001 & 1.007\\
 & 200 & 0.644 & 0.579 & 0.606 & 0.588 & 0.576 & 0.573 & 0.47 & 0.558 & 0.472 & 0.513 & 0.51 & 0.477\\
 & 1000 & 0.462 & 0.348 & 0.53 & 0.361 & 0.373 & 0.367 & 0.236 & 0.249 & 0.232 & 0.257 & 0.328 & 0.315\\
 & 5000 & 0.331 & 0.187 & 0.279 & 0.192 & 0.248 & 0.239 & 0.102 & 0.104 & 0.1 & 0.142 & 0.267 & 0.262\\
 & 25000 & 0.197 & 0.084 & 0.084 & 0.084 & 0.084 & 0.084 & 0.043 & 0.043 & 0.043 & 0.071 & 0.182 & 0.212\\
\bottomrule
\end{tabular}

    \end{adjustbox}
    \label{tab:complete_l2_risks_2}
\end{table}

\newpage
\bibliographystyle{plainnat-modified}
\bibliography{bibliography}

\end{document}